\documentclass[aps,10pt,nofootinbib,floatfix]{revtex4}
\usepackage{amsmath}
\usepackage{graphicx}
\usepackage{slashed}
\begin{document}
\title{Living Without Supersymmetry -- the Conformal Alternative and a Dynamical Higgs Boson}
\author{Philip D. Mannheim}
\affiliation{Department of Physics, University of Connecticut, Storrs, CT 06269, USA.
email: philip.mannheim@uconn.edu}
\date{May 13, 2017}
\begin{abstract}
We show that the key results of supersymmetry can be achieved via conformal symmetry instead. We propose that the Higgs boson be a dynamical fermion-antifermion bound state rather than an elementary scalar field, so that there is then no quadratically divergent self-energy problem for it and thus no need to invoke supersymmetry to resolve the problem. To obtain such a dynamical Higgs boson we study a conformal invariant gauge theory of interacting fermions and gauge bosons. The conformal invariance of the theory is realized via scaling with anomalous dimensions in the ultraviolet, and by a dynamical symmetry breaking via fermion bilinear condensates in the infrared, a breaking  in which the dynamical dimension of the composite operator $\bar{\psi}\psi$ is reduced from three to two. With this reduction in dimension we can augment the gauge theory with a  four-fermion interaction made renormalizable by this reduction, and can reinterpret the theory as a renormalizable version of the Nambu-Jona-Lasinio model, with the gauge theory sector with its now massive fermion being a mean-field theory and the four-fermion interaction being the residual interaction. It is this residual interaction and not the mean field that then generates dynamical Goldstone and Higgs states, states that, as noted by Baker and Johnson, the gauge theory sector itself  does not possess. The Higgs boson is found to be a narrow resonance just above threshold, with its width potentially being a diagnostic that could distinguish a dynamical Higgs boson from an elementary one. We couple the theory to a gravity theory, conformal gravity, that is equally conformal invariant, with the interplay between conformal gravity and the four-fermion interaction taking care of the vacuum energy problem.  With conformal gravity being a unitary and renormalizable quantum theory of gravity there is no need for string theory with its supersymmetric underpinnings. With the vacuum energy problem being resolved and with conformal gravity fits to phenomena such as galactic rotation curves and the accelerating universe not needing dark matter, there is no need to introduce supersymmetry for either the vacuum energy problem or to provide a potential dark matter candidate. We propose that it is conformal symmetry rather than supersymmetry that is fundamental, with the theory of nature being a locally conformal, locally gauge invariant, non-Abelian Nambu-Jona-Lasinio theory.
\end{abstract}
\maketitle

\section{Introduction}

The assumption of a supersymmetry between bosons and fermions has been found capable of addressing many key issues in particle physics and gravity (see e.g. \cite{Wess1992,Weinberg2000,Polchinski1998,Shifman2012}).  In flat space physics  an interplay between bosons and fermions  can render logarithmically divergent Feynman diagrams finite. Similarly,  an interplay between bosons and fermions can cancel the perturbative quadratic divergence that an elementary Higgs scalar field would possess (the hierarchy problem). In addition, the existence of fermionic supersymmetry generators allows one to evade the Coleman-Mandula theorem \cite{Coleman1967} that forbids the combining of spacetime and bosonic internal symmetry generators in a common Lie algebra. And with the inclusion of supersymmetry one can potentially achieve a unification of the coupling constants of $SU(3)\times SU(2)_{\rm L}\times U(1)$ at a grand unified energy scale. In the presence of gravity an interplay between bosons and fermions can cancel the quartic divergence in the vacuum energy.  Cancellation of perturbative infinities  can also be found in supergravity, the local version of supersymmetry. With supersymmetry one can construct a consistent candidate quantum theory of gravity, string theory, which permits a possible unification of all of the fundamental forces and a metrication (geometrization) of them. Finally, with supersymmetry one has a prime candidate for dark matter. 

Despite this quite extensive theoretical inventory, actual experimental detection of any of the required superpartners  of the standard fermions and bosons has so far proven elusive. Now until quite recently one could account for such non-detection by endowing the superpartners with ever higher masses or ever weaker couplings to ordinary matter. However, in order to cancel the quadratic self-energy divergence that an elementary Higgs field would have, one would need a supersymmetric particle with a mass reasonably close to that of the Higgs boson itself. And  now that the Higgs boson has been discovered at the Large Hadron Collider (LHC) and its 125 GeV mass has now been established \cite{ATLAS2012,CMS2012}, one should thus anticipate finding a superparticle at the LHC in the same 125 GeV mass region. However, no evidence for any such particle has emerged in an exploration of this mass region, or in decays such as $B^0_{\rm s}\rightarrow \mu^{+}+\mu^{-}$  that were thought to be particularly favorable for supersymmetry \cite{Aaij2013,Khachatryan2015}.  Not only was no evidence for supersymmetry found in the original  LHC run 1 at a beam center of mass energy of 7 to 8 TeV, at the even higher energies in the subsequent LHC run 2 at a beam center of mass energy of 13 TeV, and with a far more extensive search of possible relevant channels, no sign of any superparticles up to masses quite significantly above 125 GeV was found at all.\footnote{Run 2 data presented by the ATLAS, CMS, and LHCb collaborations at the 38th International Conference on High Energy Physics in Chicago in August 2016 may be found at www.ichep2016.org and in the conference proceedings.} And while the superparticle search at the LHC is still ongoing, and while the so far unsuccessful underground dark matter searches are continuing, nonetheless the situation is disquieting enough that one should at least contemplate whether it might be possible to dispense with supersymmetry altogether. If one is to consider doing so however, then one must seek an alternative to supersymmetry that has the potential to also achieve its key successes. In this paper we present such a candidate alternative, one that is also based on a  symmetry, namely conformal symmetry.

Since the Higgs self-energy problem is currently the most pressing concern for supersymmetry, in this paper we shall concentrate on the issue of generating the Higgs boson dynamically, since one then no longer encounters  the quadratic self-energy divergence problem that is associated with an elementary Higgs scalar field. Moreover, independent of any supersymmetry considerations, now that the Higgs boson has been shown to exist, it anyway becomes imperative to ascertain whether it is elementary or composite, and determine whether or not a fundamental, double-well Higgs potential is to actually be present in the fundamental action of nature. In the present paper we will show that in a conformal or scale invariant theory as realized via critical scaling with anomalous dimensions there is dynamical symmetry breaking via a fermion bilinear condensate, with a fermionic mass and dynamical Higgs and Goldstone particles being generated, and with the mass of the Higgs boson naturally being of the same order as that of the fermion, rather than being altogether larger. Moreover, if conformal symmetry is to be exact at the level of the Lagrangian and to only be broken in the vacuum (giving the dimensionful $\bar{\psi}\psi$ composite operator a vacuum expectation value would break both chiral and scale symmetry), the presence of any dimensionful tachyonic $-\mu^2\phi^2$ term in an elementary Higgs field Lagrangian would expressly be forbidden. With the dynamical Higgs boson that we find being not a below-threshold bound state but a narrow resonance that lies just above the threshold in the fermion-antifermion scattering amplitude, its width could potentially be a diagnostic that could distinguish a dynamical Higgs boson from an elementary  one.

In order to develop the background needed to establish our results,  given that our work is based on earlier work from quite some time ago, for the benefit of the reader and to make the present paper self-contained, in Sec. II we briefly review the antecedents of our current work, antecedents that originated in the work of Johnson, Baker, and Willey \cite{Johnson1964,Johnson1967,Baker1969,Baker1971a,Baker1971b,Johnson1973} and the present author \cite{Mannheim1974a,Mannheim1975,Mannheim1978} on critical scaling in quantum electrodynamics (QED) in the 1960s and 1970s. In these studies it was found that the bare fermion mass scaled as 
\begin{eqnarray}
m_0=m\left(\frac{\Lambda^2}{m^2}\right)^{\gamma(\alpha)/2}, 
\label{L1}
\end{eqnarray}
and would thus vanish in the limit of infinite cutoff if the $\alpha$-dependent scaling power $\gamma(\alpha)$ is negative. With the physical mass $m$ being non-zero, this would initially suggest that the fermion mass is generated via dynamical chiral symmetry breaking. However, this turned out not to be the case since as shown by Baker and Johnson \cite{Baker1971a} such a scaling behavior for $m_0$ actually corresponds  to the presence of a non-zero bare mass term $m_0(\bar{\psi}\psi)_0$ in the Lagrangian, with the chiral symmetry thus being broken ab initio at the level of the Lagrangian itself, and with there being no accompanying pseudoscalar massless Goldstone boson. This phenomenon is known in the literature as the Baker-Johnson evasion of the Goldstone theorem, with there being no Goldstone boson in a QED theory in which the fermion propagator scales asymptotically with an anomalous dimension.

Despite this, one of the key new points of this paper will be to connect the work of Johnson, Baker and Willey (JBW) to dynamical chiral symmetry breaking and massless Goldstone boson generation after all. To this end in Sec. III we discuss mass generation in the  Nambu-Jona-Lasinio (NJL) model \cite{Nambu1961}, and in Sec. IV we adapt the analysis to the JBW critical scaling case. In particular, we show that the non-chiral invariant JBW electrodynamics with its ab initio $m_0(\bar{\psi}\psi)_0$ term is actually the mean-field approximation to a theory that is chiral invariant, namely QED with a massless fermion coupled to a four-fermion interaction. In such a situation, and just as in the NJL model, the mean field possesses no Goldstone boson because a mean-field sector never does. Rather, it is the residual interaction that accompanies the mean field that generates a dynamical pseudoscalar Goldstone bound state, one that because of the underlying chiral symmetry is accompanied by a dynamical scalar Higgs boson. We thus embed massless electrodynamics in a larger theory that is chiral invariant by augmenting it with a four-fermion interaction so that the full Lagrangian is chirally symmetric, and then break the chiral symmetry dynamically so that the theory can  be decomposed into two sectors, a mean-field sector and a residual interaction sector. Neither of these  two sectors is separately chiral invariant, it is only their sum that is. The mean-field sector is thus recognized as the non-chiral-invariant JBW electrodynamics theory with its intrinsic $m_0(\bar{\psi}\psi)_0$ term, and it is the residual interaction that then generates a massless Goldstone boson. Then, because of the underlying chiral symmetry the residual interaction has to generate a massive dynamical scalar Higgs boson as well. Thus by reinterpreting JBW electrodynamics as a mean-field theory, we are not only able to connect it to massless Goldstone boson generation, we are able to provide a new calculational scheme for generating Higgs bosons dynamically as well.

In our discussion below we follow \cite{Mannheim1974a,Mannheim1975,Mannheim1978} and augment electrodynamics with a four-fermion interaction, one that we show here is made renormalizable by the very same dynamical symmetry breaking mechanism that generates the dynamical Goldstone and Higgs bosons. As originally proposed by the present author, this four-fermion interaction was introduced for other two reasons: to cancel infinities in the vacuum (zero-point) energy density, infinities that one ordinarily would normal order away, and to facilitate the development of a formalism for treating symmetry breaking by fermion bilinears. (Contemporaneous with the author equivalent results on symmetry breaking by composite operators were obtained by Cornwall, Jackiw, and Tomboulis \cite{Cornwall1974}). The central theme of the present paper is to show that this very same four-fermion interaction also serves to provide a residual interaction that then generates the Goldstone and Higgs bosons that are not present in JBW electrodynamics itself. Augmenting JBW electrodynamics with a  four-fermion interaction thus enables us to evade the Baker-Johnson evasion of the Goldstone theorem, i.e. to nonetheless obtain a Goldstone boson in a theory in which the fermion propagator does scale asymptotically with an anomalous dimension. Now from the perspective of flat space physics there is no particular need to cancel such vacuum energy infinities since energies are not observable, only energy differences. However, once one couples to gravity one cannot throw away any contribution to the vacuum energy since the hallmark of Einstein's formulation of gravity is that gravity is to couple to every form of energy and not  just to energy differences.  Thus once we extend conformal invariance to gravity, which we do, we then cannot ignore infinities in the vacuum energy, and they have to be canceled. It is thus gravity that will force the four-fermion interaction (the only choice that does not involve the introduction of fields other than those already in the QED action) and its associated Goldstone and Higgs bosons upon us, and as such  the four-fermion interaction would itself, just as we find,  need to become renormalizable so that it would not destroy renormalizability. However, to cancel the vacuum energy density  infinities completely we will need to include not just the four-fermion contribution but also the contribution of  quantum conformal gravity itself. We discuss these vacuum energy issues  in Sec. IV and in Sec. V. 

Also in Sec. IV we compare our work on dynamical symmetry breaking with other approaches that have appeared in the literature. In particular, we contrast our work with studies of the quenched, planar graph, ladder approximation to the Abelian gluon model, and show that we obtain dynamical Goldstone and Higgs bosons in the weak coupling regime where, based on these quenched ladder studies,  it is generally thought that no generation of dynamical bound states could occur.  While these same quenched ladder studies show that one can obtain dynamical Goldstone and Higgs bosons in the strong coupling limit since in going from weak to strong coupling a phase transition occurs, in Sec. IV  we call this wisdom into question by showing that it is not actually valid to use the quenched ladder approximation in the strong coupling regime as the non-planar graphs that are left out are not only just as big as the planar ones that are included, they serve to eliminate the phase transition altogether, to thereby eliminate the rationale for requiring strong coupling in the first place. Finally, in Sec. V we show that conformal symmetry can achieve all of the key results of supersymmetry, and thus essentially supplant it as a candidate fundamental symmetry of nature. In this section we present a candidate model for the action of the universe, one based on conformal gravity, a non-Abelian Yang-Mills gauge theory, and a four-fermion interaction as made renormalizable by critical scaling and anomalous dimensions. We show that the model is not only renormalizable, it is even finite.

\section{JBW Electrodynamics}

\subsection{Vanishing of the Bare Fermion Mass}

In order to explore whether the Higgs field might be dynamical we need a tractable calculational scheme in which one can study dynamical symmetry breaking  via fermion bilinear condensates non-perturbatively.  To this end we adapt some earlier work of the present author from the 1970s. This earlier work was itself based on even earlier work of Johnson, Baker, and Willey from the 1960s on quantum electrodynamics. The objective of the study of Johnson, Baker, and Willey was to determine whether it might be possible for all the renormalization constants of a quantum field theory to be finite. Quantum electrodynamics was a particularly convenient theory to study since its gauge structure meant that two of its renormalization constants (the fermion-antifermion-gauge boson vertex renormalization constant $Z_1$ and the fermion wave function renormalization constant $Z_2$ to which $Z_1$  is equal) were gauge dependent and could be made finite by an appropriate choice of gauge, with the anomalous dimension of the fermion $\gamma_{\rm F}$ consequently then being zero -- and for convenience  in the following we shall set $Z_1$ and  $Z_2$ equal to one. Johnson, Baker, and Willey were thus left with the gauge boson wave function renormalization constant $Z_3$ and the fermion bare mass $m_0$ and its shift $\delta m$ to address. 

Now if one were also to consider the coupling of electrodynamics to gravity, one would then have to address another infinity that electrodynamics possesses, namely that of the zero-point vacuum energy density, and one of the objectives of the present paper is to address this issue. Since Johnson, Baker, and Willey were considering electrodynamics in flat space, the need to address the vacuum energy density infinity did not arise in their study as it could be normal ordered away.  

As regards $Z_3$ and $m_0$, Johnson, Baker, and Willey showed that $Z_3$ would be finite if the fermion-antifermion-gauge boson coupling constant $\alpha$ was at a solution to the Gell-Mann-Low eigenvalue condition -- and for convenience  in the following we shall set $Z_3$ equal to one. At this eigenvalue they showed that the bare mass scaled as $m_0=m\left(\Lambda^2/m^2\right)^{\gamma(\alpha)/2}$, where $\Lambda$ is an ultraviolet cutoff and $m=m_0+\delta m$ is the renormalized fermion mass. Consequently if the power $\gamma(\alpha)$ is negative, which it is perturbatively [$\gamma(\alpha)=-3\alpha/2\pi-3\alpha^2/16\pi^2+{\rm O}(\alpha^3)$], the bare mass would vanish in the limit of infinite cutoff and $\delta m$ would be finite. As such, the work of Johnson, Baker, and Willey was quite remarkable since it predated the work of Wilson and of Callan and Symanzik on critical scaling and the renormalization group. 

Subsequently, Adler and Bardeen \cite{Adler1971,Adler1972} recast the work of Johnson, Baker, and Willey in the language of the renormalization group itself, and showed that  the  renormalized inverse fermion propagator $\tilde{S}^{-1}(p,m)$ and the renormalized vertex function $\tilde{\Gamma}_{\rm S}(p,p,0,m)$ associated with the insertion of  the composite operator $\theta=\bar{\psi}\psi$ with zero momentum into the fermion propagator were related by 
\begin{eqnarray}
\left[m\frac{\partial}{m}+\beta(\alpha)\frac{\partial}{\partial \alpha}\right]\tilde{S}^{-1}(p,m)=-m[1-\gamma_{\theta}(\alpha)]\tilde{\Gamma}_{\rm S}(p,p,0,m)
\label{L2}
\end{eqnarray}
in the limit in which the fermion momentum $p_{\mu}$ is deep Euclidean and the fermionic wave function anomalous dimension $\gamma_{\rm F}$ is zero. In the critical scaling situation where $\beta(\alpha)=0$ this equation admits of an exact asymptotic solution
\begin{eqnarray}
\tilde{S}^{-1}(p,m)&=& \slashed{p}-m\left(\frac{-p^2-i\epsilon}{m^2}\right)^{\gamma_{\theta}(\alpha)/2}+i\epsilon,
\nonumber\\
\tilde{\Gamma}_{\rm S}(p,p,0,m)&=&\left(\frac{-p^2-i\epsilon}{m^2}\right)^{\gamma_{\theta}(\alpha)/2}, 
\label{L3}
\end{eqnarray}
With the parameter $\gamma(\alpha)$ of (\ref{L1}) thus being identifiable as the anomalous dimension $\gamma_{\theta}(\alpha)$ of $\bar{\psi}\psi$, and with the  full dimension of $\bar{\psi}\psi$ being given by $d_{\theta}(\alpha)=3+\gamma_{\theta}(\alpha)$, the mechanism for both the finiteness and vanishing of $m_0$ is to have the dimension of $\bar{\psi}\psi$ be less than canonical.\footnote{Photons that are not dressed are referred to as being quenched. While such quenching can be obtained by $\beta(\alpha)$ having a zero away from the origin, as we discuss in Sec. IV-J, there has been much study in the literature of models in which photon dressings are not taken into account so that $\beta(\alpha)$ is then zero for any value of $\alpha$. As can be seen from (\ref{L2}), in such cases one can also obtain critical scaling with anomalous dimensions.}

\subsection{Non-Vanishing of the Physical Fermion Mass and the Baker-Johnson Evasion of the Goldstone Theorem}

In general,  a $Z_2=1$ fermion propagator $S(p,m)=(\slashed{p}-m_0-\Sigma(p,m))^{-1}$  would obey the, as yet unrenormalized, Schwinger-Dyson equation 
\begin{eqnarray}
\Sigma(p,m)=ie_0^2\int \frac{d^4k}{(2\pi)^4} D_{\mu\nu}(k)\Gamma^{\mu}(p,p-k)S(p-k,m)\gamma^{\nu},
\label{L4}
\end{eqnarray}
where $D_{\mu\nu}(k)$ is the exact photon propagator, $e_0$ is the bare charge, and $\Gamma^{\mu}(p,p-k)$ is the exact photon-fermion-anti-fermion vertex. Johnson, Baker, and Willey \cite{Johnson1964} initially studied this equation in the limit in which $D_{\mu\nu}(k)$  was taken to be the bare photon propagator, but in which $S(p,m)$ and $\Gamma^{\mu}(p,p-k)$ were otherwise fully dressed, and obtained the asymptotic scaling behavior $\Sigma (p,m)=m(-p^2/m^2)^{\gamma_{\theta}(\alpha)/2}$ of the type exhibited in (\ref{L3}). Subsequently, they found \cite{Johnson1967} that they could justify the use of an undressed photon propagator if the bare charge satisfied the Gell-Mann-Low eigenvalue equation. If we set $m_0=0$ (as would follow from (\ref{L1}) in the limit of large cutoff if $\gamma_{\theta}(\alpha)<0$), the Schwinger-Dyson equation could have both trivial and non-trivial solutions for $\Sigma(p,m)$. Then, if the non-zero solution is chosen, the fermion mass would behave just as dynamical masses behave in self-consistent theories of mass generation, and so one initially would expect the presence of a massless pseudoscalar Goldstone boson associated with the generation of such a fermion mass. However, this turned out not to be the case since there was a hidden renormalization effect in the theory, one associated with the renormalization constant $Z^{-1/2}_{\theta}= (\Lambda^2/m^2)^{\gamma_{\theta}(\alpha)/2}$ that renormalizes $\bar{\psi}\psi$ according to $Z^{-1/2}_{\theta}(\bar{\psi}\psi)_0=\bar{\psi}\psi$  \cite{Adler1971}. ($Z^{-1/2}_{\theta}$ is also equal to $Z_S$ the vertex renormalization constant for $\Gamma_{\rm S}(p,p,0,m)$.) With the product $m_0(\bar{\psi}\psi)_{0}$ being equal to $m_0Z^{1/2}_{\theta}\bar{\psi}\psi$, and thus equal to $m\bar{\psi}\psi$ on setting  $m$ to be the finite but non-zero $m=m_0Z^{1/2}_{\theta}$, the mass term $m_0(\bar{\psi}\psi)_{0}=m\bar{\psi}\psi$ is then a renormalization group invariant, with a non-zero $m_0(\bar{\psi}\psi)_{0}$ term thus being present in the bare Lagrangian from the outset. Consequently, the chiral symmetry is already broken in the Lagrangian itself and the Goldstone theorem does not apply. This then is the Baker-Johnson \cite{Baker1971a} evasion of the Goldstone theorem.

To understand the role played by the bare mass, it is instructive to look not at the Schwinger-Dyson equation, but at the Bethe-Salpeter equation for the anticommutator of $\gamma_5$ with $\Sigma(p,m)$, viz. \cite{Johnson1968}
\begin{eqnarray}
\{\gamma_5,\Sigma(p,m)\}&=&\int d^4k K(p,k,0)S(k,m)\{\gamma_5,\Sigma(k,m)\}S(k,m) 
+2m_0\int d^4k K(p,k,0)S(k,m)\gamma_5S(k,m).
\label{L5}
\end{eqnarray}
where $K(p,k,0)$ is the Bethe-Salpter kernel.\footnote{Without the bare mass term, the relation of this equation to dynamical symmetry breaking is discussed in \cite{Baker1964}.} To obtain an asymptotic solution to (\ref{L5}) Johnson first improved the convergence properties of the Bethe-Salpeter kernel integral by noting that the subtracted $\int d^4k K(p,k,0)S(k,m)\gamma_5S(k,m)-\int d^4k K(p^{\prime},k,0)S(k,m)\gamma_5S(k,m)$ at two values of the momenta was well-defined. Thus if either the bare mass is identically zero or if it vanishes in the limit of infinite cutoff, the subtracted  $\{\gamma_5,\Sigma(p,m)\}-\{\gamma_5,\Sigma(p^{\prime},m)\}=\int d^4k K(p,k,0)S(k,m)\{\gamma_5,\Sigma(k,m)\}S(k,m) -\int d^4k K(p^{\prime},k,0)S(k,m)\{\gamma_5,\Sigma(k,m)\}S(k,m)$ is well-defined. From this subtracted equation Johnson was then able to extract out an asymptotic solution for $\{\gamma_5,\Sigma(p,m)\}$, and it was found to precisely be of the scaling form.\footnote{A summary of the calculation may be found in \cite{Mannheim1975}.} However, since the $m_0$-dependent term had dropped out, we see that one can get asymptotic scaling for the fermion propagator without needing to require that $m_0$ be zero identically. A form in which $m_0$ only vanishes in the limit of infinite cutoff, but is not zero identically,  is thus compatible with a fermion propagator that satisfies a homogeneous equation, with inspection of (\ref{L4}) or the subtracted version of (\ref{L5}) in and of themselves not immediately indicating what the relevant situation for $m_0$ might be. Rather one needs to actually determine $m_0$ to see whether or not it behaves as in (\ref{L1}), and, as we explain below,  one needs to check whether or not there actually is a pole in the pseudoscalar sector. Below in Sec. IV-J we shall elucidate the role played by $m_0$ by discussing an approximation to the full JBW calculation, the so-called quenched ladder approximation, in which one restricts to a bare (i.e. quenched) photon, and only keeps planar diagrams (the ladder or rainbow approximation).

Now as originally noted in \cite{Baker1964}, in order  to establish the presence of a Goldstone boson it is not sufficient to look at the self-consistent equation for the fermion mass alone. Rather, one must look at the fermion-antifermion scattering amplitude, to see whether there might actually be a massless pole in it, or whether the renormalization procedure might prevent this from occurring. And when Johnson, Baker, and Willey did this in their study of electrodynamics, they found that there was no massless Goldstone pole, with the kernel of the Bethe-Salpeter equation for the scattering amplitude being found to be non-compact, so that no pole was generated. Thus having a non-trivial solution to the self-consistent equation for the mass is a necessary but not sufficient condition to secure a Goldstone pole. (The self-consistent mass equation is essentially the self-consistent equation for the residue at the Goldstone pole, and from a study of the equation for the would-be residue alone one cannot establish the presence of the pole itself.) The cause of this lack of a Goldstone boson in the presence of dynamical mass generation was explored by Baker and Johnson and is known as the Baker-Johnson evasion of the Goldstone theorem.

To further illustrate the issues involved we note that if there is to be a Goldstone boson it must also appear in $\tilde{\Gamma}_{\rm P}(p,p+q,q,m)$, the insertion of the pseudoscalar $\bar{\psi}i\gamma_5\psi$ into the fermion propagator. This Green's function 
obeys
\begin{eqnarray}
&&\tilde{\Gamma}_{\rm P}(p,p+q,q,m)=Z_{\rm P}i\gamma_5
+\int d^4k \tilde{K}(p,k,q)\tilde{S}(k,m)\tilde{\Gamma}_{\rm P}(k,k+q,q,m)\tilde{S}(k+q,m). 
\label{L6}
\end{eqnarray}
Here the tilde symbol indicates that everything is  renormalized, with $Z_{\rm P}$ renormalizing the pseudoscalar vertex function. Since the large $p^2$ behavior of the theory is not sensitive to mass, this $Z_{\rm P}$ is equal to the previously introduced $Z_{\rm S}$, with $Z_{\rm P}$ vanishing as $(\Lambda^2/m^2)^{\gamma_{\theta}(\alpha)/2}$. To see if there is a pole we note that we can rewrite (\ref{L6}) by inserting its left-hand side into its right-hand side iteratively, to symbolically then obtain  $-i\gamma_5\tilde{\Gamma}_{\rm P}=Z_{\rm P}+Z_{\rm P}\Pi Z_{\rm P}+Z_{\rm P}\Pi Z_{\rm P}\Pi Z_{\rm P}+...$, where $\Pi$ is an appropriate vacuum polarization term. Thus we obtain
\begin{eqnarray}
-i\gamma_5\tilde{\Gamma}_{\rm P}=\frac{Z_{\rm P}}{1-Z_{\rm P}\Pi}=\frac{1}{Z_{\rm P}^{-1}-\Pi}.
\label{L7}
\end{eqnarray}
Then with $Z_{\rm P}^{-1}$ diverging much faster than $\Pi$, no pole is generated. Thus one again obtains the Baker-Johnson evasion of the Goldstone theorem. And not only would this imply that there is no  dynamical pseudoscalar bound state Goldstone particle in the theory, implicit in the analysis is that there would be no dynamical scalar bound state Higgs particle either.

Now, in and of itself, the fact that (\ref{L6}) becomes homogeneous when $Z_{\rm P}$ vanishes does not automatically exclude the possible presence of a pole, since an analogous situation is met in the non-renormalizable but cut-off NJL model. As we discuss in more detail in Sec. III, there one introduces a four-fermion coupling $(g/2)(\bar{\psi}i\gamma_5\psi)^2$, and in the pseudoscalar channel of the fermion-antifermion scattering amplitude one obtains a $T$-matrix of the form\footnote{In general the Bethe-Salpeter equation for the scattering amplitude $T$ is of the  symbolic form $T=K-\int KSST$, with the kernel being given here by $g$ to lowest order in the four-fermion interaction.}
\begin{eqnarray}
T_{\rm P}=\frac{g}{1-g\Pi}=\frac{1}{g^{-1}-\Pi}.
\label{L8}
\end{eqnarray}
Now in the  NJL  case both $g^{-1}$ and $\Pi$ are divergent in the limit of infinite cutoff and $g$ is zero. However, even though  both $g^{-1}$ and $\Pi$ are divergent, they both diverge at the precisely the same rate, with there then indeed being a massless pole in $T_{\rm P}$. While one would automatically obtain a pole if $g$ and $\Pi$ are both finite (given dynamical mass generation of course), one could also obtain a pole if $g^{-1}$ and $\Pi$ diverge, provided they diverge at the same rate. In the renormalizable model we discuss in Sec. IV, we will see that both Goldstone and Higgs bosons will be generated by such a mechanism. 

\subsection{Non-Zero Vacuum Expectation Value for $\bar{\psi}\psi$ and the condition $\gamma_{\theta}(\alpha)=-1$}

Now even though the non-trivial solution to the self-consistent fermion mass generating equation given in (\ref{L4}) might not require a Goldstone boson, there was still the issue of determining what would oblige the theory to actually choose the non-trivial solution to it rather than the trivial one in the first place. To this end the present author compared the energy densities of the two solutions to find \cite{Mannheim1974a,Mannheim1975} that if $\gamma_{\theta}(\alpha)$ took the special value 
\begin{eqnarray}
\gamma_{\theta}(\alpha)=-1, 
\label{L9}
\end{eqnarray}
the infrared divergences that would then follow (the theory having  been softened so much in the ultraviolet and thus made more and more divergent in the infrared) would then drive the theory into a spontaneously broken vacuum $|\Omega_{\rm m}\rangle$ in which $\langle \Omega_{\rm m}|\bar{\psi}\psi|\Omega_{\rm m}\rangle \neq 0$. In order to take care of the infinities that the energy density contained the present author chose not to normal order them away, but rather to cancel them by a counterterm, with the appropriate one being a four-fermion interaction with coupling constant $g$. Now for a point-coupled such interaction this counterterm would itself generate new infinities. However with the dimension of $\bar{\psi}\psi$ having been reduced from $d_{\theta}(\alpha)=3$ to $d_{\theta}(\alpha)=3+\gamma_{\theta}(\alpha)=2$, the four-fermion theory becomes renormalizable, something we expound on in detail in Secs. IV-F and V-A below. With this specific counterterm the then finite energy density was found to have none other than a double-well potential structure  in which $\langle \Omega_{\rm m}|\bar{\psi}\psi|\Omega_{\rm m}\rangle=m/g$ was non-zero. Specifically, in terms of   a renormalization group subtraction point $\mu^2$ that we elaborate on in Sec. IV below, the renormalized energy density was given as the double-welled
\begin{eqnarray}
\tilde{\epsilon}(m)=\frac{m^2\mu^2}{16\pi^2}\left[{\rm ln}\left(\frac{m^2}{M^2}\right)-1\right],
\label{L10}
\end{eqnarray}
with a local maximum at  $m=0$ where $\langle \Omega_{\rm 0}|\bar{\psi}\psi|\Omega_{\rm 0}\rangle=0$ and a degenerate global minimum at $m=M$ where $\langle \Omega_{\rm M}|\bar{\psi}\psi|\Omega_{\rm M}\rangle=M/g$ is non-zero. Mass generation in JBW electrodynamics is thus associated with a vacuum in which  $\langle \Omega_{\rm M}|\bar{\psi}\psi|\Omega_{\rm M}\rangle$ is non-zero.\footnote{While a reader might initially baulk at the notion that a vacuum could be degenerate in a non-chirally-invariant theory such as JBW electrodynamics, on reinterpreting JBW electrodynamics as the mean field sector of  a chiral-invariant massless fermion QED theory coupled to a four-fermion interaction, in Sec. IV we will be able to identify the vacuum $|\Omega_{\rm M}\rangle$ as the Hartree-Fock, mean-field vacuum to the full chiral-invariant massless QED plus four-fermion theory.}

As originally introduced by Kadanoff and Wilson, critical scaling described the behavior of a crystal at the critical phase transition temperature where the correlation length is infinite. However at the same critical temperature the order parameter is zero, with it only being non-zero in the ordered phase below the critical temperature.  In the case of critical scaling  in a quantum field theory, when $\gamma_{\theta}(\alpha)=-1$ we can have both scaling with anomalous dimensions and  a non-zero value for the order parameter $\langle \Omega_M|\bar{\psi}\psi|\Omega_M\rangle$ occur simultaneously. This happens because in a massless theory (analogous to an infinite correlation length) there is no scale, so infrared divergences (needed to generate long range order and an order parameter) are also present; with the effect of $\gamma_{\theta}(\alpha)=-1$ being to soften the theory so much in the ultraviolet that it becomes sufficiently divergent in the infrared to cause dynamical symmetry breaking to take place. Our work thus provides a framework in which aspects of critical phenomena both at and below the critical temperature are simultaneously present. And in fact one of the motivations for the work of the present author in the 1970s was to try to find such a framework, with it being  through the condition $\gamma_{\theta}(\alpha)=-1$ that it was achieved.

To underscore and illuminate the interplay between mass generation and the spontaneously broken vacuum, a second, independent derivation  of the $\gamma_{\theta}(\alpha)=-1$ condition was also provided in \cite{Mannheim1975}. In this derivation the fermion propagator was derived in two separate ways, via the Wilson operator product expansion and  via a renormalization group analysis, and compatibility between the two was sought. The Wilson expansion describes the short distance behavior of a massless theory as constructed in a non-spontaneously broken normal vacuum $|\Omega_{\rm 0}\rangle$. The renormalization group describes the short-distance behavior of a theory in which the fermion mass is non-zero. In such a theory we have seen that since the mass is non-zero, at critical scaling the renormalization group describes fluctuations around a spontaneously broken vacuum $|\Omega_{\rm m}\rangle$. To compare the two we thus take matrix elements of the Wilson expansion in $|\Omega_{\rm m}\rangle$. Specifically, in the Wilson operator product  expansion at a critical point  the leading behavior at short distance of the massless fermion two point function is given by
\begin{eqnarray}
T(\psi(x)\bar{\psi}(0))&=&\langle \Omega_{\rm 0}|T(\psi(x)\bar{\psi}(0))|\Omega_{\rm 0}\rangle
+(\mu^2x^2)^{\gamma_{\theta}(\alpha)/2}:\psi(0)\bar{\psi}(0):
\label{L11}
\end{eqnarray}
where the dots indicate normal ordering with respect to the massless vacuum $|\Omega_{\rm 0}\rangle$ according to $:\psi(0)\bar{\psi}(0):=\psi(0)\bar{\psi}(0)- \langle \Omega_{\rm 0}|\psi(0)\bar{\psi}(0)|\Omega_{\rm 0}\rangle$, $\langle \Omega_{\rm 0}|:\psi(0)\bar{\psi}(0):|\Omega_{\rm 0}\rangle=0$, and $\mu^2$ is an off-shell Green's function subtraction point. If we now take the matrix element of this expansion in the degenerate vacuum $|\Omega_{\rm m}\rangle$ we obtain an asymptotic propagator and inverse propagator that up to coefficients behave as 
\begin{eqnarray}
\tilde{S}(p,m)&=&\frac{1}{\slashed{p}}+(-p^2)^{(-\gamma_{\theta}(\alpha)/2-2)},
\nonumber\\
\tilde{S}^{-1}(p,m)&=&\slashed{p}-(-p^2)^{(-\gamma_{\theta}(\alpha)/2-1)}.
\label{L12}
\end{eqnarray}
On comparing with (\ref{L3}), (\ref{L9}) follows. (In \cite{Mannheim1975} it was shown that the coefficients match too.) Moreover, not only do we recover (\ref{L9}), we confirm that the relevant vacuum for JBW electrodynamics is indeed a spontaneously broken one.\footnote{Some separate discussion of an interplay of the Wilson operator product expansion and vacuum condensates may be found in \cite{Shifman1979}. The condition $\gamma_{\theta}(\alpha)=-1$ is also encountered in the quenched ladder approximation to an Abelian gluon model when the coupling constant is given by $\alpha=\pi/3$, and will be discussed in Sec. IV-J below.}

\subsection{Evasion of the Baker-Johnson Evasion of the Goldstone Theorem}

As we see, the JBW theory has much of the structure of dynamical symmetry breaking and yet has no dynamical Goldstone boson, and thus no dynamical Higgs boson either. Moreover it has much of the structure of the NJL model. The essence of the NJL model is to rewrite the four-fermion Lagrangian with a strictly massless fermion in terms of a mean-field sector action with a massive fermion term that is not in the initial Lagrangian, together with a compensating residual interaction sector action according to  
\begin{eqnarray}
{\cal L}=i\bar{\psi}\gamma^{\mu}\partial_{\mu}\psi -\frac{g}{2}(\bar{\psi}\psi)^2-\frac{g}{2}(\bar{\psi}i\gamma_5\psi)^2={\cal L}_{\rm MF}+{\cal L}_{\rm RI},
\label{L13}
\end{eqnarray}
where 
\begin{eqnarray}
{\cal L}_{\rm MF}&=&i\bar{\psi}\gamma^{\mu}\partial_{\mu}\psi -m\bar{\psi}\psi+\frac{m^2}{2g},\qquad
\nonumber\\
{\cal L}_{\rm RI}&=&-\frac{g}{2}\left(\bar{\psi}\psi-\frac{m}{g}\right)^2-\frac{g}{2}(\bar{\psi}i\gamma_5\psi)^2.
\label{L14}
\end{eqnarray}
Even though the full Lagrangian ${\cal L}$ is globally chiral symmetric under $\psi\rightarrow e^{i\alpha_5\gamma_5}\psi$ with spacetime independent phase $\alpha_5$, neither ${\cal L}_{\rm MF}$ nor ${\cal L }_{\rm RI}$ are themselves separately chirally symmetric. Thus in dynamical symmetry breaking one produces a mean-field theory in which the chiral symmetry is expressly broken at the level of the mean-field Lagrangian. Moreover, no Goldstone boson is present in the mean-field Lagrangian, as could not be the case since the mean-field Lagrangian is expressly not chiral symmetric. Rather, one is generated not by the mean field at all but by the residual interaction, and it is the residual interaction that is needed in order to restore the chiral symmetry that the mean-field sector itself does not possess.

Now, as had been noted by the present author in \cite{Mannheim1974a,Mannheim1975}, study of symmetry breaking in JBW electrodynamics can be obtained from the NJL model by replacing the point vertex for the insertion of a zero-momentum scalar $\bar{\psi}\psi$ operator into the fermion propagator (viz. $\tilde{\Gamma}_{\rm S}(p,p,0)=1$) by $\tilde{\Gamma}_{\rm S}(p,p,0)=(-p^2/m^2)^{\gamma_{\theta}(\alpha)/2}$ as given above by the renormalization group equation. Thus, as we show in detail in Sec. IV, we can reinterpret JBW electrodynamics as coupled to a four-fermion interaction to be a mean-field theory, one associated with  Lagrangian of the form:
\begin{eqnarray}
{\cal L}_{\rm QED}&=&-\frac{1}{4}F_{\mu\nu}F^{\mu\nu}+\bar{\psi}\gamma^{\mu}(i\partial_{\mu}-eA_{\mu})\psi 
-m\bar{\psi}\psi+\frac{m^2}{2g},
\label{L15}
\end{eqnarray}
where the $m^2/2g$ term is just a constant. And as such, this theory should not contain a Goldstone boson since mean-field theory never does, with the Baker-Johnson evasion of the Goldstone theorem then necessarily having to hold in the mean-field sector. And indeed, for a critical scaling electrodynamics to admit of a mean-field theory structure in the first place, the mean-field theory must thus necessarily be of the Baker-Johnson-evasion type. Nonetheless, as we show in Secs. IV-E and IV-G, the related residual interaction  will generate a massless pseudoscalar Goldstone boson, and will do so while generating a massive scalar Higgs boson at the same time. Thus by enlarging electrodynamics to include a  dynamical four-fermion interaction we are able to evade the Baker-Johnson evasion of the Goldstone theorem, and relate the dynamically generated mass and the spontaneously broken symmetry in the mean-field sector to a Goldstone boson after all. Then, as a very welcome bonus, we obtain a dynamical Higgs boson as well.

While we shall present our derivation of these results below, we note here that all we actually need for the discussion is the behavior of the pure fermion Green's functions containing fermion fields and fermion $\bar{\psi}\psi$  insertions, and for them we only need the assumption of scaling with anomalous dimensions (or equivalently conformal invariance with anomalous dimensions). We do not actually need to specify how the critical scaling was brought about, and thus do not actually need to introduce any explicit coupling to gauge bosons whose associated dynamics could cause coupling constant renormalization beta functions (such as $\beta(\alpha)$) to actually vanish. Our results are thus quite generic, and will continue to hold even if there are many species of fermion (assuming critical scaling of course), so that it thus suffices to discuss a single species of fermion and a single type of symmetry (in our case chiral symmetry) alone. Also we note that we only need to discuss spontaneous breakdown of a global symmetry, since once we have generated a massless Goldstone boson by some dynamical means, Jackiw and Johnson showed \cite{Jackiw1973}  that such a  dynamically generated Goldstone boson would automatically couple to a massless  external gauge field with the relevant quantum numbers, and would put a massless pole in the gauge boson vacuum polarization. This would then cause the gauge boson to become massive, to thus provide an explicit dynamical realization of the Higgs mechanism that  was presented in \cite{Englert1964,Higgs1964a,Higgs1964b,Guralnik1964}.

\section{The Nambu-Jona-Lasinio Chiral Four-Fermion Model}

\subsection{The Nambu-Jona-Lasinio Model as a Mean-Field Theory}

The NJL  model is a chirally-symmetric four-fermion model of interacting massless fermions with action
\begin{eqnarray}
I_{\rm NJL}=\int d^4x \left[i\bar{\psi}\gamma^{\mu}\partial_{\mu}\psi-\frac{g}{2}[\bar{\psi}\psi]^2-\frac{g}{2}[\bar{\psi}i\gamma_5\psi]^2\right].
\label{L16}
\end{eqnarray}
As such it is a relativistic generalization of the BCS (Bardeen-Cooper-Schrieffer) model \cite{Cooper1956,Bardeen1957}. In the mean-field, Hartree-Fock  approximation one introduces a trial wave function parameter $m$ that is not in the original action, and then decomposes  the action into two pieces,  a mean-field piece and a residual interaction according to:
\begin{eqnarray}
I_{\rm NJL}&=&\int d^4x \left[i \bar{\psi}\gamma^{\mu}\partial_{\mu}\psi-m\bar{\psi}\psi +\frac{m^2}{2g}\right]
+\int d^4x \left[-\frac{g}{2}\left(\bar{\psi}\psi-\frac{m}{g}\right)^2-\frac{g}{2}\left(\bar{\psi}i\gamma_5\psi\right)^2\right]
\nonumber\\
&=&I_{\rm MF}+I_{\rm RI},
\label{L17}
\end{eqnarray}
where $I_{\rm MF}$ contains the kinetic energy of a now massive fermion and a self-consistent $m^2/2g$ term. This $m^2/2g$ term acts like a cosmological constant and contributes to the mean-field vacuum energy density. In the mean-field Hartree-Fock approximation one sets
\begin{eqnarray}
&&\langle \Omega_{\rm m}|\left[\bar{\psi}\psi-\frac{m}{g}\right]^2|\Omega_{\rm m}\rangle=\left(\langle \Omega_{\rm m}|\left[\bar{\psi}\psi-\frac{m}{g}\right]|\Omega_{\rm m}\rangle\right)^2=0,
\nonumber\\
&&\langle \Omega_{\rm m}|\bar{\psi}\psi|\Omega_{\rm m}\rangle=\frac{m}{g},\qquad \langle \Omega_{\rm m}|\bar{\psi}i\gamma_5\psi|\Omega_{\rm m}\rangle=0.
\label{L18}
\end{eqnarray}
\begin{figure}[htb]
\begin{center}
\includegraphics[scale=0.20]{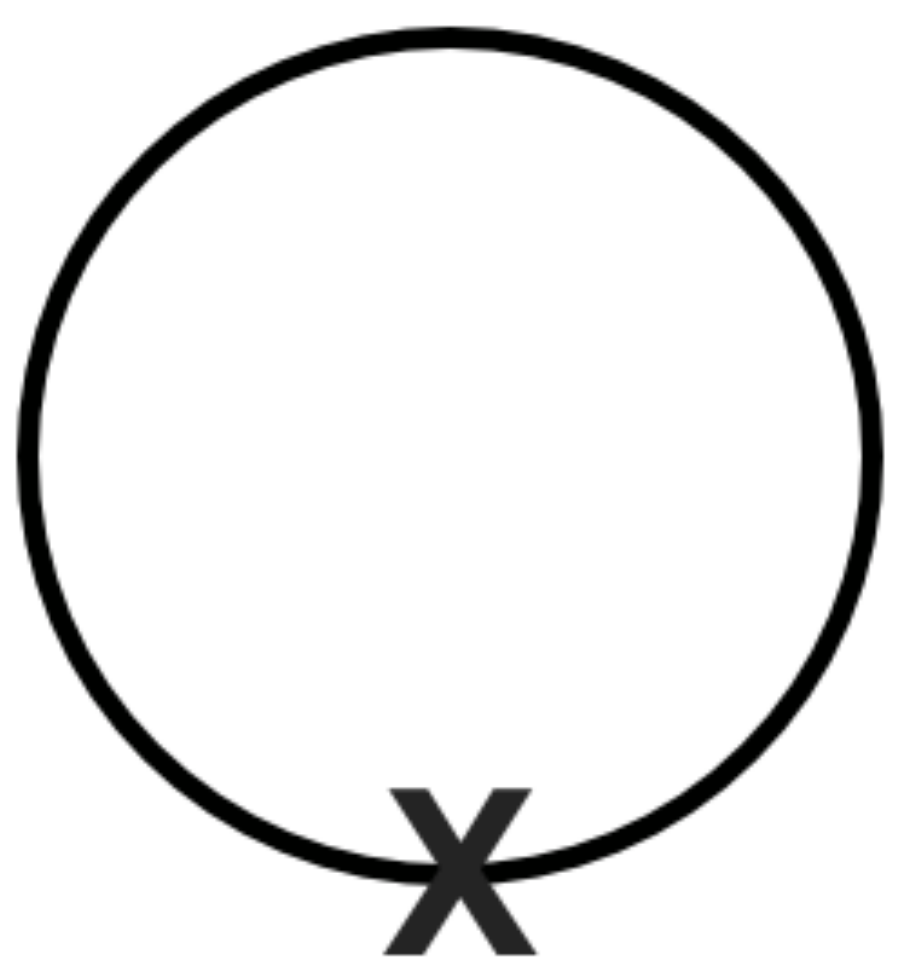}
\caption{$\langle \Omega_{\rm m}|\bar{\psi}\psi|\Omega_{\rm m}\rangle$ as evaluated in the mean-field approximation to the point-coupled NJL model.}
\label{baretadpole}
\end{center}
\end{figure}
In this approximation one evaluates the one fermion loop contribution to $\langle \Omega_{\rm m}|\bar{\psi}\psi|\Omega_{\rm m}\rangle$ using $I_{\rm MF}$ alone to give the tadpole diagram of Fig. (\ref{baretadpole}), with the physical fermion mass $M$ then being the value of $m$ that satisfies 
\begin{eqnarray}
\langle \Omega_{\rm m}|\bar{\psi}\psi|\Omega_{\rm m}\rangle=-i\int \frac{d^4p}{(2\pi)^4} {\rm Tr}\left[\frac{1}{\slashed{p}-m+i\epsilon}\right]=\frac{m}{g}, 
\label{L19}
\end{eqnarray}
viz. the gap equation
\begin{eqnarray}
-\frac{M\Lambda^2}{4\pi^2}+\frac{M^3}{4\pi^2}{\rm ln}\left(\frac{\Lambda^2}{M^2}\right)=\frac{M}{g},
\label{L20}
\end{eqnarray}
where $\Lambda$ is an ultraviolet cutoff, as needed since the NJL model is not renormalizable.

For (\ref{L20}) to have a non-trivial solution the four-fermion coupling constant $g$ must be negative (viz. attractive with our definition of $g$ as given in (\ref{L16})), and the combination $-g\Lambda^2/4\pi^2$ must be greater than one. This condition, which imposes a minimum value on $-g\Lambda^2$, is quite different from that found in the BCS theory, since in the BCS case there is a bound state Cooper pair no matter how weak the coupling $\lambda$ might be so long as it is attractive. And with the BCS gap parameter behaving as $\exp(1/\lambda)$, the gap parameter has an essential singularity at $g=0$, to thus exhibit its non-perturbative nature.  No such essential singularity is manifest in (\ref{L20}). Dynamical symmetry breaking in the NJL model thus has some intrinsic differences compared to dynamical symmetry breaking in BCS. However as we shall see below in (\ref{L49}), when we study JBW QED at $\gamma_{\theta}(\alpha)=-1$, we shall find dynamical symmetry breaking no matter how weak the four-fermion coupling constant  $g$ might be as long as it is attractive, and shall find an essential singularity at $g=0$.

Given this gap equation we can calculate the one loop mean-field vacuum energy density $\tilde{\epsilon}(m)=\epsilon(m)-m^2/2g$ as a function of $m$ to obtain
\begin{eqnarray}
\tilde{\epsilon}(m)&=&i\int \frac{d^4p}{(2\pi)^4}{\rm Tr~ln}\left[\frac{\slashed{p}-m+i\epsilon}{\slashed{p}+i\epsilon}\right]-\frac{m^2}{2g}
\nonumber\\
&=&-\frac{m^2\Lambda^2}{8\pi^2}
+\frac{m^4}{16\pi^2}{\rm ln}\left(\frac{\Lambda^2}{m^2}\right)+\frac{m^4}{32\pi^2}-\frac{m^2}{2g}
\nonumber\\
&=&\frac{m^4}{16\pi^2}{\rm ln}\left(\frac{\Lambda^2}{m^2}\right)-\frac{m^2M^2}{8\pi^2}{\rm ln}\left(\frac{\Lambda^2}{M^2}\right)+\frac{m^4}{32\pi^2}.
\label{L21}
\end{eqnarray}
As we explain below, $\epsilon(m)$ can be constructed as the infinite summation of massless graphs with zero-momentum point $m\bar{\psi}\psi$ insertions (see Fig. (\ref{lw1}) below).

We thus see that while the vacuum energy density $i\int d^4p/(2\pi)^4{\rm Tr~ln}[\slashed{p}-m+i\epsilon]$ has quartic, quadratic and logarithmically divergent pieces, the subtraction of the massless vacuum energy density $i\int d^4p/(2\pi)^4{\rm Tr~ln}[\slashed{p}+i\epsilon]$ removes the quartic divergence, with the subtraction of the self-consistent induced mean-field term $m^2/2g$ then leaving $\tilde{\epsilon}(m)$ only logarithmically divergent. We shall return to the quartic divergence in Sec. V-C below when we couple the theory to gravity, but since we are for the moment doing a flat space calculation where only energy differences matter, use of this $\tilde{\epsilon}(m)$ suffices to show that the massive vacuum lies lower than the massless one where $m=0$. I.e. we recognize the logarithmically divergent  $\tilde{\epsilon}(m)$ as having a local maximum at $m=0$, and a global minimum at $m=M$ where $M$ itself is finite.  We thus induce none other than a dynamical double-well potential, and identify $M$ as the matrix element of a fermion bilinear according to $M/g=\langle \Omega_{\rm M}|\bar{\psi}\psi|\Omega_{\rm M}\rangle$.

\subsection{Higgs-Like Lagrangian}

While $\tilde{\epsilon}(m)$ has a double-well form familiar from a Higgs model as built out of a Higgs field that is an elementary, and thus quantum, field, $m$ itself is not a quantum field. Rather, it is only a c-number matrix element, with $\tilde{\epsilon}(m)$ having a Higgs potential structure even though  no elementary Higgs field is present.  As regards a kinetic energy term, we look not at matrix elements in the translationally-invariant vacuum $|\Omega_{M}\rangle$ but instead at matrix elements in coherent states $|C\rangle$  where  $m(x)=\langle C|\bar{\psi}(x)\psi(x)|C\rangle$ is now spacetime dependent. Then we find \cite{Eguchi1974,Mannheim1976} that the resulting mean-field effective action has a logarithmically divergent part  of the form
\begin{eqnarray}
I_{\rm EFF}&=&\int \frac{d^4x}{8\pi^2}{\rm ln}\left(\frac{\Lambda^2}{M^2}\right)\bigg[
\frac{1}{2}\partial_{\mu}m(x)\partial^{\mu}m(x)
+m^2(x)M^2-\frac{1}{2}m^4(x)\bigg].
\label{L22}
\end{eqnarray}

If we go further and introduce a coupling $g_{\rm A}\bar{\psi}\gamma_{\mu}\gamma_5A^{\mu}_{5}\psi $ to an axial gauge field $A^{\mu}_{5}(x)$, on setting $\phi=\langle C|\bar{\psi}(1+\gamma_5)\psi|C\rangle$ the effective action becomes 
\begin{eqnarray}
I_{\rm EFF}&=&\int \frac{d^4x}{8\pi^2}{\rm ln}\left(\frac{\Lambda^2}{M^2}\right)\bigg[
\frac{1}{2}|(\partial_{\mu}-2ig_{\rm A}A_{\mu 5})\phi(x)|^2
+|\phi(x)|^2M^2-\frac{1}{2}|\phi(x)|^4-\frac{g_{\rm A}^2}{6}F_{\mu\nu 5}F^{\mu\nu 5}\bigg].
\label{L23}
\end{eqnarray}
We recognize this action as a double-well Ginzburg-Landau type Higgs Lagrangian with order parameter $\phi(x)$, only now generated dynamically. We thus generalize to the relativistic chiral case Gorkov's derivation of the Ginzburg-Landau order parameter action starting from the BCS four-fermion theory, and see that just as in the theory of superconductivity, there is no need for the Higgs Lagrangian to be built out of a quantized scalar field. In the $I_{\rm EFF}$ effective action associated with the NJL model there is a double-well Higgs potential, but since $m(x)=\langle C|\bar{\psi}(x)\psi(x)|C\rangle$ is a c-number, $m(x)$ does not itself represent a q-number scalar field. Rather, as we now show, the q-number fields are to be found as collective modes generated by the residual interaction, with no elementary scalar field being needed at all. 

\subsection{The Collective Tachyon Modes when the Fermion is Massless}

To find the collective modes we need to evaluate the vacuum polarizations
\begin{eqnarray}
\Pi_{\rm S}(x)&=&\langle \Omega|T(\bar{\psi}(x)\psi(x)\bar{\psi}(0)\psi(0))|\Omega\rangle, 
\nonumber\\
\Pi_{\rm P}(x)&=&\langle \Omega|T(\bar{\psi}(x)i\gamma_5\psi(x)\bar{\psi}(0)i\gamma_5\psi(0))|\Omega\rangle
\label{L24}
\end{eqnarray}
associated with the scalar and pseudoscalar sectors, as is appropriate to a chiral-invariant theory. To see why, from the perspective of  $\Pi_{\rm S}(x)$ and $\Pi_{\rm P}(x)$, the symmetry needs to be broken, we first evaluate  $\Pi_{\rm S}(x)$ and $\Pi_{\rm P}(x)$ on the assumption that the fermion is massless. If we take the fermion to be massless (i.e. setting $|\Omega\rangle=|\Omega_{\rm 0}\rangle$ where $\langle \Omega_{\rm 0}|\bar{\psi}\psi|\Omega_{\rm 0}\rangle=0$) to one loop order as evaluated using the original $I_{\rm NJL}$ action we obtain
\begin{eqnarray}
&&\Pi_{\rm S}(q^2,m=0)=-i\int\frac{d^4p}{(2\pi)^4}{\rm Tr}\left[\frac{1}{\slashed{p}+i\epsilon}\frac{1}{\slashed{p}+\slashed{q}+i\epsilon}\right],
\nonumber\\
&&\Pi_{\rm P}(q^2,m=0)=-i\int\frac{d^4p}{(2\pi)^4}{\rm Tr}\left[i\gamma_5\frac{1}{\slashed{p}+i\epsilon}i\gamma_5\frac{1}{\slashed{p}+\slashed{q}+i\epsilon}\right],
\nonumber\\
\label{L25}
\end{eqnarray}
to thus yield
\begin{eqnarray}
\nonumber\\
\Pi_{\rm S}(q^2,m=0)=\Pi_{\rm P}(q^2,m=0)=-\frac{\Lambda^2}{4\pi^2}-\frac{q^2}{8\pi^2}{\rm ln}\left(\frac{\Lambda^2}{-q^2}\right)-\frac{q^2}{8\pi^2}.
\label{L26}
\end{eqnarray}
The scattering matrices in the two channels are given by iterating the vacuum polarizations according to $T=g+g\Pi g+g\Pi g\Pi g+...$, to yield
\begin{eqnarray}
T_{\rm S}(q^2,m=0)&=&\frac{g}{1-g\Pi_{\rm S}(q^2,m=0)}=\frac{1}{g^{-1}-\Pi_{\rm S}(q^2,m=0)},
\nonumber\\
T_{\rm P}(q^2,m=0)&=&\frac{g}{1-g\Pi_{\rm P}(q^2,m=0)}=\frac{1}{g^{-1}-\Pi_{\rm P}(q^2,m=0)}.
\label{L27}
\end{eqnarray}
With $g^{-1}$ being given by the gap equation above, near $q^2=-2M^2$ both scattering matrices behave as
\begin{eqnarray}
T_{\rm S}(q^2,m=0)=T_{\rm P}(q^2,m=0)=\frac{Z^{-1}}{(q^2+2M^2)},\qquad
Z=\frac{1}{8\pi^2}{\rm ln}\left(\frac{\Lambda^2}{M^2}\right)
\label{L28}
\end{eqnarray}
to leading order in the cutoff. We thus obtain degenerate (i.e. chirally symmetric) scalar and pseudoscalar tachyons at $q^2=-2M^2$ (just like fluctuating around the local maximum in a double-well potential, except that these tachyons are dynamically induced and not put in by hand), with $|\Omega_{\rm 0}\rangle$ thus being unstable. Hence, before determining which vacuum is stable, already we see that if the fermion is massless the theory is unstable.

\subsection{The Collective Goldstone and Higgs Modes when the Fermion is Massive}

To find a stable vacuum, we now take the fermion to have non-zero mass $M$ (i.e. we set $|\Omega\rangle=|\Omega_{\rm M}\rangle$). Now we obtain
\begin{eqnarray}
\Pi_{\rm S}(q^2,M)
&=&-i\int\frac{d^4p}{(2\pi)^4}{\rm Tr}\left[\frac{1}{\slashed{p}-M+i\epsilon}\frac{1}{\slashed{p}+\slashed{q}-M+i\epsilon}\right]
\nonumber\\
&=&
-\frac{\Lambda^2}{4\pi^2}
+\frac{M^2}{4\pi^2}{\rm ln}\left(\frac{\Lambda^2}{M^2}\right)
+\frac{(4M^2-q^2)}{8\pi^2} 
+\frac{(4M^2-q^2)}{8\pi^2}{\rm ln}\left(\frac{\Lambda^2}{M^2}\right)
\nonumber\\
&-&\frac{1}{8\pi^2}\frac{(4M^2-q^2)^{3/2}}{(-q^2)^{1/2}}
{\rm ln}\left(\frac{(4M^2-q^2)^{1/2}+(-q^2)^{1/2}}{(4M^2-q^2)^{1/2}-(-q^2)^{1/2}}\right).
\label{L29}
\end{eqnarray}
and
\begin{eqnarray}
\Pi_{\rm P}(q^2,M)&=&-i\int\frac{d^4p}{(2\pi)^4}{\rm Tr}\left[i\gamma_5\frac{1}{\slashed{p}-M+i\epsilon}i\gamma_5\frac{1}{\slashed{p}+\slashed{q}-M+i\epsilon}\right]
\nonumber\\
&=&-\frac{\Lambda^2}{4\pi^2}
+\frac{M^2}{4\pi^2}{\rm ln}\left(\frac{\Lambda^2}{M^2}\right) 
-\frac{q^2}{8\pi^2}{\rm ln}\left(\frac{\Lambda^2}{M^2}\right) 
+\frac{(4M^2-q^2)}{8\pi^2}
\nonumber\\
&+&\frac{(8M^4-8M^2q^2+q^4)}{8\pi^2 (-q^2)^{1/2}(4M^2-q^2)^{1/2}}{\rm ln}
\left(\frac{(4M^2-q^2)^{1/2}+(-q^2)^{1/2}}{(4M^2-q^2)^{1/2}-(-q^2)^{1/2}}\right).
\label{L30}
\end{eqnarray}
As we see, both $\Pi_{\rm S}(q^2,M)$ and $\Pi_{\rm P}(q^2,M)$ have a branch point at $q^2=4M^2$, viz. at the threshold for the creation of a fermion and antifermion pair each with mass $M$. Given the form for $g^{-1}$,  we find a dynamical pseudoscalar Goldstone boson bound state at $q^2=0$ and a  dynamical scalar Higgs boson bound state at $q^2=4M^2$ ($=-2\times  M^2({\rm tachyon})$), with the two scattering amplitudes behaving near their poles as
\begin{eqnarray}
T_{\rm S}(q^2,M)&=&\frac{R_{\rm S}^{-1}}{(q^2-4M^2)},\qquad T_{\rm P}(q^2,M)=\frac{R_{\rm P}^{-1}}{q^2},
\label{L31}
\end{eqnarray}
where\footnote{We have labelled the residues $R^{-1}_{\rm S}$ and $R^{-1}_{\rm P}$ to indicate that they are not the previously introduced $Z^{-1}_{\rm S}$ and $Z^{-1}_{\rm P}$.}
\begin{eqnarray}
R_{\rm S}&=&R_{\rm P}=\frac{1}{8\pi^2}{\rm ln}\left(\frac{\Lambda^2}{M^2}\right).
\label{L32}
\end{eqnarray}
As we see, the two dynamical bound states are not degenerate in mass (spontaneously broken chiral symmetry), and the dynamical Higgs scalar mass $2M$ is twice the induced mass of the fermion, to thus lie right at the threshold of the fermion-antifermion scattering amplitude. In addition, we note that despite the fact that there is a cutoff $\Lambda$ in the theory, the Higgs mass is not at that scale but at the finite fermion mass scale $M$ instead. Thus unlike the elementary Higgs case, in the dynamical case the Higgs mass is fully under control.

\subsection{Fixing the Wick Contour for the Vacuum Energy Density}

For what is to follow below, we need to make one further comment regarding the evaluation of the vacuum energy density. As discussed for instance in \cite{Mannheim1975}, what we have been calling $\epsilon(m)$ is not the energy density of the vacuum. Rather, according to the Gell-Mann-Low adiabatic switching procedure, it is actually an energy density difference
\begin{eqnarray}
\epsilon(m)&=&\frac{1}{V}[\langle \Omega_m|H_m|\Omega_m\rangle -\langle \Omega_0|H_0|\Omega_0\rangle],
\label{L33}
\end{eqnarray}
in a volume $V$, where $H_m=H_0+m\bar{\psi}\psi$ is the Hamiltonian density in the presence of the $m\bar{\psi}\psi$ term, while $H_0$ is the Hamiltonian density in its absence. Specifically, in the adiabatic switching procedure one starts with a Hamiltonian $H_0$ at time $t=-\infty$ and ground state $|\Omega^{-}_{\rm 0}\rangle$, switches on a source term such as $m\bar{\psi}\psi$, and then switches the source off at $t=+\infty$, to then return $H_0$ to its ground state, only now in a state $|\Omega^{+}_{\rm 0}\rangle$ that can only differ from  $|\Omega^{-}_{\rm 0}\rangle$ by a phase. That phase is  given by the energy density difference exhibited in (\ref{L33}). As constructed, this phase could not know what the ground state energy density of $H_{0}$ itself might be (it is only gravity that could know), and thus $\epsilon(m)$ could only be an energy density difference. Given (\ref{L33}), we note that since $\langle \Omega_0|\bar{\psi}\psi|\Omega_0\rangle=0$, we can rewrite $\epsilon(m)$ as 
\begin{eqnarray}
\epsilon(m)=\frac{1}{V}[\langle \Omega_m|H_m|\Omega_m\rangle -\langle \Omega_0|H_m|\Omega_0\rangle],
\label{L34}
\end{eqnarray}
to put it in the form relevant to the dynamical symmetry breaking of interest to us here.

\begin{figure}[htb]
\begin{center}
\includegraphics[scale=0.3]{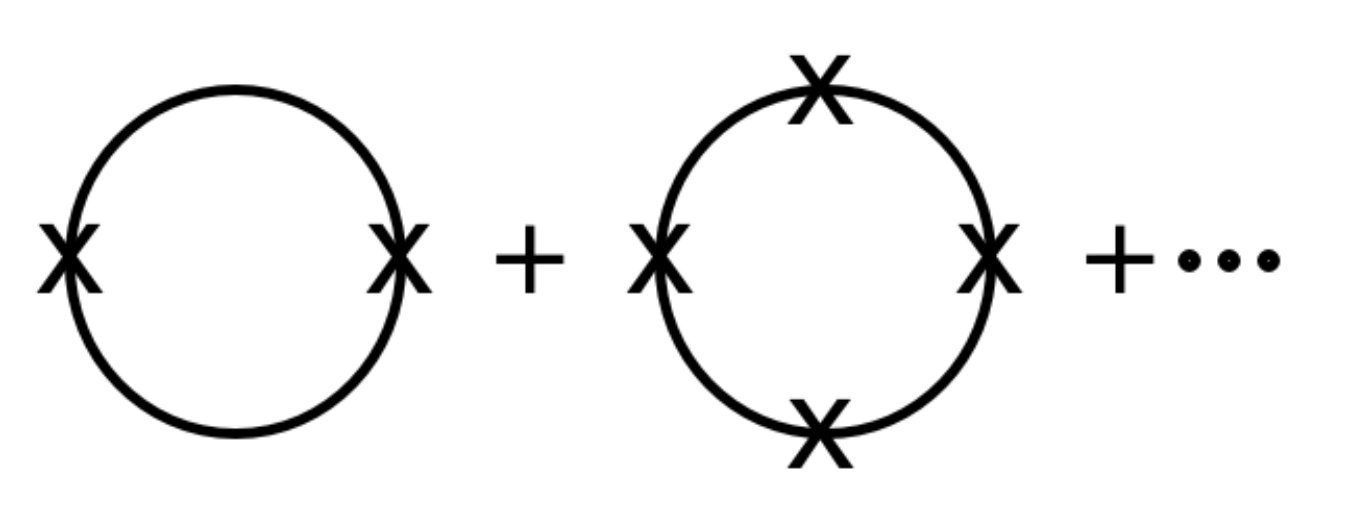}
\caption{Vacuum energy density $\epsilon(m)$ via an infinite summation of massless graphs with zero-momentum point $m\bar{\psi}\psi$  insertions.}
\label{lw1}
\end{center}
\end{figure}

Diagramatically, $\epsilon(m)$ generates the Green's functions associated with zero-momentum insertions of $\bar{\psi}\psi$, and can be written as
\begin{eqnarray}
\epsilon(m)=\sum \frac{1}{n!}G^{(n)}_0(q_{\mu}=0,m=0)m^n
\label{L35}
\end{eqnarray}
Here  $G^{(n)}_0$ is the $\bar{\psi}\psi$ Green's function with $n$ insertions as calculated in the massless $H_0$ theory, with the $G^{(2)}_0$ and $G^{(4)}_0$ terms for instance being given by
\begin{eqnarray}
&&G^{(2)}(q_{\mu}=0,m=0)=-i\int\frac{d^4p}{(2\pi)^4}{\rm Tr}\left[\frac{1}{\slashed{p}+i\epsilon}\frac{1}{\slashed{p}+i\epsilon}\right],
\nonumber\\
&&G^{(4)}(q_{\mu}=0,m=0)=
-i\int\frac{d^4p}{(2\pi)^4}{\rm Tr}\left[\frac{1}{\slashed{p}+i\epsilon}\frac{1}{\slashed{p}+i\epsilon}\frac{1}{\slashed{p}+i\epsilon}\frac{1}{\slashed{p}+i\epsilon}\right]
\label{L36}
\end{eqnarray}
in the NJL mean-field case. Formally, the infinite series for $\epsilon(m)$ given in Fig. (\ref{lw1}) can be summed, to give 
\begin{eqnarray}
\epsilon(m)&=&i\int\frac{d^4p}{(2\pi)^4}\sum_{n=1}^{\infty}\frac{(-1)}{2n}{\rm Tr}\left[(-i)^2\left(\frac{i}{\slashed{p}+i\epsilon}\right)^2m^2\right]^n
\nonumber\\
&=&i\int \frac{d^4p}{(2\pi)^4}{\rm Tr~ln}\left[\frac{\slashed{p}-m+i\epsilon}{\slashed{p}+i\epsilon}\right],
\label{L37}
\end{eqnarray}
just as needed for (\ref{L21}). 

In (\ref{L35}) we note that the contour for the $p_0$ integration in each of the $G^{(n)}_0$ is that associated with a massless Feynman propagator, and not that associated with a massive one. However, for both the $1/(\slashed{p}+i\epsilon)$ and $1/(\slashed{p}-m+i\epsilon)$ propagators the poles are located below the real $p_0$ axis when $p_0$ is positive, and above the real $p_0$ axis when $p_0$ is negative (upper left and lower right quadrants of the complex $p_0$ plane). Thus in both the cases we can make the same Wick rotation along quarter circles in the upper right and lower left quadrants, to obtain a Wick contour loop that contains no poles within it, to thus yield 
\begin{eqnarray}
\int_{-\infty}^{\infty}dp_0+\int_{\infty}^{i\infty}dp_0+\int_{i\infty}^{-i\infty}dp_0+\int_{-i\infty}^{-\infty}dp_0=0.
\label{L38}
\end{eqnarray}
Then, with the two quarter circle at infinity terms being well-enough behaved that we are able to drop them, we obtain
\begin{eqnarray}
-i\int_{-\infty}^{\infty}dp_0=-i\int_{-i\infty}^{i\infty}dp_0=\int_{-\infty}^{\infty}dp_4
\label{L39}
\end{eqnarray}
where $p_4=-ip_0$. 

As constructed, for determining $\epsilon(m)$ we should in general use (\ref{L35}) with its massless fermion propagator contour, and even if we can do the infinite sum and obtain some function of a massive fermion propagator, we should continue to use the same massless fermion contour, i.e. we should Wick rotate every term in (\ref{L35}) before doing the summation. In fact, we already did so to derive (\ref{L37}). Specifically, we note that we can actually only set $x-x^2/2+x^3/3+...={\rm ln}(1+x)$ if $x <1$. With $x=m^2/p^2$, we first Wick rotate each term in the sum in (\ref{L37}), then do the integration from $p^2=a^2$ to $p^2=\Lambda^2$ for each term where we must take $a^2>m^2$, and finally then do the summation. The expression that results is then proportional to  the Wick rotated $I(a^2)=\int_{a^2}^{\Lambda^2} dp^2 p^2 {\rm ln}(1+p^2/m^2)$. Since $I(a^2)$ is found to be an analytic function of $a^2$ all the way to $a^2=0$ (where there is a branch point), $I(a^2)$  can be continued to $a^2=0$, from which the Wick rotated form of (\ref{L37}), an integral over all $p^2$, then follows.

Now for the NJL case it does not actually  matter whether we use the massless or the massive Wick contours since they happen to coincide, with neither containing any poles. However, as we show below, in the JBW electrodynamics case the two Wick rotations do not coincide (for general $\Sigma(p)$ the $1/(\slashed{p}-\Sigma(p)+i\epsilon)$ propagator can have a much more complicated structure in the complex $p_0$ plane), and we must use the massless Wick contour loop since that is what (\ref{L35}) requires. Moreover, for the JBW case the great utility of  (\ref{L35}) is that while the scaling solution given in (\ref{L3}) only applies for $p^2 \gg m^2$, if there is to be critical scaling in the massless theory,  then scaling forms would hold at all momenta as there is no mass scale in the massless theory. Thus, even if a theory with a mass is only scale invariant for large momenta, its $\bar{\psi}\psi$ Green's functions can be constructed by an infinite summation of graphs all of which are scale invariant for all momenta, and all of which use the massless theory Feynman propagator contour.

\subsection{General Requirements for the Generation of Goldstone and Higgs Bosons}

To summarize, given the Baker-Johnson evasion of the Goldstone theorem and the constraints that the renormalization process can produce, we see that in order to generate a Goldstone boson in a renormalizable quantum field theory via dynamical symmetry breaking four conditions need to be met. First, we need to show that the unbroken vacuum possesses a tachyonic instability. Second, we need to show that the fermion mass obeys a self-consistent gap type equation. Third, we need to show that the vacuum associated with the non-trivial solution to the self-consistent gap type equation has lower energy density than the vacuum associated with the trivial solution. And fourth, we need to show that there is in fact a massless pole in the fermion-antifermion scattering amplitude. In Sec. IV we shall show that in JBW electrodynamics coupled to a four-fermion interaction all four of these criteria are met when $\gamma_{\theta}(\alpha)=-1$.

As regards the Higgs boson,  if we can produce a pseudoscalar bound state at all, then in a chirally symmetric theory we must get a scalar bound state as well. The two states will necessarily be degenerate in mass if the symmetry is unbroken. However, when the symmetry is broken, the mass degeneracy of the two states will be lifted, with the scalar bound state necessarily acquiring a mass of order the symmetry breaking scale, so that there is then no hierarchy problem for it, and no need to utilize an alternative such as the breaking of scale invariance for instance in order to control its mass.

\section{JBW Electrodynamics Coupled to a Four-Fermion Interaction}

\subsection{Vacuum Energy Density for Arbitrary $\gamma_{\theta}(\alpha)$}

As described above, we decompose the  ${\cal{L}}_{\rm QED-FF}$ Lagrangian associated with a massless fermion QED coupled to a four-fermion interaction into mean-field and residual interaction pieces according to 

\begin{eqnarray}
{\cal {L}}_{\rm QED-FF}&=&-\frac{1}{4}F_{\mu\nu}F^{\mu\nu}+\bar{\psi}\gamma^{\mu}(i\partial_{\mu}-eA_{\mu})\psi 
-\frac{g}{2}[\bar{\psi}\psi]^2-\frac{g}{2}[\bar{\psi}i\gamma_5\psi]^2
\nonumber\\
&=&-\frac{1}{4}F_{\mu\nu}F^{\mu\nu}+\bar{\psi}\gamma^{\mu}(i\partial_{\mu}-eA_{\mu})\psi 
-m\bar{\psi}\psi +\frac{m^2}{2g}
-\frac{g}{2}\left(\bar{\psi}\psi-\frac{m}{g}\right)^2-\frac{g}{2}\left(\bar{\psi}i\gamma_5\psi\right)^2\
\nonumber\\
&=&{\cal{L}}_{\rm QED-MF}+{\cal{L}}_{\rm QED-RI}.
\label{L40}
\end{eqnarray}
According to (\ref{L35}), in order to determine the $\epsilon(m)$ associated with ${\cal{L}_{\rm QED-MF}}$ we need to sum an infinite number of massless theory graphs. With our assumption of critical scaling, as noted in \cite{Mannheim1974a,Mannheim1975} these massless graphs can be obtained from the NJL point vertex graphs by replacing point vertices with $\Gamma_{\rm S}(p,p,0)=1$ by the fully dressed and renormalized $\tilde{\Gamma}_{\rm S}(p,p,0)$. However, since these needed vacuum energy density graphs are massless theory graphs we need the massless theory $\tilde{\Gamma}_{\rm S}(p,p,0)$, and in order to renormalize it we shall use an off-shell renormalization with a parameter $\mu^2$. Then, since there is no scale in the massless theory, the assumption of critical  scaling with anomalous dimensions allows us to set
\begin{eqnarray}
\tilde{S}^{-1}(p,m=0)&=& \slashed{p}+i\epsilon,
\nonumber\\
\tilde{\Gamma}_{\rm S}(p,p,0,m=0)&=&\left(\frac{-p^2-i\epsilon}{\mu^2}\right)^{\gamma_{\theta}(\alpha)/2} 
\label{L41}
\end{eqnarray}
for all momenta in the massless theory. Moreover, a critical scaling massless theory will not be just scale invariant, it will be conformal invariant too. Thus as well as requiring all massless theory Green's functions to be constrained by scale invariance and the renormalization group equations, for  two- and three-point functions conformal invariance  fixes their form completely. Thus in the massless theory we can write the exact relation
\begin{eqnarray}
&&\langle \Omega_0|T(\psi(x):\bar{\psi}(z)\psi(z):\bar{\psi}(y))|\Omega_0\rangle
= \frac{\mu^{-\gamma_{\theta}}(\slashed{y} -\slashed{z})(\slashed{z} -\slashed{x})}{[(y-z)^2(z-x)^2]^{(1+d_{\theta})/2}[(x-y)^2]^{(3-d_{\theta})/2}},
\label{L42}
\end{eqnarray}
with the form for $\tilde{\Gamma}_{\rm S}(p,p,0,m=0)$ given in (\ref{L41}) then following upon a Fourier transform and an amputation of the external fermion legs \cite{Mannheim1975}. In (\ref{L42}) we should note that the normal ordering is with respect to $|\Omega_0\rangle$, and we can use this normal ordering prescription for all Green's functions other than those related to the vacuum energy density, as the vacuum energy density plays no role in the standard Dyson-Wick expansion of Green's functions. Now we could normalize $\mu$ so that it is equal to the eventual dynamical mass $M$ right away, but for tracking where everything comes from  it is more convenient to keep it as is until the end. 

\begin{figure}[htpb]
\begin{center}
\includegraphics[scale=0.3]{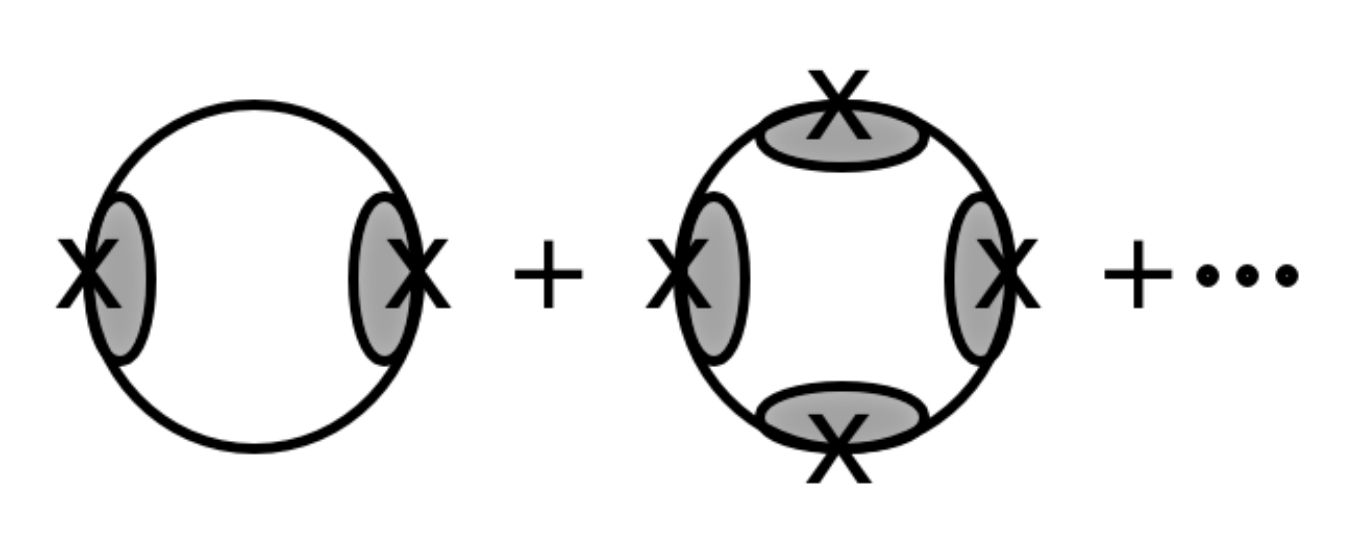}
\caption{Vacuum energy density  $\epsilon(m)$ via an infinite summation of massless graphs with zero-momentum dressed $m\bar{\psi}\psi$ insertions.}
\label{lw2}
\end{center}
\end{figure}

Given (\ref{L41}), we replace (\ref{L37}) by
\begin{eqnarray}
\epsilon(m)&=&i\int\frac{d^4p}{(2\pi)^4}\sum_{n=1}^{\infty}\frac{(-1)}{2n}
{\rm Tr}\left[(-i)^2\left(\frac{-p^2-i\epsilon}{\mu^2}\right)^{\gamma_{\theta}(\alpha)}\left(\frac{i}{\slashed{p}+i\epsilon}\right)^2m^2\right]^n
\nonumber\\
&=&\frac{i}{2}\int \frac{d^4p}{(2\pi)^4}{\rm Tr~ln}\left[1-\frac{m^2}{p^2+i\epsilon}\left(\frac{-p^2-i\epsilon}{\mu^2}\right)^{\gamma_{\theta}(\alpha)}\right],
\nonumber\\
\label{L43}
\end{eqnarray}
with the infinite summation of massless graphs in Fig. (\ref{lw1}) being replaced by the infinite summation in Fig. (\ref{lw2}).

In terms of the quantity
\begin{eqnarray}
\tilde{S}^{-1}_{\mu}(p)&=& \slashed{p}-m\left(\frac{-p^2-i\epsilon}{\mu^2}\right)^{\gamma_{\theta}(\alpha)/2}+i\epsilon,\label{L44}
\end{eqnarray}
we can rewrite $\epsilon(m)$ as 
\begin{eqnarray}
\epsilon(m)=i\int \frac{d^4p}{(2\pi)^4}\left[{\rm Tr~ln}(\tilde{S}^{-1}_{\mu}(p))-{\rm Tr~ln}(\slashed{p}+i\epsilon)\right].
\label{L45}
\end{eqnarray}
Given the form of (\ref{L45}), on comparing with (\ref{L3}) it is suggestive to think of $\tilde{S}^{-1}_{\mu}(p)$ as the massive theory propagator. However, it cannot be, since, as constructed, (\ref{L3}) only gives the asymptotic form for the massive propagator. Moreover, in the massive theory  the renormalization group equation for  the Green's function involving a further $\bar{\psi}\psi$ insertion is of the form \cite{Adler1971}
\begin{eqnarray}
&&\left[m\frac{\partial}{m}+\beta(\alpha)\frac{\partial}{\partial \alpha}+\gamma_{\theta}(\alpha)\right]\tilde{\Gamma}_{\rm S}(p,p,0,m)
=m(1-\gamma_{\theta}(\alpha)]\tilde{\Gamma}_{\rm SS}(p,p,0,m),
\label{L46}
\end{eqnarray}
where $\tilde{\Gamma}_{\rm SS}$ contains two zero-momentum $\bar{\psi}\psi$ insertions. Since  $\tilde{\Gamma}_{\rm SS}$ is not zero identically,  $\tilde{\Gamma}_{\rm S}$ and thus $\tilde{S}^{-1}(p,m)$ of (\ref{L3}) must have non-leading terms beyond those exhibited in (\ref{L3}). Since on power counting grounds  $[m\partial_m+\beta(\alpha)\partial_{ \alpha}+2\gamma_{\theta}(\alpha)]\tilde{\Gamma}_{\rm SS}(p,p,0,m)$ will be related to a $\tilde{\Gamma}_{\rm SSS}$ that contains three  zero-momentum insertions, $\tilde{\Gamma}_{\rm SS}(p,p,0,m)$  will  acquire a leading term of the form $(-p^2/m^2)^{\gamma_{\theta}(\alpha)}$, while $\tilde{\Gamma}_{\rm S}(p,p,0,m)$ will acquire a non-leading term of the form $m(-p^2/m^2)^{\gamma_{\theta}(\alpha)}$. Consequently, $\tilde{S}^{-1}(p,m)$ will then behave  as $\tilde{S}^{-1}(p,m)= \slashed{p}-m(-p^2/m^2)^{\gamma_{\theta}(\alpha)/2}-m^2(-p^2/m^2)^{\gamma_{\theta}(\alpha)}$. Further non-leading terms would then be generated via the renormalization group for Green's functions with even more insertions. Thus $\tilde{S}^{-1}_{\mu}(p)$ is not the exact fermion propagator for JBW electrodynamics. And while we can treat $m$ as a dynamically generated parameter, one which is associated with chiral symmetry breaking and which sets the mass scale, we cannot identify it with the position of the pole in the exact fermion propagator, even though it would set the scale for it. 

Despite this, $\tilde{S}^{-1}_{\mu}(p)$  can still serve our purposes here not as the exact propagator but as the mean-field theory propagator needed for evaluating the mean-field theory Feynman graphs that we study, since the vertex function $\tilde{\Gamma}_{\rm S}(p,p,0,m=0)=(p^2/\mu^2)^{\gamma_{\theta}(\alpha)/2} $ is the exact, all momentum,  vertex function in a critical scaling massless theory. However, even in this mean-field theory we cannot evaluate Feynman graph contours using the pole structure of $\tilde{S}^{-1}_{\mu}(p)$ since its poles are not located on the real  $p_0$ axis and would therefore contribute in a Wick rotation. However, as we noted above, the $p_0$ contour is fixed not by the massive theory but by the underlying massless one. Thus, as we elaborate on in more detail below, we must evaluate (\ref{L43}) using the massless theory Wick contour just as given in (\ref{L39}). Nonetheless,  we will continue to utilize the $\tilde{S}^{-1}_{\mu}(p)$ propagator since it is very convenient for bookkeeping purposes. (To actually determine the full fermion propagator in our theory, we would need to include the effect of the residual four-fermion interaction on $\tilde{S}^{-1}_{\mu}(p)$, something that is beyond the scope of the present paper and anyway not needed for the study of the collective Goldstone and Higgs bosons that are of interest to us here.)

\subsection{Vacuum Energy Density for $\gamma_{\theta}(\alpha)=-1$}

\begin{figure}[htpb]
\begin{center}
\includegraphics[scale=0.4]{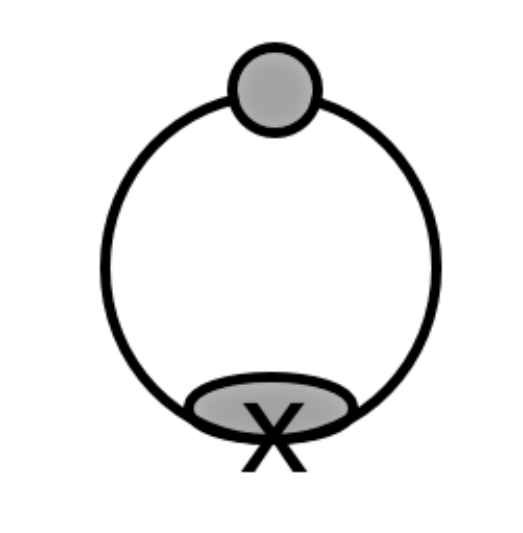}
\caption{The  $\gamma_{\theta}(\alpha)=-1$ tadpole graph for $\langle \Omega_m|\bar{\psi}\psi|\Omega_m\rangle$ with a zero-momentum dressed $m\bar{\psi}\psi$ insertion and a dressed $\tilde{S}_{\mu}(p)$ propagator.}
\label{lw3}
\end{center}
\end{figure}

When $\gamma_{\theta}(\alpha)=-1$, evaluation of $\epsilon(m)$ is straightforward, and yields \cite{Mannheim1974a,Mannheim1975}
\begin{eqnarray}
\epsilon(m)=-\frac{m^2\mu^2}{8\pi^2}\left[ {\rm ln}\left(\frac{\Lambda^2}{m\mu}
\right)+\frac{1}{2}\right]. 
\label{L47}
\end{eqnarray}
With $\epsilon^{\prime}(m)$ being equal to $\langle \Omega_m|\bar{\psi}\psi|\Omega_m\rangle$, in the Hartree-Fock approximation we obtain (Fig. (\ref{lw3}))
\begin{eqnarray}
\langle \Omega_m|\bar{\psi}\psi|\Omega_m\rangle&=&\epsilon^{\prime}(m)
=-i\int \frac{d^4p}{(2\pi)^4}{\rm Tr}[\tilde{\Gamma}_{\rm S}(p,p,0,m=0)\tilde{S}_{\mu}(p)]
\nonumber\\
&=&4i\int \frac{d^4p}{(2\pi)^4}\frac{m\mu^2}{(p^2+i\epsilon)^2+m^2\mu^2}
=-\frac{m\mu^2}{4\pi^2}{\rm ln}\left(\frac{\Lambda^2}{m\mu}
\right)=\frac{m}{g}. 
\label{L48}
\end{eqnarray}

We thus identify the physical mass as the one that satisfies (\ref{L48}) according to the manifestly non-perturbative gap-type equation
\begin{eqnarray}
-\frac{\mu^2}{4\pi^2}{\rm ln}\left(\frac{\Lambda^2}{M\mu}
\right)=\frac{1}{g},\qquad M=\frac{\Lambda^2}{\mu}\exp\left(\frac{4\pi^2}{\mu^2g}\right). 
\label{L49}
\end{eqnarray}
Thus just as in BCS, dynamical symmetry breaking will occur no matter how weak the four-fermion coupling constant  $g$ might be as long as it is attractive, and again like in BCS, there is an essential singularity at $g=0$.  Since the general wisdom on dynamical symmetry breaking is that it requires strong coupling (see Sec. IV-J below), our study here provides a counterexample to this wisdom, something we discuss in detail in Sec. IV-J. Moreover, we should note that not only do we not need $-g$ to be large, if the solution to $\beta(\alpha)=0$ is given not by the bare charge but by the physical fine-structure constant (the possibility studied in \cite{Adler1972}), then the QED sector would be weakly coupled too.

\begin{figure}[htpb]
\begin{center}
\includegraphics[scale=0.3]{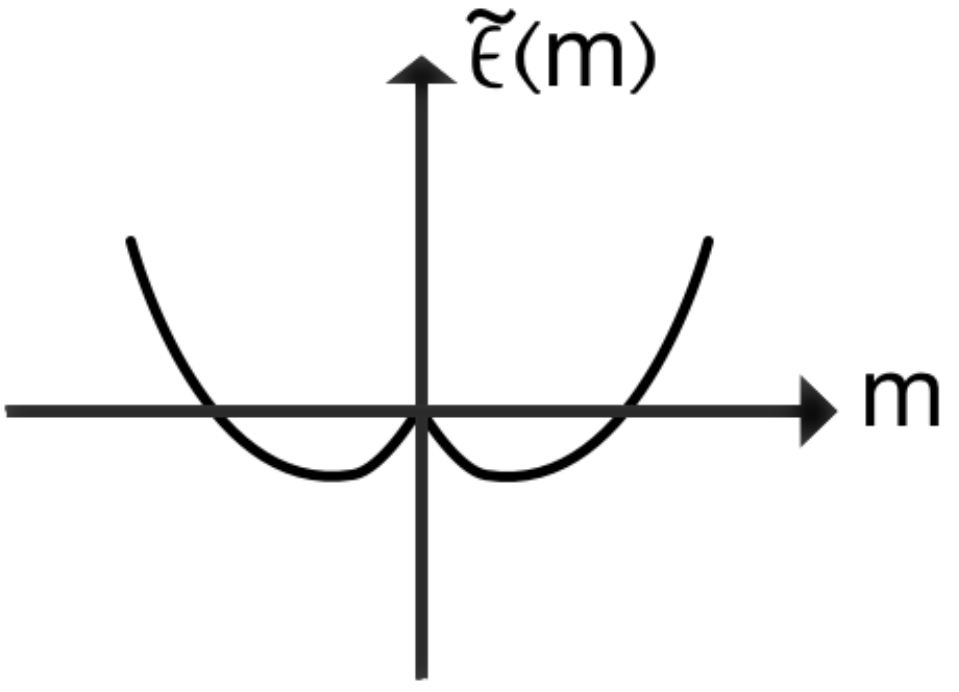}
\caption{Dynamically generated double-well potential for the renormalized vacuum energy density when $\gamma_{\theta}(\alpha)=-1$.}
\label{lw4}
\end{center}
\end{figure}
Finally, recalling the $m^2/2g$ counterterm in (\ref{L40}), we can write the renormalized mean-field vacuum energy density just as previously given in (\ref{L10}), viz. (see Fig. (\ref{lw4}))
\begin{eqnarray}
\tilde{\epsilon}(m)=\epsilon(m)-\frac{m^2}{2g}=\frac{m^2\mu^2}{16\pi^2}\left[{\rm ln}\left(\frac{m^2}{M^2}\right)-1\right],
\label{L50}
\end{eqnarray}
with its local maximum at $m=0$ and its global minimum at $m=M$. Quite remarkably, with only the one counterterm, $m^2/2g$, as expressly provided by the mean-field theory, we find that $\tilde{\epsilon}(m)$ is completely finite.  This then is the power of dynamical symmetry breaking, it generates appropriate counterterms automatically.

\subsection{Higgs-Like Lagrangian}

To develop an analog of a kinetic energy term to add on to $\tilde{\epsilon}(m)$, we need to determine 
the massive theory $\Pi_{\rm S}(x,m)$ as defined in (\ref{L24}). In the massless theory first, we can use conformal invariance to determine $\Pi_{\rm S}(x,m=0)$ exactly. Thus we set
\begin{eqnarray}
&&\langle \Omega_0|T(:\bar{\psi}(x)\psi(x)::\bar{\psi}(y)\psi(y):)|\Omega_0\rangle
=\frac{\mu^{-2\gamma_{\theta}}{\rm Tr}[(\slashed{x}-\slashed{y})(\slashed{y}-\slashed{x})]}{[(x-y)^2]^{(d_{\theta}+1)/2}[(y-x)^2]^{(d_{\theta}+1)/2}}
=-\frac{4\mu^{-2\gamma_{\theta}}}{[(x-y)^2]^{d_{\theta}}}.
\label{L51}
\end{eqnarray}
With an appropriate normalization Fourier transforming then gives 
\begin{eqnarray}
\Pi_{\rm S}(q^2,m=0)&=&-i\int \frac{d^4p}{(2\pi)^4}{\rm Tr}\bigg[[p^2(p+q)^2]^{\gamma_{\theta}(\alpha)/4}
\frac {1}{\slashed{p}}[p^2(p+q)^2]^{\gamma_{\theta}(\alpha)/4}\frac {1}{\slashed{p} +\slashed{q}}\bigg].
\label{L52}
\end{eqnarray}
As well as construct $\Pi_{\rm S}(x,m=0)$ via conformal invariance we can start with its definition as $\langle \Omega_0|T(:\bar{\psi}(x)\psi(x)::\bar{\psi}(y)\psi(y):)|\Omega_0\rangle$ and make a Dyson-Wick contraction between the fields at $x_{\mu}$ and $y_{\mu}$. At the one-loop level this then yields 
\begin{eqnarray}
&&\Pi_{\rm S}(q^2,m=0)=-i\int \frac{d^4p}{(2\pi)^4}{\rm Tr}\bigg[\tilde{\Gamma}_{\rm S}(p+q,p,-q,m=0)
 \tilde{S}(p,m=0)\tilde{\Gamma}_{\rm S}(p,p+q,q,m=0)\tilde{S}(p+q,m=0)\bigg],
\label{L53}
\end{eqnarray}
where the massless  $\tilde{S}(p,m=0)$ is given in (\ref{L41}), and where we have introduced
\begin{eqnarray}
\tilde{\Gamma}_{\rm S}(p,p+q,q,m=0)=\left[\frac{(-p^2)}{\mu^2}\frac{(-(p+q)^2)}{\mu^2}\right]^{\gamma_{\theta}(\alpha)/4}
=\tilde{\Gamma}_{\rm S}(p+q,p,-q,m=0).
\label{L54}
\end{eqnarray}

Now that we know the $\tilde{\Gamma}_{\rm S}(p,p+q,q,m=0)$ vertex needed for $\Pi_{\rm S}(q^2,m=0)$, just as with the infinite summation of massless theory graphs associated with the generation of the massive theory $\epsilon(m)$, the massive theory $\Pi_{\rm S}(q^2,m)$ is also given by an infinite summation. In this summation, apart from the two $\bar{\psi}(x)\psi(x)$ insertions that carry momentum $q_{\mu}$, all other insertions carry zero-momentum and couple with vertices that are given by (\ref{L41}). The summation thus results in massless fermion propagators being replaced by massive ones according to \cite{Mannheim1978}
\begin{eqnarray}
&&\Pi_{\rm S}(q^2,m)=-i\int \frac{d^4p}{(2\pi)^4}{\rm Tr}\bigg[\tilde{\Gamma}_{\rm S}(p+q,p,-q,m=0)
 \tilde{S}_{\mu}(p)\tilde{\Gamma}_{\rm S}(p,p+q,q,m=0)\tilde{S}_{\mu}(p+q)\bigg],
\label{L55}
\end{eqnarray}
where the massive theory $\tilde{S}_{\mu}(p)$ is given in (\ref{L44}). 

\begin{figure}[htpb]
\begin{center}
\includegraphics[scale=0.3]{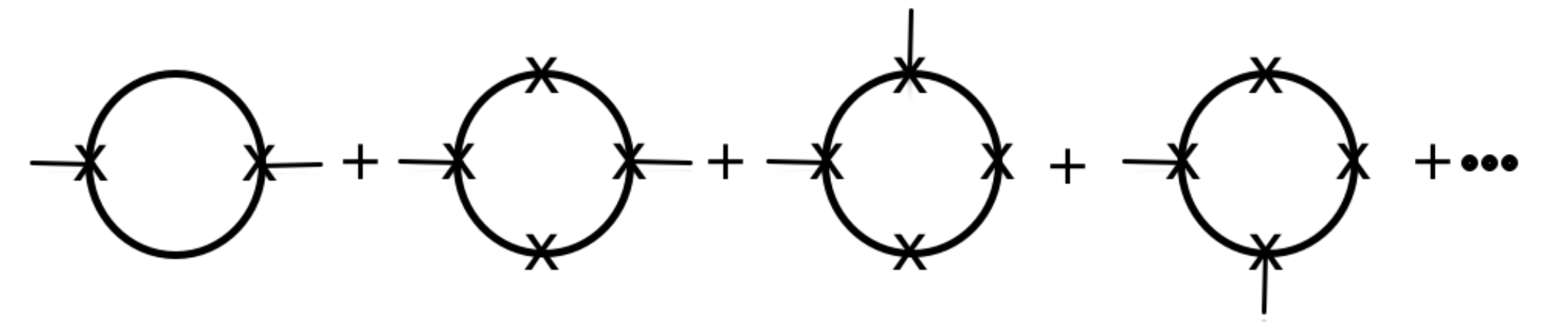}
\caption{$\Pi_{\rm S}(q^2,m(x))$ developed as an infinite summation of massless graphs, each with two point $m\bar{\psi}\psi$ insertions carrying momentum $q_{\mu}$ (shown as external lines), with all other point $m\bar{\psi}\psi$ insertions carrying zero momentum.}
\label{lw5}
\end{center}
\end{figure}

\begin{figure}[htpb]
\begin{center}
\includegraphics[scale=0.3]{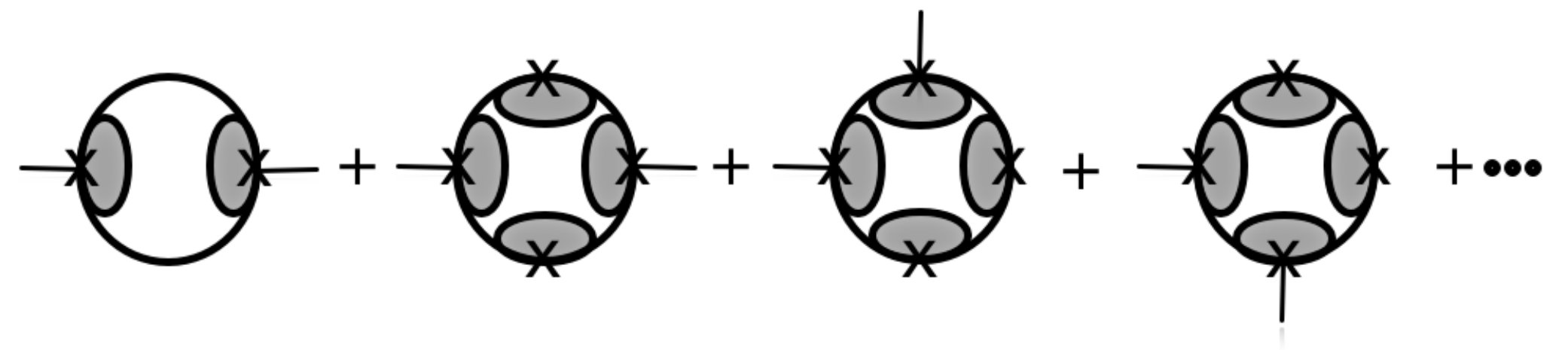}
\caption{$\Pi_{\rm S}(q^2,m(x))$ developed as an infinite summation of massless graphs, each with two dressed $m\bar{\psi}\psi$  insertions carrying momentum $q_{\mu}$ (shown as external lines), with all other dressed $m\bar{\psi}\psi$  insertions carrying zero momentum.}
\label{lw6}
\end{center}
\end{figure}

As discussed in \cite{Mannheim1976}, to now get the coefficient of the kinetic energy associated with a coherent state in which $\langle C|\bar{\psi}\psi|C\rangle =m(x)$ with spacetime dependent $m(x)$, we need to calculate the derivative of $\Pi_{\rm S}(q^2,m(x))$ at $q^2=0$. Now even though the massive $\Pi_{\rm S}(-\partial_{\mu}\partial^{\mu},m(x))$ depends on the spacetime coordinates when $m(x)$ depends on the spacetime coordinates,  we note that if we develop $\Pi_{\rm S}(-\partial_{\mu}\partial^{\mu},m(x))$ as an infinite sum of massless graphs, for each of those graphs there are no spacetime dependent $m(x)$  factors inside the graphs themselves, and so we can evaluate the graphs using standard momentum space Feynman diagram techniques. Graphically, in the NJL case first we evaluate $\Pi_{\rm S}(-\partial_{\mu}\partial^{\mu},m(x))$ using the summation in Fig. (\ref{lw5}), and then in the JBW case we use the summation given in Fig. (\ref{lw6}).

Via  the summation in the NJL case  the kinetic energy terms given in (\ref{L22}) and (\ref{L23}) were obtained in \cite{Mannheim1976}. In the JBW case with $\gamma_{\theta}(\alpha)=-1$, an expansion of $\Pi_{\rm S}(q^2=0,m)$ around $q^2=0$ is algebraically found to give the $q^2$ derivative $\Pi^{\prime}_{\rm S}(q^2,m)=-3\mu/128\pi m$ at $q^2=0$. This then yields an effective Higgs-like Lagrangian of the form \cite{Mannheim1978}
\begin{eqnarray}
{\cal{L}}_{\rm EFF}&=&-\tilde{\epsilon}(m(x))
-\frac{1}{2}m(x)[\Pi_{\rm S}(-\partial_{\mu}\partial^{\mu},m(x))-\Pi_{\rm S}(0,m(x))
]m(x)+...
\nonumber\\
&=&-\frac{m^2(x)\mu^2}{16\pi^2}\left[{\rm ln}\left(\frac{m^2(x)}{M^2}\right)-1\right]
+\frac{3\mu}{256\pi m(x)}\partial_{\mu}m(x)\partial^{\mu}m(x)+....
\label{L56}
\end{eqnarray}
Here the dots denote higher gradient terms, and there is no reason to be concerned about their presence since $m(x)$ is only a c-number, and thus (\ref{L56}) would not be associated with a non-renormalizable or non-local field theory if the higher gradient  terms are included. Rather, (\ref{L56}) is generated by dynamical symmetry breaking in a local, renormalizable field theory, one which leads to the expansion given in (\ref{L56}) in which every term is automatically finite. In this way, without ever introducing any input fundamental tachyonic mass term, we can generate an effective double-well Higgs Lagrangian, one which could readily be coupled to a gauge field just as in (\ref{L23}).

\subsection{The Collective Tachyon Modes when the Fermion is Massless}

To test for tachyons we need to evaluate the massless theory $\Pi_{\rm S}(q^2,m=0)$ given above and also the pseudoscalar $\Pi_{\rm P}(q^2,m=0)$, which is given by
\begin{eqnarray}
\Pi_{\rm P}(q^2,m=0)&=&-i\int \frac{d^4p}{(2\pi)^4}{\rm Tr}\bigg[[p^2(p+q)^2]^{\gamma_{\theta}(\alpha)/4}i\gamma_5
\frac {1}{\slashed{p}}[p^2(p+q)^2]^{\gamma_{\theta}(\alpha)/4}i\gamma_5\frac {1}{\slashed{p} +\slashed{q}}\bigg].
\label{L57}
\end{eqnarray}
When $\gamma_{\theta}(\alpha)=-1$ a straightforward Wick rotation with spacelike $q^2$ yields 
\begin{eqnarray}
&&\Pi_{\rm S}(q^2,m=0)=\Pi_{\rm P}(q^2,m=0)
=-\frac{\mu^2}{4\pi^2}\bigg[{\rm ln}\left(\frac{\Lambda^2}{(-q^2)}\right)-3+4~{\rm ln}2\bigg].
\label{L58}
\end{eqnarray}
With $g^{-1}$ being given in (\ref{L49}), we thus see that both $g^{-1}-\Pi_{\rm S}(q^2,m=0)$ and $g^{-1}-\Pi_{\rm P}(q^2,m=0)$ are finite, with the four-fermion interaction thus supplying just the needed counterterm to make both the massless $\Pi_{\rm S}(q^2,m=0)$ and the massless $\Pi_{\rm P}(q^2,m=0)$ be finite.

With the $T$ matrix being given by 
\begin{eqnarray}
T_{\rm S}(q^2,m=0)&=&\frac{g}{1-g\Pi_{\rm S}(q^2,m=0)}=\frac{1}{g^{-1}-\Pi_{\rm S}(q^2,m=0)},
\nonumber\\
T_{\rm P}(q^2,m=0)&=&\frac{g}{1-g\Pi_{\rm P}(q^2,m=0)}=\frac{1}{g^{-1}-\Pi_{\rm P}(q^2,m=0)},
\label{L59}
\end{eqnarray}
we see that both the scalar and pseudoscalar scattering matrices have a spacelike pole at
\begin{eqnarray}
q^2=-M\mu e^{4{\rm ln}2-3}=-0.797M\mu,
\label{L60}
\end{eqnarray}
with both amplitudes behaving as 
\begin{eqnarray}
T_{\rm S}(q^2,m=0)=T_{\rm P}(q^2,m=0)=\frac{31.448M\mu}{(q^2+0.797M\mu)}
\label{L61}
\end{eqnarray}
near the tachyonic poles. We thus confirm that the massless vacuum is unstable.

\subsection{The Collective Goldstone Mode when the Fermion is Massive}

With the massive theory  $\Pi_{\rm S}(q^2,m)$ being given by (\ref{L55}), because of the chiral invariance of the massless theory vertices the analogous massive theory $\Pi_{\rm P}(q^2,m)$ is given by
\begin{eqnarray}
&&\Pi_{\rm P}(q^2,m)=-i\int \frac{d^4p}{(2\pi)^4}{\rm Tr}\bigg[\tilde{\Gamma}_{\rm S}(p+q,p,-q,m=0)
i\gamma_5\tilde{S}_{\mu}(p)\tilde{\Gamma}_{\rm S}(p,p+q,q,m=0)i\gamma_5\tilde{S}_{\mu}(p+q)\bigg].
\label{L62}
\end{eqnarray}
Both the massive $\Pi_{\rm S}(q^2,m)$ and the massive $\Pi_{\rm P}(q^2,m)$ are logarithmically divergent  when $\gamma_{\theta}(\alpha)=-1$, with the divergence being the same as that of the massless $\Pi_{\rm S}(q^2,m=0)$ and $\Pi_{\rm P}(q^2,m=0)$ since the large momentum behavior of the Green's functions is not sensitive to the fermion mass. Consequently, both $g^{-1}-\Pi_{\rm S}(q^2,m)$ and $g^{-1}-\Pi_{\rm P}(q^2,m)$ are finite, with the four-fermion interaction thus supplying just the needed counterterm to make both the massive $\Pi_{\rm S}(q^2,m)$ and the massive $\Pi_{\rm P}(q^2,m)$ be finite.

When $\gamma_{\theta}(\alpha)=-1$, $\Pi_{\rm S}(q^2,m)$ and $\Pi_{\rm P}(q^2,m)$ evaluate to
\begin{eqnarray}
&&\Pi_{\rm S}(q^2,m)=-4i\mu^2\int \frac{d^4p}{(2\pi)^4}\frac{N(q,p)+m^2\mu^2}{D(q,p,m)},
\nonumber\\
&&\Pi_{\rm P}(q^2,m)=-4i\mu^2\int \frac{d^4p}{(2\pi)^4}\frac{N(q,p)-m^2\mu^2}{D(q,p,m)},
\label{L63}
\end{eqnarray}
where
\begin{eqnarray}
N(q,p)&=&(p^2+i\epsilon-q^2/4)
(-(p-q/2)^2-i\epsilon)^{1/2}(-(p+q/2)^2-i\epsilon)^{1/2},
\nonumber\\
D(q,p,m)&=&(((p-q/2)^2+i\epsilon)^2+m^2\mu^2)
(((p+q/2)^2+i\epsilon)^2+m^2\mu^2).
\label{L64}
\end{eqnarray}
(In (\ref{L63}) and (\ref{L64}) we have conveniently translated $p_{\mu}$ to $p_{\mu}-q_{\mu}/2$.)

On now evaluating $\Pi_{\rm P}(q^2,m)$ at $q^2=0$ we obtain
\begin{eqnarray}
\Pi_{\rm P}(q^2=0,m)&=&-4i\mu^2\int \frac{d^4p}{(2\pi)^4}\frac{(p^2)(-p^2)-m^2\mu^2}{((p^2+i\epsilon)^2+m^2\mu^2)^2}.
\nonumber\\
&=&4i\mu^2\int \frac{d^4p}{(2\pi)^4}\frac{1}{(p^2+i\epsilon)^2+m^2\mu^2}.
\label{L65}
\end{eqnarray}
On comparing with (\ref{L48}) and (\ref{L49}) we see that when $m$ is equal to $M$, $\Pi_{\rm P}(q^2=0,M)$ is equal to none other than $g^{-1}$. In the pseudoscalar $T_{\rm P}(q^2,M)=[g^{-1}-\Pi_{\rm P}(q^2,M)]^{-1}$ channel we thus obtain our sought-after massless pseudoscalar Goldstone boson. Finally, with an expansion of $\Pi_{\rm P}(q^2=0,M)$ around $q^2=0$ algebraically being found to give the $q^2$ derivative $\Pi^{\prime}_{\rm P}(q^2,M)=-7\mu/128\pi M$ at $q^2=0$, near the Goldstone pole $T_{\rm P}(q^2,M)$ is found to evaluate to 
\begin{eqnarray}
T_{\rm P}(q^2,M)=\frac{128\pi M}{7\mu q^2}=\frac{57.446 M}{\mu q^2}.
\label{L66}
\end{eqnarray}
Now at $\gamma_{\theta}(\alpha)=-1$ the quantity $g^{-1}$ is infinite (as counterterms need to be) and $g$ itself is zero. Thus even though the $\Pi_{\rm P}$-independent homogeneous term in $T_{\rm P}=g+g\Pi_{\rm P} g+g\Pi_{\rm P} g\Pi_{\rm P} g+...=g/(1-g\Pi_{\rm P})$ would be zero, nonetheless the interplay between the numerator and the denominator still enables a pole to be generated. Thus, as we had noted above, even if the homogeneous term in a scattering amplitude iteration vanishes there still could be a pole. Thus in conclusion we note that even though JBW electrodynamics does not on its own have a Goldstone boson pole, when it is coupled to the four-fermion interaction it then does.

\subsection{Renormalizability of the Four-Fermion Interaction}

The fact that the $\Pi_{\rm S}(q^2,M)$ and $\Pi_{\rm P}(q^2,M)$ Green's functions are only logarithmically divergent, and the fact that accordingly the $T_{\rm S}(q^2,M)=[g^{-1}-\Pi_{\rm S}(q^2,M)]^{-1}$ and $T_{\rm P}(q^2,M)=[g^{-1}-\Pi_{\rm P}(q^2,M)]^{-1}$ scattering amplitudes are both finite is due to two separate effects, one an ultraviolet effect and the other an infrared one. From the perspective of the ultraviolet structure of the theory alone, one finds that with $\gamma_{\theta}(\alpha)=-1$, the divergences in $\Pi_{\rm S}(q^2,m=0)$ and $\Pi_{\rm P}(q^2,M)$ are only logarithmic, and not the quadratic ones that  they would have been with $\gamma_{\theta}(\alpha)=0$ (the NJL situation). Thus given this, one is free to choose what is initially an  arbitrary $g^{-1}$ so that it diverges logarithmically too, and one is free to pick its coefficient so that both $T_{\rm S}(q^2,M)$ and $T_{\rm P}(q^2,M)$ then become finite. In this sense $g^{-1}$ acts as a renormalization counterterm, with the $\gamma_{\theta}(\alpha)=-1$ condition making $T_{\rm S}(q^2,M)$ and $T_{\rm P}(q^2,M)$ renormalizable to lowest order in the four-fermion coupling constant $g$. 

However, when one introduces the infrared (long range order) Hartree-Fock self-consistent condition for $g^{-1}$ as given in (\ref{L48}), we find that its structure is such that $g^{-1}$ also diverges logarithmically, and does so  with a coefficient that precisely cancels the log divergences in  $\Pi_{\rm S}(q^2,M)$ and $\Pi_{\rm P}(q^2,M)$, so that $T_{\rm S}(q^2,M)$ and $T_{\rm P}(q^2,M)$ are then automatically rendered finite. The condition $\gamma_{\theta}(\alpha)=-1$ thus softens the ultraviolet behavior of the theory to make $T_{\rm S}(q^2,M)$ and $T_{\rm P}(q^2,M)$  be renormalizable, while also causing dynamical symmetry breaking to occur in the infrared,  to thus automatically make  both $T_{\rm S}(q^2,M)$ and $T_{\rm P}(q^2,M)$ be finite. This then is the power of dynamical symmetry breaking.

The fact that $T_{\rm S}(q^2,M)$ and $T_{\rm P}(q^2,M)$ are both automatically made finite is not by accident. Rather, the theory has no choice. Specifically, with dynamical symmetry breaking there must be a massless Goldstone pole in $T_{\rm P}(q^2,M)$ at $q^2=0$. Since $T_{\rm P}^{-1}(q^2,M)$ is to vanish at $q^2=0$, $T_{\rm P}(q^2=0,M)$ must be finite, and thus the log divergences in $g^{-1}$ and $\Pi_{\rm S}(q^2,M)$ must cancel each other identically because of the Goldstone theorem. Then, since $T_{\rm S}(q^2,M)$ and $T_{\rm P}(q^2,M)$ have the same ultraviolet behavior because of the underlying chiral symmetry (the ultraviolet behavior being independent of the fermion mass), $T_{\rm S}(q^2,M)$ must automatically be rendered finite also. (This does not require that $T_{\rm S}^{-1}(q^2,M)$ also vanish at $q^2=0$, and we shall see below that $T_{\rm S}(q^2,M)$ has a pole at a finite location elsewhere.)

While the above analysis shows that $T_{\rm S}(q^2,M)$  and $T_{\rm P}(q^2,M)$  are both finite to lowest order in $g$, the argument immediately generalizes to all higher orders in $g$ as well, since as noted in \cite{Mannheim2016c}, to any order in $g$ there must still be a Goldstone boson. $T_{\rm P}(q^2,M)$  and $T_{\rm P}(q^2,M)$ must thus automatically be finite to all orders in $g$. In \cite{Mannheim2016c} it is shown that the way that this occurs in practice is because the higher order in $g$ corrections are found to only lead to a single log divergence in $\Pi_{\rm S}(q^2,M)$, in $\Pi_{\rm P}(q^2,M)$, and in $\langle \Omega_M|\bar{\psi}\psi|\Omega_M\rangle=M/g$, with there being no log squared terms or higher. These single log terms then cancel in  $T_{\rm P}(q^2,M)$  and $T_{\rm P}(q^2,M)$, to render both $T_{\rm P}(q^2,M)$  and $T_{\rm P}(q^2,M)$ finite to all orders in the four-fermion coupling constant $g$. Thus by dressing the $\bar{\psi}\psi$ and $\bar{\psi}i\gamma^5\psi$ vertices to all orders in QED with $\gamma_{\theta}(\alpha)=-1$ one is, for the first time as far as we know, able  to obtain a completely renormalizable scalar plus pseudoscalar $[\bar{\psi}\psi]^2+[\bar{\psi}i\gamma_5\psi]^2$ four-fermion interaction. Such a renormalizabilty is different from that employed in  $\bar{\psi}\gamma_{\mu}\psi\bar{\psi}\gamma^{\mu}\psi$ and $\bar{\psi}\gamma_{\mu}\gamma^5\psi\bar{\psi}\gamma^{\mu}\gamma^5\psi$ interactions, with the vector and axial-vector interactions instead being mediated by intermediate gauge bosons that acquire masses by the Higgs mechanism. We shall have occasion to return to the $[\bar{\psi}\psi]^2+[\bar{\psi}i\gamma_5\psi]^2$ theory when we discuss the vacuum energy density in the presence of gravity in Sec. V-G, but first we need to address the Higgs pole in the scalar scattering channel $T_{\rm S}(q^2,M)$.

\subsection{The Collective Higgs Mode when the Fermion is Massive -- the Needed Contour}

Because we were able to show that $\Pi_{\rm P}(q^2=0,M)$ and $g^{-1}$ were identically equal, we did not actually need to explicitly evaluate either quantity, and thus to establish the presence of a Goldstone pole we did not need to explicitly specify the contour needed for the $p_0$ integration. To show that there is a Higgs boson pole in the scalar channel we will need to specify the contour and will need to evaluate the $\Pi_{\rm S}(q^2,M)$ integral explicitly, since, unlike in the Goldstone case where there is an axial-vector Ward identity, there appears to be no general theorem or relevant Ward identity that would tell us a priori what value the mass of a dynamical Higgs boson  should be. Since each massive fermion graph is an infinite sum of massless fermion graphs, as noted above, the massive theory inherits its contour from the massless one. For the massless case we note that on translating $p_{\mu}$ to $p_{\mu}-q_{\mu}/2$ the massless $\Pi_{\rm S}(q^2,m=0)$ given in (\ref{L52}) evaluates to 
\begin{eqnarray}
\Pi_{\rm S}(q^2,m=0)=-4i\mu^2\int \frac{d^4p}{(2\pi)^4}\frac{N(q,p)}{D(q,p,m=0)},
\label{L67}
\end{eqnarray}
when $\gamma_{\theta}(\alpha)=-1$, with $N(q,p)$ and $D(q,p)$ being given in (\ref{L64}). The integrand in $\Pi_{\rm S}(q^2,m=0)$ has both poles and branch points, the poles coming  from the zeroes of $D(q,p,m=0)$ and the branch points from the zeroes of $N(q,p)$. 

For spacelike $q_{\mu}$ we set $q_{\mu}=(0,0,0,q_3)$ and find that all poles and branch points are in the lower right and upper left quadrants in the complex $p_0$ plane. Consequently, for spacelike $q_{\mu}$ we can use the Wick contour loop given in (\ref{L39}) as is since there are no poles within the loop, and indeed we already did so when we tested for tachyons.

For timelike $q_{\mu}$ we set $q_{\mu}=(q_0,0,0,0)$ with $q_0\geq 0$, to find poles at 
\begin{eqnarray}
&&p_0=q_0/2+p-i\epsilon,\qquad p_0=-q_0/2+p-i\epsilon,
\nonumber\\
&&p_0=q_0/2-p+i\epsilon,\qquad p_0=-q_0/2-p+i\epsilon. 
\label{L68}
\end{eqnarray}
The $p_0=q_0/2+p-i\epsilon$ pole is always in the lower right quadrant in the complex $p_0$ plane, and the $p_0=-q_0/2-p+i\epsilon$ pole is always in the upper left quadrant. If $p>q_0/2$ the $p_0=-q_0/2+p-i\epsilon$ pole is in the lower right quadrant  and the $p_0=q_0/2-p+i\epsilon$ pole is  in the upper left quadrant. However, if $p<q_0/2$ the $p_0=-q_0/2+p-i\epsilon$ pole migrates  to the lower left quadrant  and the $p_0=q_0/2-p+i\epsilon$ pole migrates to the upper right quadrant. 

\begin{figure}[htpb]
\begin{center}
\includegraphics[scale=0.3]{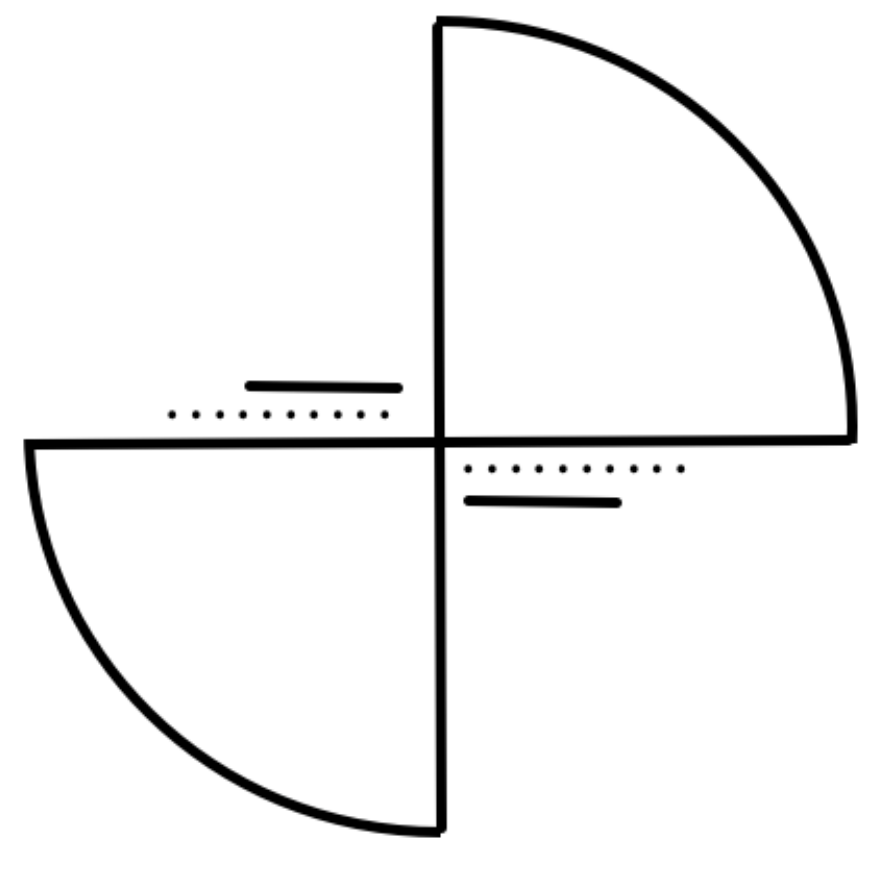}
\caption{The Wick contour for $\Pi_{\rm S}(q^2,m)$ in the complex $p_0$ plane when $q^2$ is spacelike. The branch cuts are shown as lines and the poles as dots.}
\label{lw7}
\end{center}
\end{figure}

\begin{figure}[htpb]
\begin{center}
\includegraphics[scale=0.3]{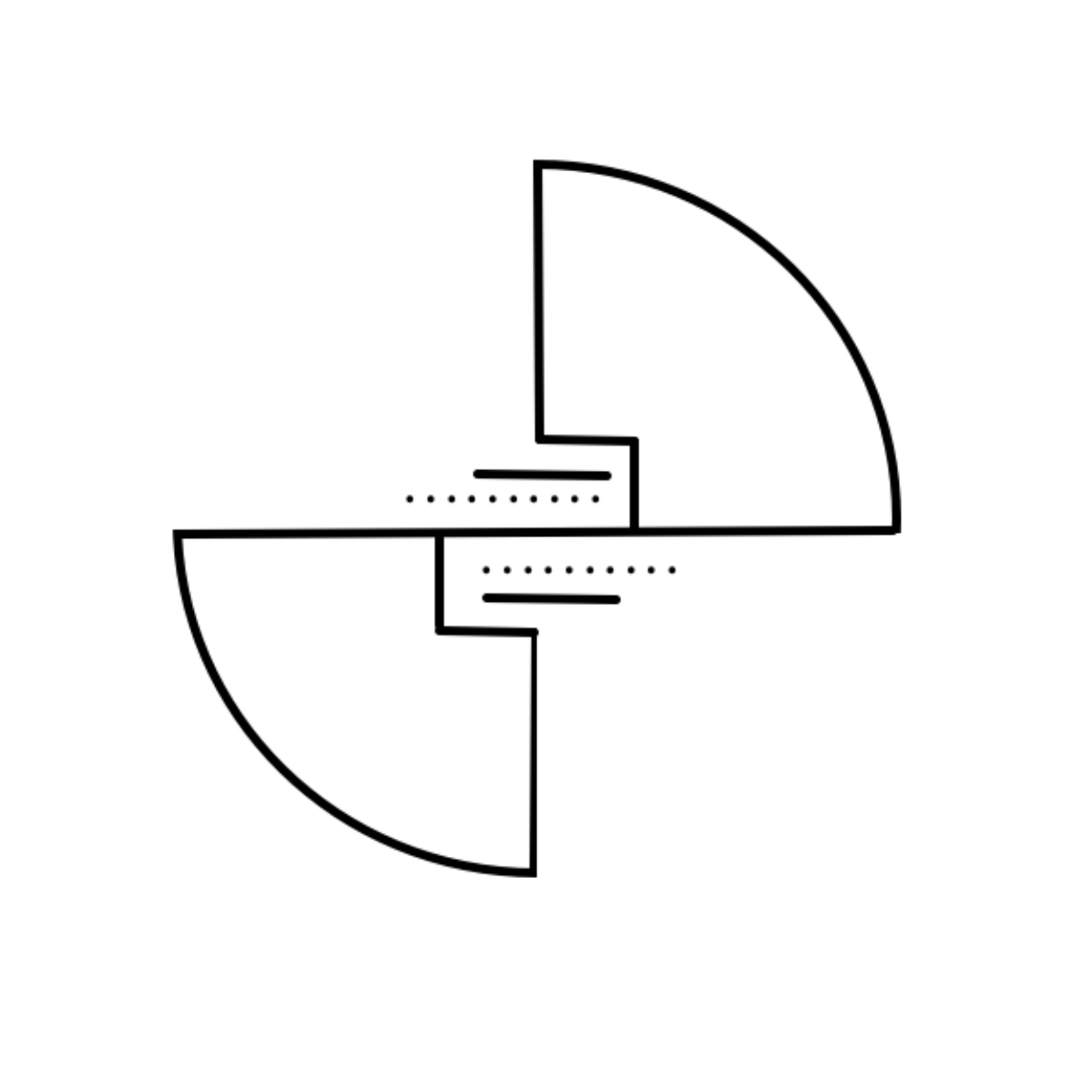}
\caption{The migrated Wick contour for $\Pi_{\rm S}(q^2,m)$ in the complex $p_0$ plane when $q^2$ is timelike. The branch cuts are shown as lines and the poles as dots.}
\label{lw8}
\end{center}
\end{figure}

The pattern of  $N(q,p)=0$ branch points completely follows the same pattern as that of the poles. By taking branch cuts to terminate at either end at branch points, we will have two branch cuts in total. We shall take one branch cut to run between the two branch points in the upper half $p_0$ plane and the other to run between the two branch points in the lower half plane. Thus for $p>q_0/2$ all poles and branch cuts are in the upper left and lower right quadrants, and so we can make the standard Wick rotation given in (\ref{L39}) as is as per Fig. (\ref{lw7}). However, for $p<q_0/2$ we will in addition need to circumnavigate the branch points and poles that have migrated to the upper right and lower left half planes. Since the branch points and poles have migrated from the upper left and lower right planes into the upper right and lower left planes, as they migrate we must deform the Wick contour loop so that no singularities enter the loop as per Fig. (\ref{lw8}). 


For timelike $q_{\mu}$ the full Wick contour loop is then the standard one given in (\ref{L38}) and (\ref{L39}), as augmented  with an integration above the cut from $p_0=0$ to the branch point at $p_0=q_0/2-p$, then round this branch point followed by an integration to the branch point at $p_0=-q_0/2+p$, then round this branch point and back to $p_0=0$. This contour does not enclose any of the poles (they have also been circumnavigated), and thus we can write
\begin{eqnarray}
-i\int_{-\infty}^{\infty}dp_0=\int_{-\infty}^{\infty}dp_4+ I_{\rm cut}.
\label{L69}
\end{eqnarray}
Consequently, the full cut contribution is given by four times the first section, viz.
\begin{eqnarray}
I_{\rm cut}=-\frac{4i\mu^2}{\pi^3}\int_0^{q_0/2}dp p^2 \int_0 ^{q_0/2-p}dp_0\frac{N(q_0,p,p_0)}{D(q_0,p,p_0,m)}.
\label{L70}
\end{eqnarray}
The imaginary $p_0$ axis contribution, $I_{\rm Wick}$, is given by
\begin{eqnarray}
I_{\rm Wick}=\frac{\mu^2}{\pi^3}\int_0^{\infty} dp p^2\int_{\infty}^{\infty}dp_4\frac{N(q_0,p,p_4)+m^2\mu^2}{D(q_0,p,p_4,m)}.
\label{L71}
\end{eqnarray}
Now while $p^2$ is spacelike along the $p_4$ axis, in the cut region both $(p_0-q_0/2)^2-p^2$ and $(p_0+q_0/2)^2-p^2$ are timelike. Thus while we recognize both $[-((p_0-q_0/2)^2-p^2)]^{1/2}$ and $[-((p_0+q_0/2)^2-p^2)]^{1/2}$ as being  real and positive on the $p_4$ axis, given that $\tilde{\Gamma}_{\rm S}(p,p,0,m=0)=(-p^2/\mu^2)^{\gamma_{\theta}(\alpha)/2}$, each of the two square roots should be interpreted with an extra factor of $i$ in the timelike case. Thus while the net square root factor  in $N(q_0,p,p_4)$ is positive definite, the net square root factor  in $N(q_0,p,p_0)$  possesses an overall minus sign.

\begin{figure}[htpb]
\begin{center}
\includegraphics[scale=0.3]{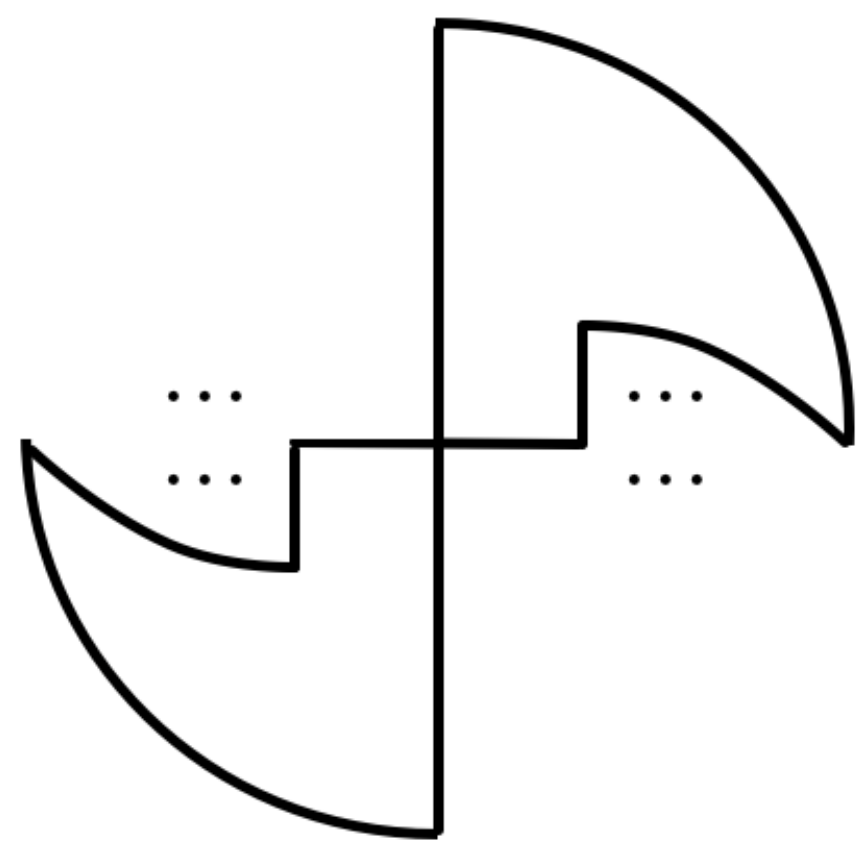}
\caption{The deformed, complex $p_0$ plane Wick contour needed for $\tilde{S}_{\mu}(p)$. Poles are shown as dots.}
\label{lw9}
\end{center}
\end{figure}

To appreciate the nature and sense of the contour it is instructive to change the location of the branch cut in  the massless theory $\tilde{\Gamma}_{\rm S}(p,p,0,m=0)$ as given in  (\ref{L41}) by replacing it by 
\begin{eqnarray}
\tilde{\Gamma}_{\rm S}(p,p,0,m=0)&=&\left(\frac{p^2+i\epsilon}{\nu^2}\right)^{\gamma_{\theta}(\alpha)/2} 
\label{L72}
\end{eqnarray}
In this case the mean-field theory effective propagator given in  (\ref{L44}) would be replaced by
\begin{eqnarray}
\tilde{S}^{-1}_{\nu}(p)&=& \slashed{p}-m\left(\frac{p^2+i\epsilon}{\nu^2}\right)^{\gamma_{\theta}(\alpha)/2}+i\epsilon.
\label{L73}
\end{eqnarray}
As a function of a complex variable, $\tilde{S}_{\nu}(p)$ has poles at $p_0^2-p^2+i\epsilon=m\nu$ and at the tachyonic $p_0^2-p^2+i\epsilon=-m\nu$ when $\gamma_{\theta}(\alpha)=-1$. All the poles in $p_0^2-p^2+i\epsilon=m\nu$ lie in the lower right and upper left quadrants in the complex $p_0$ plane, as do all the poles in $p_0^2-p^2+i\epsilon=-m\nu$ if $p>(m\nu)^{1/2}$. However for $p<(m\nu)^{1/2}$ the poles migrate to  $p_0=\pm i(m\nu -p^2)^{1/2}\mp \epsilon$. While these poles lie on the imaginary axis they are slightly displaced from it into the upper left and lower right quadrants. Consequently, none of the poles in $\tilde{S}_{\nu}(p)$ lie inside the standard Wick contour loop. For this propagator we can thus Wick rotate as per (\ref{L38}) and (\ref{L39}). Suppose we now continue back from $\nu^2$ to $\mu^2$. When we do so the poles in  $\tilde{S}_{\nu}(p)$ will move into the complex plane according to $p_0^2-p^2+i\epsilon \pm im\mu=0$, and in particular some will move into the upper right and lower left quadrants. Thus when we make this continuation we must at the same time deform the Wick contour loop so that it continues to contain no poles. We thus consider the upper right and left quadrant poles to be in a zone of avoidance. To specify this zone exactly we note that the poles of $p_0^2=p^2\pm im\mu$ are given as
\begin{eqnarray}
p_0&=&\frac{1}{2^{1/2}}\left[(p^4+m^2\mu^2)^{1/2}+p^2)\right]^{1/2}
\pm \frac{i}{2^{1/2}}\left[(p^4+m^2\mu^2)^{1/2}-p^2)\right]^{1/2},
\nonumber\\
p_0&=&-\frac{1}{2^{1/2}}\left[(p^4+m^2\mu^2)^{1/2}+p^2)\right]^{1/2}
\mp \frac{i}{2^{1/2}}\left[(p^4+m^2\mu^2)^{1/2}-p^2)\right]^{1/2}
\label{L74}
\end{eqnarray}
The poles in the upper right quadrant thus lie in a region that begins at $p=0$ where $p_0=(m\mu)^{1/2}(1+i)/2^{1/2}$, with an imaginary part that falls off as $p$ increases, reaching zero at $p=\infty$ where the real part of the location of the pole becomes infinite, with the zone of avoidance thus being wedge shaped. An analogous situation exists for the poles in the lower left quadrant. Thus if we want to define a contour for the massive theory with the  $\tilde{S}_{\mu}(p)$ propagator, for Green's functions such as $\Pi_{\rm S}(q^2,m)$ we must define the $p_0$ integration to run not along the real axis, but rather to skirt the zones of avoidance in the lower left and upper right quadrants as per Fig. (\ref{lw9}) by going around them so that no poles are then picked up in the Wick contour loop. In this way the complex $p_0$ plane poles in $\tilde{S}_{\mu}(p)$ do not play a physical role in the Wick contour loop needed for $\Pi_{\rm S}(q^2,m)$. 

The complex $p_0$  plane poles in $\tilde{S}_{\mu}(p)$ would however play a role if we want to integrate using a Feynman contour. For $\tilde{S}_{\nu}(p)$ first, the Feynman contour is obtained by closing below the real $p_0$ axis and integrating along a semicircle in the lower half plane. This contour would then include all poles with ${\rm Re}[p_0]>0$, and for $\tilde{S}_{\nu}(p)$ all would have ${\rm Im}[p_0]<0$. If we now continue to $\tilde{S}_{\mu}(p)$ we would continue to include all poles with ${\rm Re}[p_0]>0$. This would require us to include the zone of avoidance in ${\rm Re}[p_0]>0$ but not include the zone of avoidance with ${\rm Re}[p_0]<0$. Thus for ${\rm Re}[p_0]>0$ the  Feynman contour is the compliment of the Wick contour, while for ${\rm Re}[p_0]<0$ the  Feynman contour is the same as the Wick contour

The general rule then for all of the cases described above is that the Wick contour loop integration is always to be defined as being the contour  that contains no poles and circumnavigates all upper right and lower left quadrant cuts. For all the cases this will always yield (\ref{L69}). Similarly, the Feynman contour is to always be defined as the contour that includes all poles with ${\rm Re}[p_0]>0$.

Finally, since the $p_0$  contours are different for spacelike and timelike $q_{\mu}$, we cannot first evaluate $\Pi_{\rm S}(q^2,m)$ for spacelike $q_{\mu}$ (say using Feynman parameters for amplitudes with Euclidean $p_{\mu}$ and $q_{\mu}$) and then continue the resulting answer to timelike $q_{\mu}$ since we would miss the migrated cuts, with the spacelike and timelike $q_{\mu}$ Wick contour loops being different. We will thus need to evaluate the timelike $q_{\mu}$ case directly.

\subsection{The Collective Higgs Mode when the Fermion is Massive -- Results}

For timelike $q_{\mu}=(q_0,0,0,0)$, we shall explicitly evaluate $I_{\rm Wick}$ and $I_{\rm cut}$ in detail in the appendix, and will show there that as a function of $q_0^2=q^2$, $I_{\rm Wick}$ has a branch point at $q^2=2m\mu$. While we will show thus explicitly in the appendix, it  may be understood heuristically by noting that at  $p=0$ the massive $\tilde{S}_{\mu}(p)$ propagator has poles at  $p_0=(1+i)(m\mu)^{1/2}/2^{1/2}$ and  at $p_0=(1-i)(m\mu)^{1/2}/2^{1/2}$, and thus a particle-antiparticle threshold at $q^2=((1+i)(m\mu)^{1/2}/2^{1/2}+(1-i)(m\mu)^{1/2}/2^{1/2})^2=2m\mu$. For $q^2$ below this threshold the integrands in both $I_{\rm Wick}$ and $I_{\rm cut}$ as given in (\ref{L71}) and (\ref{L70}) are real. Consequently, with $g^{-1}$ being real and with $I_{\rm cut}$ itself possessing an overall factor of $i$, there cannot be any bound state Higgs boson pole at or below $q^2= 2m\mu$ (unless both $I_{\rm Wick}-1/g$ and $I_{\rm cut}$ just happen to vanish at some common value of $q^2$ in that region -- though this turns out not to be the case). However, while the integrand in $I_{\rm cut}$ remains real above the $q^2= 2m\mu$ threshold so that $I_{\rm cut}$ itself remains pure imaginary, the integrand in $I_{\rm Wick}$  becomes complex, and then one can find a pole.   Any solution above the threshold must thus satisfy $g^{-1}-{\rm Re}[I_{\rm Wick}]-{\rm Im}[I_{\rm Wick}]-I_{\rm cut}=0$. The Higgs boson must thus be a resonance, with its width then being fixed by $I_{\rm cut}$. With the actual integrals only being doable numerically, in the appendix we show that there is an explicit solution, with our sought-after dynamical massive scalar Higgs boson being a narrow resonance lying below the real axis in the complex $q^2$ plane, with parameters
\begin{eqnarray}
q_0(\rm Higgs)&=&(1.480-0.017i)(M\mu)^{1/2},
\nonumber\\
q^2(\rm Higgs)&=&(2.189-0.051i)M\mu.
\label{L75}
\end{eqnarray}
We had noted above that we always had the freedom to normalize $\mu$ to $M$. On now doing so, the Higgs boson parameters become
\begin{eqnarray}
q_0(\rm Higgs)&=&(1.480-0.017i)M,
\nonumber\\
q^2(\rm Higgs)&=&(2.189-0.051i)M^2,
\label{L76}
\end{eqnarray}
to thus naturally be of order the fermion mass scale. Thus even though the $\Pi_{\rm S}$-independent homogeneous term in $T_{\rm S}=g+g\Pi_{\rm S} g+g\Pi_{\rm S} g\Pi_{\rm S} g+...=g/(1-g\Pi_{\rm S})$ is zero ($g^{-1}$ being divergent according to (\ref{L49})), nonetheless we again see that the vanishing of the homogeneous term in the scattering amplitude need not prevent the presence of a pole.

As well as measure the mass of the Higgs boson one can also measure its width, and from a direct measurement at the Higgs peak the width was found to be no greater than 3.4 GeV \cite{Chatrchyan2014}, to thus give a small width to mass ratio of no more than 0.027, a value that compares well with the width to mass ratio that we have found above, viz. 0.017/1.480=0.011. Recently, the Higgs boson width has been measured using an indirect, off-peak technique and an analysis that involves some dynamical, standard-model-based (viz. elementary Higgs field) assumptions, with it then being found \cite{CMS2014} that the width is no bigger than 22 MeV. Consequently, the width to mass ratio is reduced to $0.00017$. Now while this ratio is smaller than the small value for the width to mass ratio that we have found above, and while it is not yet clear how many of the dynamical assumptions involved will actually hold in our case, nonetheless it is of interest that the experimental data do support a narrow width Higgs boson rather than a broad one. With the standard model expectation for the Higgs width being just 4 MeV \cite{CMS2014},  the measured value of the Higgs width could eventually prove to be a diagnostic for distinguishing a dynamical Higgs boson from an elementary one.

Finally, we note that in the literature attention has focussed on the fact that in the NJL model  the dynamical Higgs boson is a stable bound state that lies right at the particle-antiparticle threshold with a mass twice that of the dynamical fermion. However, as we see, this is not a generic feature of dynamical symmetry breaking, and in fact it could only possibly occur if the scattering amplitude is purely real at the threshold. For a point coupled theory such as the NJL model, this is in fact the case. However, once we give the coupling some momentum dependence the dynamical Higgs boson could move away from the particle-antiparticle threshold, and could potentially become a resonance rather than a stable bound state. The Higgs boson width could thus be a diagnostic for distinguishing our approach to dynamical symmetry breaking from various other approaches to dynamical symmetry breaking that we describe in Sec. IV-J below, approaches in which the Higgs boson is not typically found to be a resonance above threshold. Moreover, and also in contrast, for an elementary Higgs boson, the mass is given by the second derivative of the potential at the minimum, and is thus real if the potential is real.

\subsection{Distinguishing a Dynamical Higgs Boson from an Elementary One}

If the Higgs boson is to be dynamical, it would be very instructive to identify some ways to distinguish it from an elementary Higgs boson. Also we would need to account for the fact that  an elementary Higgs field theory works so well in weak interactions. To this end let us consider the path integral representation of the generator $Z(\bar{\eta},\eta)$ of fermion Green's functions associated with the fermion sector of the ${{\cal L}}_{\rm QED-FF}$ Lagrangian given in (\ref{L40}), viz.
\begin{eqnarray}
Z(\bar{\eta}, \eta)&=&\int D[\bar{\psi}]D[\psi]D[A_{\mu}]\exp\bigg{[}i\int d^4x \bigg{(}-\frac{1}{4}F_{\mu\nu}F^{\mu\nu}+\bar{\psi}\gamma^{\mu}(i\partial_{\mu}-eA_{\mu})\psi
-\frac{g}{2}(\bar{\psi}\psi)^2+\bar{\eta}\psi+\bar{\psi}\eta\bigg{)}\bigg{]},
\label{L77}
\end{eqnarray}
with Grassmann sources $\eta$ and $\bar{\eta}$. (For simplicity we have left out the $(g/2)(\bar{\psi}i\gamma_5\psi)^2$ term present in (\ref{L40}), though it could be incorporated via a dummy pseudoscalar field if desired. Also we have left out a $J^{\mu}(x)A_{\mu}(x)$ source term for $A_{\mu}$.) Via Gaussian path integration on a dummy scalar field variable $\sigma(x)$, $Z(\bar{\eta}, \eta)$  can be rewritten as
\begin{eqnarray}
Z(\bar{\eta}, \eta)&=&\int D[\bar{\psi}]D[\psi]D[A_{\mu}]D[\sigma]\exp\bigg{[}i\int d^4x \bigg{(}-\frac{1}{4}F_{\mu\nu}F^{\mu\nu}+\bar{\psi}\gamma^{\mu}(i\partial_{\mu}-eA_{\mu})\psi
\nonumber\\
&-&\frac{g}{2}(\bar{\psi}\psi)^2+\frac{g}{2}\left(\frac{\sigma}{g}-\bar{\psi}\psi\right)^2+\bar{\eta}\psi+\bar{\psi}\eta\bigg{)}\bigg{]},
\label{L78}
\end{eqnarray}
and thus as
\begin{eqnarray}
Z(\bar{\eta}, \eta)=\int D[\bar{\psi}]D[\psi]D[A_{\mu}]D[\sigma]\exp\bigg{[}i\int d^4x \bigg{(}-\frac{1}{4}F_{\mu\nu}F^{\mu\nu}+\bar{\psi}\gamma^{\mu}(i\partial_{\mu}-eA_{\mu})\psi
-\sigma\bar{\psi}\psi +\frac{\sigma^2}{2g}
+\bar{\eta}\psi+\bar{\psi}\eta\bigg{)}\bigg{]}.
\label{L79}
\end{eqnarray}
We recognize (\ref{L79}) as having the same structure as the mean-field Lagrangian ${{\cal L}}_{\rm QED-MF}$ given in (\ref{L40}). Thus the fermion Green's functions of the ${{\cal L}}_{\rm QED-FF}$ theory of interest to us in this paper  are given as the fermion Green's functions of a Yukawa-coupled scalar field theory. In consequence, diagramatically the perturbative expansions associated with (\ref{L79}) and with a theory with an elementary scalar field are in one to one correspondence. 

While $Z(\bar{\eta}, \eta)$ as given in (\ref{L79}) looks very much like the generating functional of an elementary Higgs theory, it differs from it in three ways: there is no kinetic energy term for the $\sigma(x)$ field, no double-well potential energy term for it  either, and most crucially as we shall see, no $J(x)\sigma(x)$ source term for it. To generate kinetic energy and potential energy terms for $\sigma(x)$, we now require that there be critical scaling in the QED sector with the dynamical dimension of $\bar{\psi}\psi$ being reduced from three to two. Thus path integration on $A_{\mu}$ serves to replace point couplings by dressed couplings, with figures such as Figs. (\ref{baretadpole}), (\ref{lw1}), and (\ref{lw5}) being replaced by Figs. (\ref{lw3}), (\ref{lw2}), and (\ref{lw6}). Path integration in the fermion sector is straightforward since all the terms in (\ref{L79}) are linear in $\bar{\psi}$ and $\psi$, with the path integration being equivalent to a one-loop Feynman diagram (as evaluated with dressed vertices). Following path integration in the fermion sector, on introducing $\tilde{\Gamma}_{\rm S}(x,m=0)$ as the Fourier transform of $\tilde{\Gamma}_{\rm S}(p,p,0,m=0)$,\footnote{ $\int d^4p\exp(ip\cdot x)(-p^2)^{-\lambda}=i\pi^2 2^{4-2\lambda}\Gamma(2-\lambda)(-x^2)^{\lambda-2}/\Gamma(\lambda)$, with $\lambda=-\gamma_{\theta}(\alpha)/2$.} we obtain an effective action in the $\sigma$ sector, which with $\gamma_{\theta}(\alpha)=-1$ is of the form \cite{Mannheim1978}  
\begin{eqnarray}
Z(\bar{\eta},\eta,J_{\mu})=&=&i{\rm Tr}{\rm ln}\left[\frac{i\slashed{\partial}_x-\int d^4x^{\prime}\sigma(x^{\prime})\tilde{\Gamma}_{\rm S}(x-x^{\prime},m=0)}{i\slashed{\partial}_x}\right]
\nonumber\\
&=&\int D[\sigma]\exp[iI_{\rm EFF}(\sigma)]=\int D[\sigma]\exp\left[i\int d^4 x\left(-\tilde{\epsilon}(\sigma)+\frac{Z(\sigma)}{2}\partial_{\mu}\sigma\partial^{\mu}\sigma+...\right)\right], 
\label{M80}
\end{eqnarray}
where according to (\ref{L56}) 
\begin{eqnarray}
I_{\rm EFF}(\sigma)=\int d^4 x\left[-\frac{\sigma^2(x)\mu^2}{16\pi^2}\left[{\rm ln}\left(\frac{\sigma^2(x)}{M^2}\right)-1\right]
+\frac{3\mu}{256\pi \sigma(x)}\partial_{\mu}\sigma(x)\partial^{\mu}\sigma(x)+....\right].
\label{M81}
\end{eqnarray}
We recognize $I_{\rm EFF}(\sigma)$ as being in the form of none other than a Higgs field action with both a double-well potential and a kinetic energy term for $\sigma(x)$.

Despite this, we note that in (\ref{L79}) there is no source term $J(x)\sigma(x)$ for the scalar field (in a true fundamental Higgs Lagrangian there would be such a source term), and thus (\ref{L79})  only generates Green's functions with external fermion legs and does not generate any Green's functions with external scalar field legs. Thus in the dynamical Higgs case one can generate the fermion Green's functions using a scalar field theory in which the only role of the scalar field is to contribute internally in Feynman diagrams and to never appear in any external legs. From the perspective of (\ref{L79}) it would be the all-order iteration of internal $\sigma$ exchange diagrams in $Z(\bar{\eta}, \eta)$ that then generates the dynamical Higgs and Goldstone poles that we have found in $T_{\rm S}(q^2,M)$ and $T_{\rm P}(q^2,M)$. The only distinction between (\ref{L79}) and an elementary Higgs field theory would be in those weak interaction processes in which the Higgs boson goes on shell, as expressed through branching ratios and, as noted above, the Higgs boson width. While beyond the scope of the present paper, it would  be very instructive to determine what such differences might then look like.

\subsection{Comparison with other Dynamical Symmetry Breaking Studies}

\begin{figure}[htpb]
\begin{center}
\includegraphics[scale=0.5]{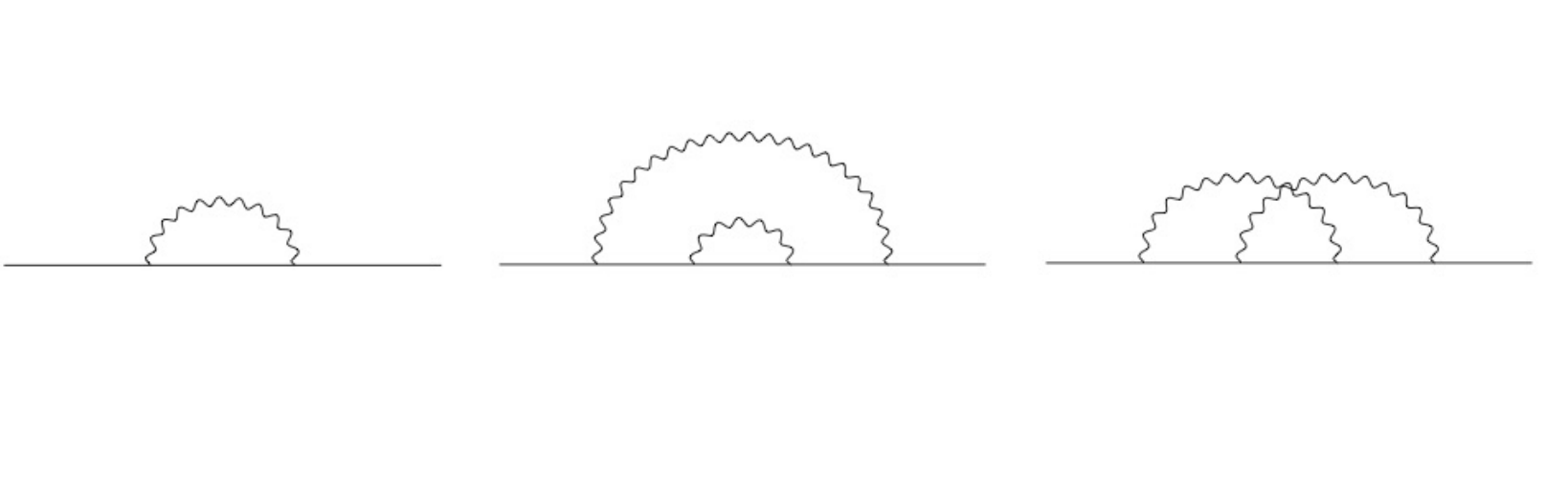}
\caption{The first few graphs in the fermion self-energy Schwinger-Dyson equation.}
\label{SD1}
\end{center}
\end{figure}

Ever since the work of Nambu and Jona-Lasinio, study of dynamical symmetry breaking has been an abiding theme in the literature, especially in Abelian and non-Abelian gauge theories, sometimes as coupled to a four-fermion interaction (see e.g.  \cite{Miransky1993,Hill2003,Contino2010,Miransky2010,Appelquist2010b,Yamawaki2010,Panico2015,
Mannheim2017} and references therein). For Abelian gauge theories, much focus has been placed on solving the Schwinger-Dyson equation for a dynamical fermion mass, with particular emphasis on the quenched, planar graph approximation in which one uses an undressed, viz. quenched,   photon propagator $(q_{\mu}q_{\nu}/q^2-\eta_{\mu\nu})/q^2$, and restricts to planar, so-called ladder or rainbow, photon exchange diagrams (i.e. one keeps the first two diagrams in Fig. (\ref{SD1}) and their higher order analogs, but leaves out the third figure in Fig. (\ref{SD1}) and its higher order analogs).  In the quenched ladder approximation a critical point has been found at which the coupling constant $\alpha$ takes the value $\alpha =\pi/3$  \cite{Maskawa1974,Maskawa1975,Fukuda1976,Fomin1978}. In the subcritical region at or below this critical point one finds that the self-consistent fermion self energy $\Sigma(p^2)$ scales asymptotically as $(p^2)^{(d_{\theta}-3)/2}$, where the dimension of the composite operator $\theta=\bar{\psi}\psi$ is given by $d_{\theta}=3+\gamma_{\theta}=2+(1-3\alpha/\pi)^{1/2}$, to thus be given by $d_{\theta}=2$ at $\alpha=\pi/3$. On comparing with (\ref{L3}) above, we see that this scaling behavior is analogous to the behavior found in JBW electrodynamics at a Gell-Mann-Low eigenvalue (where the photon is also canonical), and the quenched ladder approximation critical coupling constant condition that $d_{\theta}=2$ at $\alpha=\pi/3$ is reminiscent of the $\gamma_{\theta}(\alpha)=-1$ (viz. $d_{\theta}(\alpha)=2$) condition that we found in our study of a critical scaling QED.\footnote{In the critical scaling literature the dimension and anomalous dimension of $d_{\theta}(\alpha)$ are sometimes denoted by $d_m$ and $\gamma_m$, and defined via $d_m=3-\gamma_{m}$, with $d_m=2$ corresponding to $\gamma_m=+1$, rather than the $\gamma_{\theta}(\alpha)=-1$ that we use here.}  Hence in the $\alpha \leq \pi/3$ regime of the quenched ladder approximation the bare mass will not be zero identically but will vanish in the limit of infinite cutoff, the Baker-Johnson evasion of the Goldstone theorem will apply, and there will be no Goldstone boson when the coupling is weak. However, in the supercritical region above this critical point the authors of  \cite{Maskawa1974,Maskawa1975,Fukuda1976,Fomin1978} showed that the fermion bare mass would be zero identically, that there then would be a gauge-boson-exchange-generated Goldstone boson, with the discontinuity in $d_{\theta}$ at $\alpha=\pi/3$ indicating that in the quenched, planar graph approximation Abelian gauge theories undergo a phase transition at $\alpha=\pi/3$.

To appreciate how this comes about in more detail, we note that in general because of its Lorentz structure one can set $S^{-1}(p)=A(p^2)\slashed{p}-B(p^2)$ where $A(p^2)$ and $B(p^2)$ are Lorentz scalars. In the quenched planar approximation one can find a gauge (the Landau gauge) in which one can set $A(p^2)=1$ and $D_{\mu\nu}(q)=(q_{\mu}q_{\nu}/q^2-\eta_{\mu\nu})/q^2$, and with  $\Gamma^{\mu}(p,p-k)$ being equal to $ \gamma^{\mu}$ in this approximation, the Schwinger-Dyson equation takes the form
\begin{eqnarray}
B(p^2)=m_0+ie^2\int \frac{d^4q}{(2\pi)^4} D_{\mu\nu}(p-q)\gamma^{\mu}\frac{1}{\slashed{q}-B(q^2)}\gamma^{\nu}.
\label{M82}
\end{eqnarray}
In (\ref{M82}) we have replaced the bare charge $e_0$ by the physical charge $e$ since there is no charge renormalization in the quenched approximation. After taking the anticommutator of both sides of (\ref{M82}) with respect to $\gamma_5$, in Euclidean space (\ref{M82}) takes the form  
\begin{eqnarray}
B(p^2)=m_0+3e^2\int \frac{d^4q}{(2\pi)^4} \frac{B(q^2)}{(p-q)^2(q^2+B^2(q^2))},
\label{M83}
\end{eqnarray}
with the angular integration bringing (\ref{M83}) to the form (see e.g. \cite{Miransky1993}) 
\begin{eqnarray}
B(p^2)=m_0+\frac{3\alpha}{4\pi}\bigg{[}\int_0^{p^2}dq^2 \frac{q^2B(q^2)}{p^2(q^2+B^2(q^2))}
+\int_{p^2}^{\infty}dq^2 \frac{B(q^2)}{(q^2+B^2(q^2))}\bigg{]},
\label{M84}
\end{eqnarray}
where $\alpha=e^2/4\pi$.

On cutting off the $q^2$ integration at $\Lambda^2$, a convergent asymptotic solution  of the form $B(p^2)=m(p^2/m^2)^{(\nu-1)/2}$  is found to obey (\ref{M84}), while fixing $\nu$ and $m_0$ to be of the form
\begin{eqnarray}
\nu=\pm\left(1-\frac{3\alpha}{\pi}\right)^{1/2},\qquad m_0=\frac{3\alpha m}{2\pi(1-\nu)}\frac{\Lambda^{\nu-1}}{m^{\nu-1}}.
\label{M85}
\end{eqnarray}
From  (\ref{M85}) we see that  the solution for $B(p^2)$ is convergent when $Re[\nu]<1$, with $\nu$ being real and less than one if $\alpha \leq \pi/3$.\footnote{The power solution to (\ref{M84}) of the form $B(p^2)=(p^2)^{(\nu-1)/2}$ with $\nu$ as given in (\ref{M85}) was first presented  in \cite{Johnson1964}.} With $\nu$ having a branch point at $\alpha=\pi/3$, we can thus anticipate that a phase transition might occur at that value of the coupling constant. While the above solution gives a real $\nu$ if $\alpha \leq \pi/3$, if $\alpha>\pi/3$ we would instead obtain $B(p^2)\sim (p^2)^{-1/2}\exp[\pm i(\mu/2){\rm ln}(p^2/m^2)]$ where $\mu=(3\alpha/\pi-1)^{1/2}$. This then gives two classes of real solutions, viz.  $(p^2)^{-1/2}\cos[(\mu/2){\rm ln}(p^2/m^2]$ and $(p^2)^{-1/2}\sin[(\mu/2){\rm ln}(p^2/m^2)]$. Combining them gives the $\alpha> \pi/3$ asymptotic solution
\begin{eqnarray}
B(p^2)=\frac{m\cos[((3\alpha/\pi-1)^{1/2}/2){\rm ln}(p^2/m^2)+\sigma]}{(p^2/m^2)^{1/2}},\qquad m_0=\frac{3m\alpha\cos[(3\alpha/\pi-1)^{1/2}{\rm ln}(\Lambda/m)+\tau]}{2\pi(\mu^2+1)^{1/2}(\Lambda/m)},
\label{M86}
\end{eqnarray}
where $\sigma$  is a  (possibly $\Lambda^2/m^2$ but not $p^2/m^2$ dependent) phase and $\tau=\sigma+{\rm arctan}\mu$. As required, we see that for both $\alpha \leq \pi/3$ and $\alpha>\pi/3$ the bare mass vanishes in the limit in which the cutoff goes to infinity. However, that does not mean that the bare mass is identically zero, only that it vanishes in the limit of large cutoff. For  $\alpha \leq \pi/3$ this is the only option for the bare mass. However for  $\alpha>\pi/3$ there is a second option for the bare mass, since it  will vanish identically  if we set 
\begin{eqnarray}
\left(\frac{3\alpha}{\pi}-1\right)^{1/2}{\rm ln}\left(\frac{\Lambda}{m}\right)+\tau=\frac{\pi}{2}.
\label{M87}
\end{eqnarray}
Now  initially this would suggest that as we let $\Lambda$ go to infinity, the only allowed value for $\alpha$ would be $\alpha=\pi/3$. In order to be able to obtain a solution that is to hold for all $\alpha>\pi/3$, we must take $\tau$ to be of the form $\tau=\delta{\rm ln}(\Lambda/m)$ where $\delta$ is finite, so that  we obtain 
\begin{eqnarray}
\left(\frac{3\alpha}{\pi}-1\right)^{1/2}+\delta=\frac{\pi}{2{\rm ln}(\Lambda/m)},
\label{M88}
\end{eqnarray}
and thus
\begin{eqnarray}
B(p^2)=\frac{m\cos[((3\alpha/\pi-1)^{1/2}/2){\rm ln}(p^2/\Lambda^2)-{\rm arctan}\mu+\pi/2]}{(p^2/m^2)^{1/2}},\qquad m_0=0.
\label{M89}
\end{eqnarray}
Then, with $\delta$ being an appropriately chosen function of $\alpha$, all values of $\alpha$ greater than $\pi/3$ are allowed in the limit of infinite cutoff.\footnote{A ${\rm ln}(\Lambda/m)$ dependence to the phase $\tau$ seems not to have been considered in the quenched ladder approximation literature, where instead one restricts  \cite{Miransky1985} to $(3\alpha/\pi-1)^{1/2}=\pi/2{\rm ln}(\Lambda/m)$, a quantity that vanishes in the limit of infinite cutoff, to then not permit  $\alpha$ to take any value other than $\pi/3$.} Since for all such values of $\alpha$ the bare mass is identically zero, for $\alpha >\pi/3$ dynamical symmetry breaking will take place for any non-trivial solution to
\begin{eqnarray}
B(p^2)=3e^2\int \frac{d^4q}{(2\pi)^4} \frac{B(q^2)}{(p-q)^2(q^2+B^2(q^2))}
\label{M90}
\end{eqnarray}
that behaves asymptotically as in (\ref{M89}). Thus for all values of $\alpha$ greater than $\pi/3$ (i.e. strong coupling), the quenched ladder approximation has a chiral symmetry that is broken dynamically, with dynamical Goldstone boson generation taking place \cite{Maskawa1974,Maskawa1975,Fukuda1976,Fomin1978}. The differing behaviors of the theory below and above $\alpha=\pi/3$ is reflected in the fact that the asymptotic solution possesses  a branch point at $\alpha=\pi/3$.

The quenched ladder approximation is thus very instructive as it illustrates in a quite straightforward solvable model how dynamical symmetry breaking can occur in a quantum field theory, how one can distinguish between a bare mass that is identically zero and one that only vanishes of  the limit of infinite cutoff, and how one can identify what the implications of the behavior of the bare mass are for the generation of Goldstone bosons. The model also reinforces the result we obtained with the point-coupled NJL model, namely that one needs a minimum value for the coupling constant in order for dynamical symmetry breaking to take place. We will now show that we can reinforce this strong coupling paradigm even more by studying  a quenched ladder approximation Abelian gluon model coupled to a four-fermion interaction. And then we will have to reconcile these results with our own result reported above that when a critical scaling JBW QED with $\beta(\alpha)=0$, $\gamma_{\theta}(\alpha)=-1$ is coupled to a four-fermion interaction, one gets dynamical symmetry breaking no matter how weak the four-fermion coupling constant might be as long as it is attractive, with $\alpha$ equally being weak if it is the physical charge that obeys $\beta(\alpha)=0$.  As we will see, it will actually be the quenched ladder approximation wisdom that is misleading, as it turns out to be based on an unreliable extrapolation.

\begin{figure}[htpb]
\begin{center}
\includegraphics[scale=0.2]{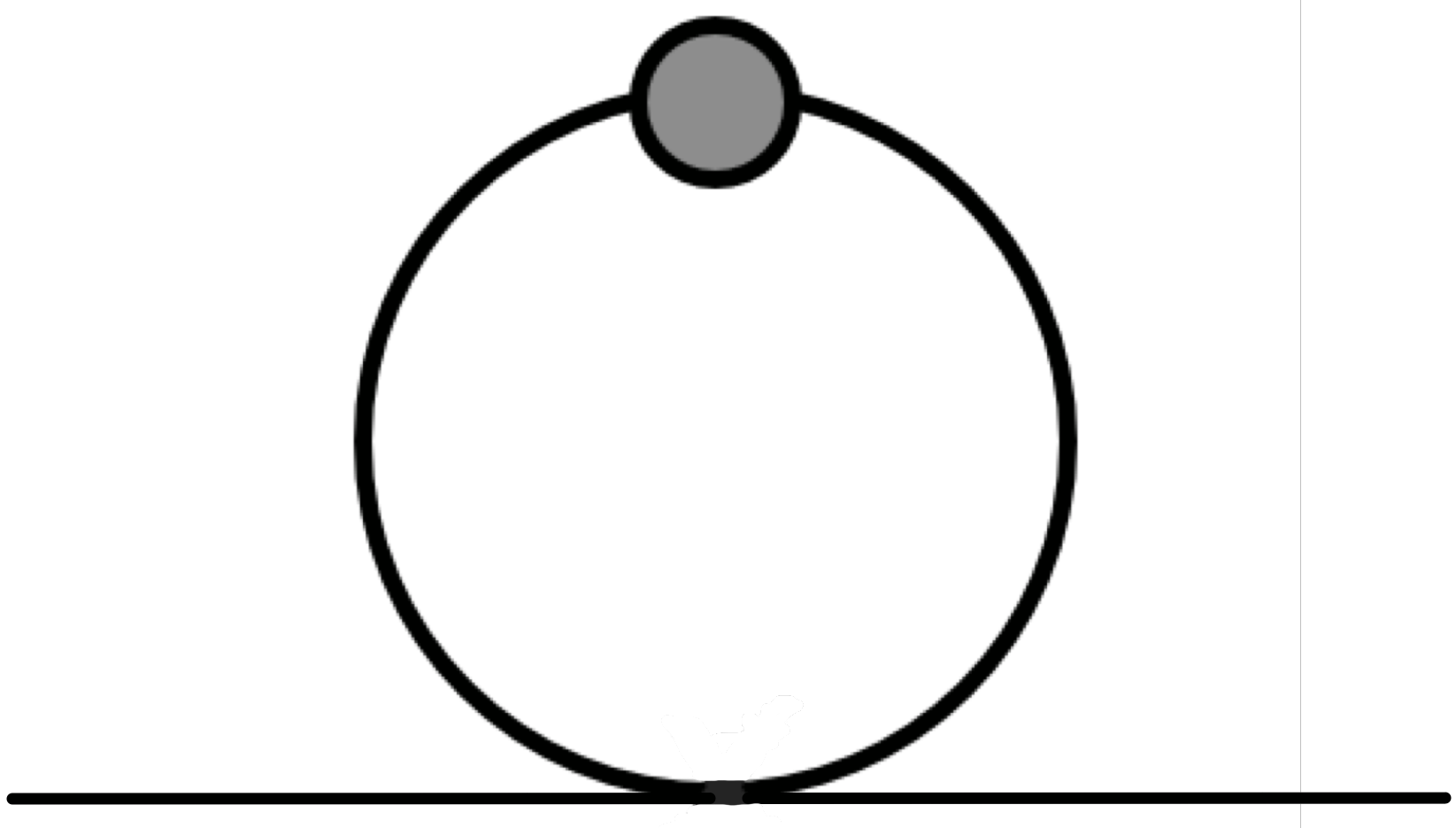}
\caption{The four-fermion interaction tadpole contribution to the fermion self energy. The blob represents the Abelian gluon quenched ladder approximation contribution to the fermion propagator.}
\label{semitadpoleselfenergy}
\end{center}
\end{figure}

But to first see how things work when a four-fermion interaction is included, it is instructive to compare and contrast  our work on JBW electrodynamics coupled to a four-fermion interaction with the studies of a quenched ladder approximation Abelian gluon model coupled to a four-fermion interaction, as discussed e.g. in \cite{Bardeen1986,Leung1986,Yamawaki1986,Kondo1989,Bardeen1990,Bardeen1992,Carena1992,Gusynin1992,Kondo1993,Harada1994,Hashimoto1998,Gusynin1998}. In the JBW case one evaluates the $\langle \Omega_m|\bar{\psi}\psi|\Omega_m\rangle$ expectation value using the fully dressed tadpole given in Fig. (\ref{lw3}), whereas, in the quenched ladder approximation one evaluates $\langle \Omega_m|\bar{\psi}\psi|\Omega_m\rangle$ using the partially dressed tadpole exhibited in Fig. (\ref{semitadpoleselfenergy}). Then, to determine the fermion propagator, in the quenched ladder approximation to the Schwinger-Dyson equation one adds on to the planar graphs contained in Fig. (\ref{SD1}) this partially dressed tadpole contribution.\footnote{In contrast, in the JBW case, to determine the fermion propagator one uses only the QED contribution to the Schwinger-Dyson equation without the addition of any tadpole graph contribution, with the tadpole contribution to the fermion mass  being generated by the residual interaction as per $m=g\langle \Omega_m|\bar{\psi}\psi|\Omega_m\rangle$. The mean field approach thus organizes the Feynman graphs very differently than the quenched ladder approximation approach.} In the quenched ladder approach one thus replaces the NJL point-coupled  bare tadpole (cf. Fig. (\ref{baretadpole})) by the quenched ladder approximation tadpole contribution to the fermion propagator exhibited in Fig. (\ref{semitadpoleselfenergy}), viz.
\begin{eqnarray}
g\langle \Omega_m|\bar{\psi}\psi|\Omega_m\rangle=-ig\int \frac{d^4q}{(2\pi)^4}{\rm Tr}\frac{1}{\slashed{q}-B(q^2)}
=-4ig\int \frac{d^4q}{(2\pi)^4}\frac{B(q^2)}{q^2-B^2(q^2)},
\label{M91}
\end{eqnarray}
where $B(p^2)$ is to be self-consistently determined from the Schwinger-Dyson equation 
\begin{eqnarray}
B(p^2)&=&g\langle \Omega_m|\bar{\psi}\psi|\Omega_m\rangle+\frac{3\alpha}{4\pi}\bigg{[}\int_0^{p^2}dq^2 \frac{q^2B(q^2)}{p^2(q^2+B^2(q^2))}
+\int_{p^2}^{\infty}dq^2 \frac{B(q^2)}{(q^2+B^2(q^2))}\bigg{]}
\nonumber\\
&=&-4ig\int \frac{d^4q}{(2\pi)^4}\frac{B(q^2)}{q^2-B^2(q^2)}+\frac{3\alpha}{4\pi}\bigg{[}\int_0^{p^2}dq^2 \frac{q^2B(q^2)}{p^2(q^2+B^2(q^2))}
+\int_{p^2}^{\infty}dq^2 \frac{B(q^2)}{(q^2+B^2(q^2))}\bigg{]}
\label{M92}
\end{eqnarray}
that is to replace (\ref{M84}) \cite{Bardeen1986,Leung1986}. With the bare mass $m_0$ now being taken to be zero identically, non-trivial solutions to (\ref{M92}) correspond to dynamical symmetry breaking. As before, we look for an asymptotic solution, and since in the quenched ladder approximation the tadpole has not been quenched enough so as to make it be only logarithmically divergent, we still need a cutoff for the four-fermion sector.\footnote{By using the partially dressed tadpole of Fig. (\ref{semitadpoleselfenergy}) rather than the fully dressed tadpole of Fig. (\ref{lw3}), one is not able to take advantage of the fact that at $\alpha=\pi/3$, the four-fermion interaction would, as per Sec. IV-F,  be renormalizable since at that value $d_{\theta}(\alpha)=2$ and $\gamma_{\theta}(\alpha)=-1$. Thus even at $\alpha=\pi/3$ one would still need a cutoff when a quenched ladder Abelian gluon model is coupled to a four-fermion interaction. However, when a critical scaling JBW electrodynamics with $d_{\theta}(\alpha)=2$ is coupled to a four-fermion interaction, no cutoff is needed.}  So this time we take the asymptotic solution to be of the form $B(p^2)=m(p^2/\Lambda^2)^{(\nu-1)/2}$, and obtain
\begin{eqnarray}
\nu=\pm\left(1-\frac{3\alpha}{\pi}\right)^{1/2},\qquad g\langle \Omega_m|\bar{\psi}\psi|\Omega_m\rangle+\frac{3\alpha m}{2\pi(\nu-1)}=0.
\label{M93}
\end{eqnarray}
With $\nu-1$ being negative, $g\langle \Omega_m|\bar{\psi}\psi|\Omega_m\rangle$ is given by the leading term in (\ref{M91}) according to 
\begin{eqnarray}
g\langle \Omega_m|\bar{\psi}\psi|\Omega_m\rangle=-\frac{mg\Lambda^2}{2\pi^2(1+\nu)},
\label{M94}
\end{eqnarray}
an expression that limits to the leading term in (\ref{L20}) as $\alpha \rightarrow 0$ if we take $\nu=+(1-3\alpha/\pi)^{1/2}$. As noted in \cite{Kondo1989,Bardeen1990} and references therein, broken symmetry solutions thus lie on the  critical surface  
\begin{eqnarray}
-g\Lambda^2=\pi^2(1+(1-3\alpha/\pi)^{1/2})^2.
\label{M95}
\end{eqnarray}
While the quenched ladder approximation on its own has no dynamical symmetry breaking solutions if $\alpha \leq \pi/3$, now we see that we can get broken symmetry solutions in the $\alpha\leq \pi/3$ region provided the Abelian gluon model is accompanied by a four-fermion interaction with an appropriately chosen value for $-g\Lambda^2$.\footnote{If we define $g^{\prime}=g\Lambda^2$, we can  rewrite (\ref{M95}) as $-g^{\prime}=\pi^2(1+(1-3\alpha/\pi)^{1/2})^2$, with $g^{\prime}$ being finite. Similarly, we can rewrite the four-fermion interaction as $-(g^{\prime}/2\Lambda^2)([\bar{\psi}\psi]^2+[\bar{\psi}i\gamma_5\psi]^2)$. In passing we note that this was the form of the interaction of the unified non-linear spinor theory of fundamental interactions that Heisenberg introduced in the 1950s,  with $g^{\prime}$ being finite and $\Lambda$ being a fundamental mass scale (see e.g. \cite{Heisenberg1957}).} As we make $\alpha$ smaller and smaller in (\ref{M95}), we have to make $-g\Lambda^2$ bigger and bigger, while at  $\alpha=\pi/3$ itself, we still need a minimum $-g\Lambda^2=\pi^2$. In contrast, in the JBW case where the coupling constant is not free to vary but must satisfy $\beta(\alpha)=0$ identically, not only is there not any quadratic $\Lambda^2$ term in $\langle \Omega_m|\bar{\psi}\psi|\Omega_m\rangle$ to begin with (cf. (\ref{L48})), as we have seen, symmetry breaking occurs no matter how small $g$ might be as long as it is negative (viz. attractive). Thus while a quenched ladder Abelian gluon model coupled to a four-fermion interaction and a critical scaling JBW electrodynamics coupled to a four-fermion interaction can both exhibit dynamical symmetry breaking at $d_{\theta}(\alpha)=2$, only the JBW case can do so for an arbitrarily weakly coupled four-fermion interaction.

In the quenched ladder approximation dynamical symmetry breaking is associated with strong coupling alone,  something that initially seems reasonable since experience with quantum mechanics indicates  that weak coupling does not lead to binding, with the 3-dimensional square well potential for instance only being able to bind above a certain minimum strength.\footnote{For a particle of mass $m$ in a 3-dimensional well of depth $V_0$ and width $a$,  binding only occurs if $V_0a^2 \geq\pi^2\hbar^2/8m$.} However, a particle in a potential is a one-body problem with a small number of degrees of freedom. The dynamics is entirely different when the system is a many-body system, as is the case in BCS theory of superconductivity. Specifically, in the BCS  theory a pair of electrons can bind into a Cooper pair  no matter how weak the coupling between them might be provided only that it is attractive \cite{Cooper1956}. It is the filling of the Fermi sea by all the other electrons in the superconductor that prevents the Cooper pairs from occupying low momentum states, to thus force the pairing wave function to be damped at large distances. Because of the effect of the filled Fermi sea, an attractive force between the electrons in a Cooper pair (as produced through the ions in the superconductor) no matter how weak it might be will lead to binding. Cooper pairing is thus an intrinsically many-body effect, with the system of interest not being just the two electrons in the Cooper pair but the two electrons plus a filled Fermi sea. Moreover, exactly the same situation is met in the JBW-NJL model with $\gamma_{\theta}(\alpha)=-1$, where the negative energy Dirac sea plays a role analogous to the Fermi sea in a metal, with the gap equation given in (\ref{L49}) requiring only that the coupling be attractive, no matter how weak it may be. Moreover, with the vacuum energy density being given by an infinite sum of massless Feynman graphs as per (\ref{L35}), every single one of these graphs would be infrared divergent. As discussed in \cite{Mannheim1974a,Mannheim1975}, when $\gamma_{\theta}(\alpha)=-1$ the theory is softened so much in the ultraviolet  that these infrared divergences become so severe that the theory is forced into a new vacuum where the fermion is no longer massless. This occurs not because the coupling is strong but because the infrared divergences are so severe. Thus strong coupling is not mandatory for dynamical symmetry breaking.

Since strong coupling is not mandatory for dynamical symmetry breaking we need to explain why study of the quenched ladder approximation suggested that it is. To this end we note that the quenched ladder approximation involves two kinds of assumptions, namely a canonical photon and a restriction to planar graphs. So we need to ask what happens when these assumptions are relaxed. The problem of including the non-planar graphs while keeping the photon canonical was actually solved exactly by Johnson, Baker, and Willey, as that was the objective of the first of their papers on electrodynamics \cite{Johnson1964}. In that paper they showed that the all order Schwinger-Dyson equation had an exact asymptotic solution in which the fermion propagator scaled as a power of the momentum of the fermion (cf. (\ref{L3})) when the photon propagator is canonical. Since the calculation only involved the asymptotic momentum properties of Feynman diagrams with none being left out (other than those involved in dressing the photon propagator), as Johnson, Baker, and Willey themselves noted, their result held no matter what the strength of the coupling constant. Their result thus holds for both weak and strong coupling, with there being no transition region in which the quenched ladder power solution given in (\ref{M85}) switches into the oscillating solution given in (\ref{M89}). Rather, no matter what the strength of the coupling constant, the solution is always in the regime in which the bare mass is not identically zero but only vanishes in the limit of large cutoff as per (\ref{L1}). Moreover, since the quenched approximation involves no charge renormalization, it corresponds to a renormalization group equation in which the $\beta(\alpha)$ term is omitted identically no matter what the value of $\alpha$, with asymptotic scaling then being the exact asymptotic solution to (\ref{L2}) for any value of $\alpha$. When treated to all orders then, the quenched, non-planar graphs completely cancel the phase transition found in the quenched, planar graph approximation, and no phase transition occurs. 

In second of their papers \cite{Johnson1967} Johnson, Baker, and Willey showed that even if one then dressed the photon propagator, it would remain canonical if the bare coupling constant obeyed the Gell-Mann-Low eigenvalue equation, with power behavior then being the exact asymptotic solution to the renormalization group equation given in (\ref{L2}).\footnote{Beyond the issue of finiteness of $Z_3$, the utility of being at a Gell-Mann-Low eigenvalue is that the residue of the deep spacelike region Landau ghost state in the photon propagator then vanishes, with the theory then being free of states of negative norm.} Thus whether the $\beta(\alpha)$ term does not contribute to (\ref{L2}) because it is taken to be zero for all $\alpha$ or because it is only zero for one value of $\alpha$, the fermion propagator is asymptotically power behaved, and the Abelian gluon model does not exhibit dynamical symmetry breaking. 

So how then do we reconcile the all-graph JBW result with the quenched planar graph result of \cite{Maskawa1974,Maskawa1975,Fukuda1976,Fomin1978}. To understand the difference we note that while the planar graphs include the first two graphs in Fig. (\ref{SD1}) and their higher-order planar analogs, left out are the third graph and its higher-order non-planar analogs. However, the third graph in Fig. (\ref{SD1}) is of the same order as the second graph in Fig. (\ref{SD1}). Thus already in second order in $\alpha$ we see a difference. And indeed the impact of non-planar graphs can already be seen in this order, since if we expand the quenched approximation $d_{\theta}=3+\gamma_{\theta}=2+(1-3\alpha/\pi)^{1/2}$ to second order according to $d_{\theta} \sim 3-3\alpha/2\pi-9\alpha^2/8\pi^2$, it differs from the exact second-order expression $d_{\theta}=3-3\alpha/2\pi-3\alpha^2/16\pi^2$ found by Johnson, Baker, and Willey under the same assumption of  a canonical photon propagator. The planar graph approximation  is thus only a weak coupling approximation\footnote{In \cite{Johnson1964} Johnson, Baker, and Willey obtained the expression $d_{\theta}=2+(1-3\alpha/\pi)^{1/2}$ for $d_{\theta}$ but noted that it was only valid for weak coupling.} and its extrapolation to strong coupling is unjustified.\footnote{In an earlier paper \cite{Mannheim1974b} the present author had presented arguments to show that dynamical symmetry breaking does not occur in an Abelian gluon model. The present manuscript completes the proof.}

Even though we have seen that the Abelian gluon model does not in and of itself exhibit a phase transition as it is always in the Baker-Johnson regime for all values of the coupling constant,  nonetheless, our interest in this paper is in working in the Baker-Johnson regime, where gauge boson exchange expressly does not generate any bound state Goldstone boson or Higgs boson in the fermion-antifermion scattering amplitude. With this regime not having been thought  to be of interest for dynamical symmetry breaking, it has essentially not been considered any further in the literature. Despite this, as we have seen in this paper, we can augment a weakly-coupled massless fermion QED with an equally weakly-coupled, non-perturbatively renormalizable, massless fermion four-fermion interaction, and then reinterpret QED in the Baker-Johnson regime as a mean-field theory with a massive fermion. And as such, there would indeed be no photon-exchange-generated Goldstone boson in the mean-field sector, since mean-field theory never does contain a Goldstone boson. Nonetheless, as we have also seen, through the four-fermion residual interaction we then do generate a dynamical pseudoscalar Goldstone boson, and do generate a dynamical scalar Higgs boson, one whose mass is naturally close to the threshold in the fermion-antifermion scattering amplitude, and not orders of magnitude larger than it. The key difference between our approach and that of the others described above is that we work in the weak coupling regime where gauge boson exchange does not generate any dynamical Goldstone or Higgs bound state, and yet still do get such bound states through an entirely different mechanism, namely that generated by a weakly coupled four-fermion residual interaction. Moreover, as we show below, such a four-fermion interaction must be included as a vacuum energy density counterterm, since once we couple the theory to gravity we can no longer ignore (i.e. normal order away) the vacuum energy density, as the hallmark of Einstein's formulation of gravity is that gravity must couple to every form of energy density whatsoever, and not to energy density difference. And if the vacuum energy density that gravity couples to is infinite, that infinity must be cancelled by an appropriate counterterm. In addition, we note that this gravitationally induced need for four-fermion counterterms applies not just to Abelian theories but to non-Abelian theories as well, with their renormalizability requiring that the relevant four-fermion counterterms have their dimension reduced from six to four. This even though it needs to be studied further, our analysis suggests that as well as Abelian theories, non-Abelian theories also be realized via critical scaling with anomalous dimensions.

If we generate a dynamical fermion mass by giving the composite operator $\bar{\psi}\psi$ a non-vanishing vacuum expectation value, we would break not just the chiral symmetry but would break scale symmetry as well, since  $\bar{\psi}\psi$ is dimensionful. However, since only one composite operator is involved, giving the operator  a non-vanishing  expectation value could only generate one Goldstone boson and not two. Hence it suffices to look at the implications of chiral symmetry breaking alone. In the chiral symmetry breaking case then, if we do generate a pseudoscalar bound state, because of the underlying chiral symmetry, we must generate a scalar bound state as well. However, the Goldstone theorem only requires that one of these bound states be massless, with the other one needing to be  massive, and to thus have a mass that would then naturally be related to the symmetry breaking scale that generated the fermion mass in the first place. 

In general, to determine how many bound states are to be generated when there is dynamical symmetry breaking, it is very instructive to look not at the bound state equations as evaluated with massive fermion propagators, but to evaluate them with massless ones instead. On doing so, this will then lead to a set of degenerate tachyons, and tell us how many states there are in that set. The number of such states will be the same as the number of bound states that will be obtained after we change the vacuum and evaluate using  massive fermion propagators. However, once the symmetry is broken the states will no longer be degenerate (or tachyonic). We carried this tachyonic analysis through for both the point-coupled NJL (Sec. III) and the dressed JBW-NJL (Sec. IV) cases, and in both cases we found just one scalar tachyon and one pseudoscalar tachyon. We did not need to appeal to the breaking of a symmetry such as scale invariance with its potential pseudo-Goldstone dilaton, and instead  identified the Higgs particle as the chiral partner of the Goldstone boson associated with chiral symmetry breaking.\footnote{We are not claiming that one could not get a dynamical dilaton via dynamical breaking of scale symmetry, but only that it appears to us that it would not be the weak interaction Higgs boson.} Moreover, we do not need to introduce a small breaking of the  scale symmetry by hand in the Lagrangian in order to give a would-be dilaton a small mass and make it a light pseudo-Goldstone boson. Rather, we can keep the chiral symmetry exact at the level of the Lagrangian (and the scale symmetry too), break the chiral symmetry  (and scale symmetry) only in the vacuum, and finish up with a necessarily massive scalar bound state Higgs particle whose mass is naturally of the order of the chiral symmetry breaking scale, and not orders of magnitude larger. 

Thus just as with the double-well potential associated with an elementary Higgs field, it is very instructive to first explore what happens if we quantize around the unbroken vacuum, and if on doing so we discover the presence of tachyons,  we know that we need to change the vacuum. Since we did find such a double-well structure in the mean-field effective Higgs Lagrangians presented in Sec. III and Sec. IV (an $SU(2)_L\times U(1)$ generalization of (\ref{L23}) may be found in \cite{Terazawa1977}), we know that in both the point-coupled NJL and the dressed JBW-NJL cases we need to change the vacuum, just as we then did in our paper. While we do find a double-well potential structure in the effective Higgs Lagrangians presented in Sec. III and Sec. IV, both of those Lagrangians are only c-number effective mean-field Lagrangians for c-number matrix elements of fermion bilinear composites. They are not to be quantized -- in fact they only arise after the fundamental fermionic theory already has been quantized and the condensates have formed. Consequently, the value of the second derivative of any such effective potential at its minimum is not a measure of the mass of  the dynamical Higgs boson that would then be generated by the residual interaction. (In fact in our case it anyway could not be since the second derivative of the effective potential given in (\ref{L56}) is real while the mass of the dynamical Higgs boson that we find is complex, as it is an above-threshold resonance with a non-zero width.) Nonetheless, the fact that such effective potentials do have well-defined non-trivial minima is an indication that the residual interaction will in fact generate dynamical bound states. 

While our work could generalize to the non-Abelian case if there is critical scaling with anomalous dimensions,\footnote{Having critical scaling in a non-Abelian gauge theory is not ordinarily considered in the literature because it would mean giving up asymptotic freedom. However, this loss of asymptotic freedom may not be as problematic as it may at first sound. As shown  in \cite{Mannheim1975}, albeit somewhat heuristically, the residual-interaction-generated fluctuations around the self-consistent Hartree-Fock vacuum turn out to be asymptotically free. So we use critical scaling with anomalous dimensions to get into the self-consistent vacuum in the first place, with the fluctuations around it then being asymptotically free.}
 in the literature most analyses of dynamical symmetry breaking have focused on the fact that non-Abelian gauge theories are asymptotically free in perturbation theory. For instance, in the $SU(3)$ QCD case with $N_f$ fermions and coupling constant $\alpha_s$, the first two terms in the expansion of the coupling constant renormalization $\beta(\alpha_s)$ function are given by $\beta(\alpha_s)=-\beta_1\alpha_s^2-\beta_2\alpha_s^3$, where $\beta_1=(33-2N_f)/6\pi$, $\beta_2=(306-38N_f)/24\pi^2$. While both $\beta_1$ and $\beta_2$ have the same positive sign for small enough $N_f$, $\beta_2$ can change sign if $N_f>8.05$. There is thus a window in which $\beta_1$ is positive and $\beta_2$ is negative and $\beta(\alpha_s)$ has a zero away from the origin, viz. $8.05\leq N_f\leq 16.5$. Within (a part of) this window it is thought that gauge-boson-exchange-generated dynamical symmetry breaking could occur.

In the literature it has been argued that if the $\beta(\alpha_s)$ function of a non-Abelian gauge theory had a zero away from the origin, one might be able to be near (the non-Abelian equivalent of) $d_{\theta}=2$ and could then implement the walking technicolor scenario in which there is a slow running of the coupling constant so as to separate the confinement and symmetry breaking phases \cite{Holdom1981,Holdom1985,Appelquist1986},  with general discussion of the non-Abelian case being found in \cite{Miransky1993,Hill2003,Contino2010,Miransky2010,Appelquist2010b,Yamawaki2010,Panico2015} and references therein.
In $SU(3)$ the anomalous dimension of the fermion mass operator is given by (see e.g. \cite{Ryttov2011} and references therein) $\gamma_{\theta}(\alpha_s)=-2\alpha_s/\pi-\alpha_s^2(303-10N_f)/36\pi^2$, and like in the Abelian case starts off negative in lowest order, and would remain negative in second order if $N_f< 30.3$, with propagators then being asymptotically damped. To have $\gamma_{\theta}(\alpha_s)$ be close to two through second order in $\beta(\alpha)$ would require a careful tuning of the number of fermions. Such a tuning would not be necessary however if instead the non-Abelian gauge theory is realized via non-perturbative critical scaling, with the dimension of $\bar{\psi}\psi$ being reduced from three to two in order to produce vacuum breaking.

Because of studies of models such as the quenched ladder approximation to the Abelian gluon theory in the $\alpha > \pi/3$ region, it had been thought that dynamical symmetry breaking can only occur for strong coupling. And with the weak interaction symmetry breaking scale (viz. the value of $\langle \Omega|\phi|\Omega\rangle$) being much bigger than the strong interaction chiral symmetry breaking scale (viz. $f_{\pi}$) that is to be produced by QCD dynamics, a non-Abelian technicolor gauge theory of strength greater than QCD has been invoked in order to break the weak interaction symmetry  dynamically. This breaking induces high mass (TeV or so region) technifermions and should lead to an equally high mass dynamical Higgs boson. While theoretical attempts to then bring the Higgs mass down to the 125 GeV value that it is now known to have are currently ongoing, no satisfactory solution to this problem has yet been found, with attempts to have the Higgs boson emerge as a relatively light (viz. pseudo) Goldstone boson that could be associated with  a spontaneous breakdown of scale symmetry (cf. a dilaton) have yet to succeed. However, as we have seen, none of this may be necessary, since our study here shows that dynamical symmetry breaking can occur even with weak coupling, to potentially make theories such as technicolor unnecessary.

In this paper we are proposing that the observed weak interaction Higgs particle be identified as the partner of the Goldstone boson associated with dynamical chiral symmetry breaking. However, the standard weak interaction is based on the breaking of the Weinberg-Salam-Glashow $SU(2)_L\times U(1)$ and is not chiral invariant. To implement our proposed status for the Higgs particle, we must thus take the weak interaction to be chiral invariant, and to be of the form $SU(2)_L\times SU(2)_R\times U(1)$ (or of the form of some even larger gauge group into which $SU(2)_L\times SU(2)_R\times U(1)$  can be embedded), and in such a case the theory would then necessarily contain right-handed neutrinos. Chiral weak interactions have been well studied in the literature (see e.g. \cite{Mannheim1980} and references therein) and the most economical way to recover $SU(2)_L \times U(1)$  is to break $SU(2)_L\times SU(2)_R\times U(1)$ via a right-handed neutrino Majorana mass, viz.  a neutrino pairing analog \cite{Mannheim1980} of Cooper pairing. Moreover, not only does a right-handed neutrino Majorana mass have just the right quantum numbers to precisely break $SU(2)_L\times SU(2)_R\times U(1)$ down to $SU(2)_L \times U(1)$ (it transforms as  an $SU(2)_R$ triplet), it is a fermion bilinear ($\psi C\psi$), and thus can naturally fit into the dynamical symmetry breaking framework presented in this paper. Thus we propose that  $SU(2)_L\times SU(2)_R\times U(1)$ be broken by a right-handed neutrino Majorana mass (at some high scale that would appropriately suppress  right-handed currents)  and then by standard fermion Dirac masses. This will generate three Goldstone bosons that will then make the $W^+$, $W^-$, and $Z$ intermediate vector bosons massive via the Higgs mechanism, and through the underlying chiral symmetry yield an observable massive scalar Higgs particle as well. Moreover, it is natural that we should associate  the three Goldstone bosons  and the single Higgs boson with the same symmetry (viz. chiral symmetry, not scale symmetry) since they all belong to the same complex doublet representation of $SU(2)_L$.

Now as well as the weak interaction having a Goldstone structure, the strong interaction does too, with quantum chromodynamics (QCD) possessing a global chiral flavor symmetry that is spontaneously broken to yield the pion family chiral multiplet. All of the members of this multiplet remain in the spectrum and none is incorporated (i.e. Higgsed) into any  massive gauge boson. There are thus two separate Goldstone boson families in nature, not one, and we would need some mechanism to distinguish them, as we would not know whether to assign quark condensate breaking to the strong or the weak interaction. Now one of the advantages of introducing technicolor is that one can address this question, with quark condensates breaking the strong interaction and technifermion condensates then breaking the weak interaction. However, suppose the weak interaction is chiral invariant. We would then need to break $SU(2)_L\times SU(2)_R\times U(1)$ down to $SU(2)_L\times U(1)$. To generate the needed $SU(2)_R$ triplet  dynamically, we would need a difermion made of a pair of fermions each one of which transforms as  a doublet under $SU(2)_R$. However, we would also need the difermion pair operator to be electrically neutral if it is to go into the vacuum. Thus absent technineutrinos, we must do the breaking via the conventional neutrinos, the only currently known electrically neutral fundamental fermions. Thus instead of technifermion condensates, we (admittedly speculatively) propose that the strong interaction chiral symmetry be broken entirely by quark condensates, and a weak interaction chiral symmetry be broken entirely by lepton condensates, i.e. by right-handed neutrino Majorana masses, and then by charged and neutral lepton Dirac masses.

Now a reader might initially baulk at a proposal such as this since weak interactions are so weak. However, they are not so weak before the symmetry is broken, as their weakness is due not to a weak coupling strength but due to the large intermediate vector boson masses that are expressly generated by the symmetry breaking mechanism itself. Also, more mass scales can be generated if the strong, electromagnetic and weak interactions are all embedded in some grand unifying gauge group. And indeed, precisely such grand unifying scales are utilized in the seesaw mechanism, which could explain the very small (milli electron Volt) masses found for the  left-handed neutrinos. In this mechanism the neutrino mass matrix is a $2\times 2$  matrix in the space of the left- and right-handed neutrinos, with matrix elements $M_{11}=0$, $M_{12}=M_{21}=M$, $M_{22}=\Delta$, where $M$ is a Dirac mass and $\Delta$ is a right-handed neutrino Majorana mass. For $\Delta \gg M$, the mass eigenvalues are given by $-M^2/\Delta$ and $\Delta$. Then for a typical $\Delta \sim 10^{15}~{\rm GeV}$, $M \sim 10^{2}~{\rm GeV}$ (i.e. $M$ of order the Higgs boson mass), $M^2/\Delta$ would be of order $10^{-2}~{\rm meV}$. The seesaw mechanism thus relies on GeV scale neutrino Dirac masses and not on meV scale ones, to thus suggest that the neutrino Dirac mass scale is much larger than the meV neutrino mass scale measured in neutrino oscillation experiments. However, while we have explained in this paper how one could generate a dynamical Higgs boson in the Baker-Johnson  regime, getting explicit numerical values for dynamical mass scales is beyond the scope of the present paper and remains to be addressed.

Having discussed how dynamical symmetry breaking works in dressed JBW-NJL type theories, and with the associated anomalous dimensions appearing because of the underlying conformal symmetry of QED, we turn now to a general comparison of conformal symmetry with supersymmetry, and shall give particular attention to the vacuum energy density, to the role of the four-fermion interaction, and to conformal gravity, a conformal invariant theory of gravity that has been advanced as a candidate alternative to the standard Einstein gravitational theory.

\section{Conformal Symmetry Challenges Supersymmetry}

\subsection{Cancellation of Infinities}

Because of the Fermi statistics of half-integer spin particles and the Bose statistics of integer spin particles, the Feynman diagrams of closed fermion loops and closed boson loops have opposite overall signs. Consequently, they are able to cancel each others' perturbative infinities to some degree. This can occur not just in supersymmetry but also in supergravity, its local extension (a recent review of cancellations in the supergravity case may be found in \cite{Bern2014}). 

To compare and contrast with conformal symmetry, we note that with critical scaling there is also a cancellation of infinities. However, it does not occur  order by order in perturbation theory. Rather, it is only achieved non-perturbatively via an  infinite summation of diagrams. In this paper we have encountered four examples of this, the finiteness of the gauge boson wave function renormalization constant $Z_3$, the form for $m_0$ given in (\ref{L1}), the structure of $T_{\rm S}(q^2)$ and $T_{\rm P}(q^2)$ in both the massless and massive cases, and the form for $\tilde{\epsilon}(m)$ as given in (\ref{L50}). 

For $Z_3$ the finiteness is achieved immediately just by being at a critical point where $\beta(\alpha)=0$. For $m_0$ it is instructive to expand (\ref{L1}) as
\begin{eqnarray}
m_0&=&m\left(\frac{\Lambda^2}{m^2}\right)^{\gamma_{\theta}(\alpha)/2}=m\bigg{[}1+\frac{\gamma_{\theta}(\alpha)}{2}{\rm ln}\left(\frac{\Lambda^2}{m^2}\right)
+\frac{\gamma_{\theta}^2(\alpha)}{8}{\rm ln}^2\left(\frac{\Lambda^2}{m^2}\right)+....\bigg{]}.
\label{M96}
\end{eqnarray}
In this expansion all the radiative-correction terms individually diverge. However because of critical scaling  the coefficients of these terms are such that their non-perturbative sum exponentiates, with the sum itself then being finite if $\gamma_{\theta}(\alpha)$ is negative. Thus if one were to write (\ref{M96}) as some low-order perturbative term plus a counterterm, the counterterm would then represent the rest of the series. Thus in the language of perturbation theory, critical scaling uniquely fixes the needed counterterm. Since the cancellation is really a cancellation of infinities in the vertex renormalization constant $Z_{\rm S}=Z^{-1/2}_{\rm \theta}=(\Lambda^2/\mu^2)^{\gamma_{\theta}(\alpha)/2}$ that multiplicatively renormalizes the massless theory $\Gamma_{\rm S}(p,p,0,m=0)= (\Lambda^2/p^2)^{-\gamma_{\theta}(\alpha)/2}$ to give $\tilde{\Gamma}_{\rm S}(p,p,0,m=0)= (p^2/\mu^2)^{\gamma_{\theta}(\alpha)/2}$, it is a purely ultraviolet effect. Thus it can occur in either a massless theory or in the short-distance behavior of a massive theory, with it not being sensitive to any mass generation that might be taking place in the infrared. 

As well as control the short-distance behavior of mass-related Green's functions such as the insertion of a single $\bar{\psi}\psi$ into the inverse fermion propagator (viz.  $\Gamma_{\rm S}(p,p+q,q,m=0)$ as per (\ref{L42})), the requirement of conformal invariance also controls the short-distance behavior of Green's functions involving more than one $\bar{\psi}\psi$, doing so even though they play no direct role in QED itself other than in its vacuum energy density. Specifically, without as yet any reference to a four-fermion interaction, consider the two-point function $\Pi_{\rm S}(x)=\langle \Omega|T(\bar{\psi}(x)\psi(x)\bar{\psi}(0)\psi(0))|\Omega\rangle$. In a free fermion theory it consists of two $\bar{\psi}\psi$ insertions in a free fermion loop, with the leading term in its  Fourier transform being the quadratic divergence given in (\ref{L26}). If we now switch on a critical scaling QED the bare vertices get dressed with anomalous dimensions  into $\tilde{\Gamma}_{\rm S}(p,p+q,q,m=0)$ as given in (\ref{L54}), with  the leading term in the Fourier transform of $\Pi_{\rm S}(x)$ now only being the logarithmic divergence given in (\ref{L58}) when $d_{\theta}=2$. Thus in a critical scaling QED with $d_{\theta}=2$, the ultraviolet structure of the $\bar{\psi}\psi$ Green's functions is softened from quadratic to logarithmic (and likewise for the $\bar{\psi}i\gamma_5\psi$ Green's functions). Hence regardless of any possible coupling of QED to a four-fermion interaction, the ultraviolet structure of the $\bar{\psi}\psi$ Green's functions in both the massless and massive fermion QED cases is no higher than a readily renormalizable logarithmic. 

In addition, we note that if there is critical scaling in QED at short distances, the short-distance behavior of $\Pi_{\rm S}(x)$ is uniquely fixed by conformal invariance, and takes the very specific form given in (\ref{L51}).   While such a form would not hold order by order in perturbation theory, it does hold non-perturbatively in QED if there is conformal invariance with anomalous dimensions, with the order by order Feynman diagrams collectively organizing themselves non-perturbatively into the scaling form given in (\ref{L51}). In (\ref{M96}) we explicitly exhibit an analogous such non-perturbative organization for the bare mass $m_0$. It is through such critical scaling organization that the four-fermion interaction is made renormalizable at $d_{\theta}=2$. Thus once the point four-fermion vertices are dressed with a $d_{\theta}=2$ dressing, they are softened enough so as to lead to diagrams that are no more than logarithmically divergent. In essence, the point four-fermion vertex is spread out just enough to make it renormalizable ($\int d^4x d^4 x^{\prime} g\bar{\psi}(x)\psi(x)V(x-x^{\prime})\bar{\psi}(x^{\prime})\psi(x^{\prime})$, with dimensionless $g$ and $V(x-x^{\prime})$ with dimension of inverse length squared), just like the introduction of intermediate vector bosons also spreads the weak interaction point four-fermion vertex out just enough to make it renormalizable.

If now we do couple a four-fermion interaction to QED, the scalar and pseudoscalar channel fermion-antifermion scattering amplitudes are given by the iteration of these very same $\Pi_{\rm S}(x)$ and $\Pi_{\rm P}(x)$ Green's functions as per the expression for $T_{\rm S}(q^2, m=0)$ given in (\ref{L59}) (or its pseudoscalar and massive fermion analogs). Moreover, if $g$ is introduced by the Hartree-Fock method as per the gap equation (\ref{L48}), the resulting $T_{\rm S}(q^2, m=0)$, $T_{\rm P}(q^2, m=0)$, $T_{\rm S}(q^2, M)$, and $T_{\rm P}(q^2,M)$ are not merely softened to logarithmic by $d_{\theta}=2$, they are actually finite. Hence with critical scaling, with $d_{\theta}=2$, and with a gap equation for $g^{-1}$,  the ultraviolet behavior of the fermion-antifermion scattering amplitude is completely under control. The four-fermion theory scattering amplitude $T$ that results (the observable quantity in the theory) is then completely finite, with the cancellation of ultraviolet divergences expressly involving an interplay between short-distance and long-distance effects that is sensitive to the mass generation mechanism. Thus the $\gamma_{\theta}(\alpha)=-1$ condition reduces the divergences in $\Pi_{\rm S}(q^2,M)$, $\Pi_{\rm P}(q^2,M)$, and $\langle \Omega_M|\bar{\psi}\psi|\Omega_M\rangle$ to logarithmic, with the mass-generating infrared Hartree-Fock condition $\langle \Omega_M|\bar{\psi}\psi|\Omega_M\rangle=M/g$ then leading to completely finite scattering amplitudes $T_{\rm S}(q^2,M)$ and $T_{\rm P}(q^2,M)$. Thus in the language of perturbation theory, critical scaling plus symmetry breaking uniquely fixes the needed counterterms. 

Exactly the same set of cancellations is found to occur for $\tilde{\epsilon}(m)$ as well.  As evidenced in (\ref{L50}), the $\gamma_{\theta}(\alpha)=-1$ condition reduces the divergence in $\epsilon(m)$ from quadratic to logarithmic, with the symmetry breaking then generating precisely the needed $m^2/2g$ counterterm  to make $\tilde{\epsilon}(m)$ completely finite. Welcome as this is, nonetheless, left out from this discussion is the vacuum energy density quartic divergence to which we had  alluded before. And so it is to this issue that we now turn.

\subsection{Supersymmetry Treatment of the Vacuum Energy Density}

There are two separate issues for the vacuum energy density. First, simply because a matter field energy-momentum tensor is composed  of products of quantum fields at the same spacetime point, there is a zero-point problem. This problem already occurs in a massless theory with a normal vacuum. And second, when one generates mass via symmetry breaking, not only does the zero-point vacuum energy density change, in addition a cosmological constant term is produced. 

To illustrate the issues that are involved, it is convenient to first look at the vacuum expectation value of the energy-momentum tensor 
\begin{eqnarray}
T^{\mu\nu}_{\rm M}=i\hbar \bar{\psi}\gamma^{\mu}\partial^{\nu}\psi 
\label{M97}
\end{eqnarray}
of a free fermion matter field of mass $m=0$ in flat, four-dimensional spacetime, with the fermion obeying the massless Dirac equation. With $k^{\mu}=(\omega_k,\bar{k})$ where $\omega_k=k$, following a Feynman contour integration in the complex frequency plane the vacuum matrix element evaluates to 
\begin{eqnarray}
\langle \Omega_0 |T^{\mu\nu}_{\rm M}|\Omega_0\rangle= -\frac{2\hbar}{(2\pi)^3}\int_{-\infty}^{\infty}d^3k\frac{k^{\mu}k^{\nu}}{\omega_k}.
\label{M98}
\end{eqnarray}
With its $k^{\mu}k^{\nu}$ structure  $\langle \Omega_0 |T^{\mu\nu}_{\rm M}|\Omega_0\rangle$ has the generic form of a perfect fluid with a timelike fluid velocity vector $U^{\mu}=(1,0,0,0)$, viz.
\begin{eqnarray}
\langle \Omega_0 |T^{\mu\nu}_{\rm M}|\Omega_0\rangle&=& (\rho_{\rm M}+p_{\rm M})U^{\mu}U^{\nu}+p\eta^{\mu\nu},
\label{M99}
\end{eqnarray}
where
\begin{eqnarray}
\rho_{\rm M}=\langle \Omega_{\rm 0}|T^{00}_{\rm M}|\Omega_0\rangle= -\frac{2\hbar}{(2\pi)^3}\int_{-\infty}^{\infty}d^3k\omega_k,
\label{M100}
\end{eqnarray}
\begin{eqnarray}
p_{\rm M}&=&\langle \Omega_0 |T^{11}_{\rm M}|\Omega_0\rangle=\langle \Omega_0 |T^{22}_{\rm M}|\Omega_0\rangle=\langle \Omega_0 |T^{33}_{\rm M}|\Omega_0\rangle
= -\frac{2\hbar}{3(2\pi)^3}\int_{-\infty}^{\infty}d^3k\frac{k^2}{\omega_k}.
\label{M101}
\end{eqnarray}
The zero-point energy density $\rho_{\rm M}$ and the zero-point pressure $p_{\rm M}$ are related by the tracelessness condition
\begin{eqnarray}
\eta_{\mu\nu}\langle \Omega_0 |T^{\mu\nu}_{\rm M}|\Omega_0\rangle&=&3p_{\rm M}-\rho_{\rm M}=0
\label{M102}
\end{eqnarray}
since $\eta_{\mu\nu}k^{\mu}k^{\nu}=0$. (We use ${\rm diag}[\eta_{\mu\nu}]=(-1,1,1,1)$ here and in the discussion of gravity below.) Since  $p_{\rm M}$ is not equal to $-\rho_{\rm M}$, the zero-point  energy-momentum tensor does not have the form of a cosmological constant term, to underscore that fact that the zero-point problem is distinct from the cosmological constant problem. 

With both $\rho_{\rm M}$ and $p_{\rm M}$ being divergent, in terms of a 3-momentum cutoff $K$ the divergences can be parametrized as the quartic divergences
\begin{eqnarray}
\rho_{\rm M}=-\frac{\hbar K^4}{4\pi^2},\qquad p _{\rm M}=-\frac{\hbar K^4}{12\pi^2}.
\label{M103}
\end{eqnarray}
Cancellation of these mass-independent quartic divergences is readily achieved in supersymmetry  since a massless boson loop has the opposite sign to a massless fermion loop.

However, the situation changes once the fermion acquires mass. For a free massive fermion in flat spacetime with vacuum $|\Omega_{\rm M}\rangle$ the form of the energy-momentum tensor remains unchanged but the Dirac equation becomes that of a massive fermion. Then, with $k^{\mu}=((k^2+m^2/\hbar^2)^{1/2},\bar{k})$, $\rho_{\rm M}$ and $p_{\rm M}$ now evaluate to
\begin{eqnarray}
\rho_{\rm M}&=&-\frac{\hbar K^4}{4\pi^2}- \frac{m^2K^2}{4\pi^2\hbar} +\frac{m^4}{16\pi^2\hbar^3}{\rm ln}\left(\frac{4\hbar^2K^2}{m^2}\right)
-\frac{m^4}{32\pi^2\hbar^3},
\nonumber \\
p _{\rm M}&=&-\frac{\hbar K^4}{12\pi^2}+ \frac{m^2K^2}{12\pi^2\hbar} -\frac{m^4}{16\pi^2\hbar^3}{\rm ln}\left(\frac{4\hbar^2K^2}{m^2}\right)
+\frac{7m^4}{96\pi^2\hbar^3},
\label{M104}
\end{eqnarray}
and while $3p_{\rm M}-\rho_{\rm M}$ is no longer zero, $p_{\rm M}$ remains unequal to $-\rho_{\rm M}$. In addition to the previous quartic divergence, in (\ref{M104}) we also encounter quadratic and logarithmic divergences. Since both of these latter two divergences are mass dependent, they cannot be canceled by an interplay between fermions and bosons unless the fermions and bosons are degenerate in mass. Since no supersymmetric partners of the ordinary particles have been detected to date, we know that the masses of the superparticles  are far from being degenerate with those of the ordinary particles, with supersymmetry thus leaving $\langle \Omega_{\rm M} |T^{\mu\nu}_{\rm M}|\Omega_{\rm M}\rangle$ quadratically  divergent. In fact the situation is similar to that met with an elementary scalar Higgs field self-energy since it too has a quadratic divergence (the contribution due to a fermion that is Yukawa-coupled to the Higgs scalar field is equal to the quantity  $\Pi_{\rm S}(q^2,M)$ given in (\ref{L29})). And it too can only be canceled via supersymmetry if there is a superparticle in the same mass region as the Higgs particle itself, and this appears not to be the case.

Finally, as regards the cosmological constant, as long as the supersymmetry is unbroken, the cosmological constant is zero. Specifically, in a supersymmetric theory one has a generic anticommutator of the form $\{Q^{\alpha},Q_{\alpha}^{\dagger}\}=H$, where the $Q_{\alpha}$ are Grassmann supercharges and $H$ is the Hamiltonian. If the supercharges annihilate $|\Omega_0\rangle$ (viz. unbroken supersymmetry), then $\langle \Omega_0|H|\Omega_0\rangle$ is zero, the energy of the vacuum is zero, and the cosmological constant is thus zero too. However if the Grassmann charges do not annihilate the vacuum $|\Omega_{\rm M}\rangle$ then $\langle \Omega_{\rm M}|H|\Omega_{\rm M}\rangle$ is non-zero and a non-zero cosmological constant is induced, one whose magnitude would be as big as the supersymmetry breaking scale. Since this scale is known to be no smaller than the largest currently accessible energy at the LHC, this would give a cosmological constant contribution to standard Einstein-gravity based cosmology that would be at least 60 or so orders of magnitude larger than allowable by current Hubble plot data.

\subsection{Conformal Gravity Treatment of the Vacuum Energy Density}

If the breaking of supersymmetry leads to uncanceled quadratic and logarithmic divergences in the vacuum energy density, then if one is not to appeal to supersymmetry one must not only seek some other mechanism to cancel the quadratic and logarithmic divergences in (\ref{M104}), one must also seek some alternate way to cancel the quartic divergence as well. Then, if the quartic divergence given in (\ref{M104}) is not to be canceled by a boson loop associated with a superparticle, then the only apparent remaining option is for it to be canceled by gravity itself, as the gravitational field $g_{\mu\nu}$ is itself bosonic. And indeed in any quantum gravitational theory one would encounter products of gravitational fields, with quantum gravity thus having a zero-point problem of its own. Now one cannot make the needed cancellation using standard Einstein gravity itself since it is not renormalizable at the quantum level. However, one can do so in conformal gravity since it is a consistent  quantum theory, being renormalizable, unitary, and ghost free \cite{Bender2008a,Bender2008b,Mannheim2011a,Mannheim2012a}. 

Conformal gravity, which the present author first thought to consider because of his familiarity with conformal invariance and critical scaling in flat spacetime theories, assumes invariance under local conformal transformations of the form $g_{\mu\nu}(x)\rightarrow e^{2\alpha(x)}g_{\mu\nu}(x)$. The pure gravitational sector of the theory is then given by the Weyl-tensor-based action (see e.g. \cite{Mannheim2012a})
\begin{eqnarray}
I_{\rm W}&=&-\alpha_g\int d^4x (-g)^{1/2}C_{\lambda\mu\nu\kappa} C^{\lambda\mu\nu\kappa}
\nonumber\\
&\equiv& -2\alpha_g\int d^4x
(-g)^{1/2}\left[R_{\mu\kappa}R^{\mu\kappa}-\frac{1}{3}
(R^{\alpha}_{\phantom{\alpha}\alpha})^2\right],
\label{M105}
\end{eqnarray}
where $\alpha_g$ is a dimensionless gravitational coupling  constant, with this action being the unique action that is locally conformal invariant in four spacetime dimensions. Functional variation of this action with respect to the metric defines a gravitational tensor
\begin{eqnarray}
W^{\mu \nu}&=&
\frac{1}{2}g^{\mu\nu}(R^{\alpha}_{\phantom{\alpha}\alpha})   
^{;\beta}_{\phantom{;\beta};\beta}+
R^{\mu\nu;\beta}_{\phantom{\mu\nu;\beta};\beta}                     
 -R^{\mu\beta;\nu}_{\phantom{\mu\beta;\nu};\beta}                        
-R^{\nu \beta;\mu}_{\phantom{\nu \beta;\mu};\beta}    
- 2R^{\mu\beta}R^{\nu}_{\phantom{\nu}\beta}                                 
+\frac{1}{2}g^{\mu\nu}R_{\alpha\beta}R^{\alpha\beta}
\nonumber\\                      
&-&\frac{2}{3}g^{\mu\nu}(R^{\alpha}_{\phantom{\alpha}\alpha})          
^{;\beta}_{\phantom{;\beta};\beta} 
+\frac{2}{3}(R^{\alpha}_{\phantom{\alpha}\alpha})^{;\mu;\nu}                           
+\frac{2}{3} R^{\alpha}_{\phantom{\alpha}\alpha}
R^{\mu\nu}                              
-\frac{1}{6}g^{\mu\nu}(R^{\alpha}_{\phantom{\alpha}\alpha})^2,
\label{M106}
\end{eqnarray}        
and a fourth-order derivative equation of motion of the form
\begin{equation}
-4\alpha_g W^{\mu\nu}+T^{\mu\nu}_{\rm M}=0,
\label{M107}
\end{equation}
when the theory is coupled to a conformal invariant matter sector. If we define $-4\alpha_g W^{\mu\nu}$ to be the energy-momentum tensor $T^{\mu\nu}_{\rm GRAV}$ of gravity (i.e. the variation with respect to the metric of the pure gravitational sector of the action), and introduce an energy-momentum tensor for the universe as a whole, we can rewrite (\ref{M107}) as
\begin{equation}
T^{\mu\nu}_{\rm UNIV}=T^{\mu\nu}_{\rm GRAV}+T^{\mu\nu}_{\rm M}=0,
\label{M108}
\end{equation}
to thus put the gravity and matter sectors on an equal footing, while showing that the  total energy-momentum tensor of the universe is zero.

Given the conformal symmetry, no dimensionful parameters are allowed in the conformal action. Thus both the Einstein-Hilbert action 
\begin{equation}
I_{\rm EH}=-\frac{1}{16 \pi G}\int d^4x (-g)^{1/2}R^{\alpha}_{\phantom{\alpha}\alpha}
\label{M109}
\end{equation}
and a cosmological constant action
\begin{equation}
I_{\Lambda}=-\int d^4x (-g)^{1/2}\Lambda
\label{M110}
\end{equation}
are forbidden. Thus just like supersymmetry, conformal symmetry forbids the presence of any fundamental cosmological constant at the level of the Lagrangian. 

The actual quantization procedure for the conformal gravity theory is somewhat unusual since it cannot follow the standard quantization prescription that is ordinarily used for fields. Specifically, for a standard matter field one obtains its equation of motion by varying the matter action with respect to the matter field, but one obtains its energy-momentum tensor $T^{\mu\nu}_{M}$ by instead varying the matter action with respect to the metric. Since $T^{\mu\nu}_{M}$ involves products of matter fields at the same point, a canonical quantization of the matter field then gives the matter energy-momentum tensor  a non-vanishing zero-point contribution. However, in a standard quantization procedure for a given matter field, the non-vanishing of $T^{\mu\nu}_{M}$ violates no constraint since  one does not simultaneously impose the equation of motion of any other field. Thus for a given matter field one does not require stationarity with respect to the metric, with $T^{\mu\nu}_{M}$ thus not being constrained to vanish. And of course, if one does not couple to gravity, one can even normal order $T^{\mu\nu}_{M}$ away.

In contrast however, for gravity the relevant field is the metric itself, and the gravitational equation of motion is then given by $T^{\mu\nu}_{GRAV}=0$, as $T^{\mu\nu}_{GRAV}$ is the variation with respect to the metric of the gravity sector action. Then, with $T^{\mu\nu}_{GRAV}$ containing products of fields at the same point, a canonical quantization of the gravitational field would give a zero-point contribution to $T^{\mu\nu}_{GRAV}$, and thus violate the stationarity condition $T^{\mu\nu}_{GRAV}=0$ that $T^{\mu\nu}_{GRAV}$ would have to obey. Hence, unlike the matter fields for which there is no constraint on $T^{\mu\nu}_{M}$ in the absence of any coupling of matter to gravity, gravity itself is always coupled to gravity, with its own stationarity condition not permitting it to consistently be quantized on its own. Since  $T^{\mu\nu}_{GRAV}$ will be non-zero if $g_{\mu\nu}$ is a quantum field, one cannot impose stationarity for $T^{\mu\nu}_{GRAV}$ on its own. Rather, one must impose stationarity on the total $T^{\mu\nu}_{\rm UNIV}$, i.e. on $T^{\mu\nu}_{\rm GRAV}+T^{\mu\nu}_{\rm M}$. Then, with $T^{\mu\nu}_{\rm M}$ already containing a zero-point contribution, the vanishing of $T^{\mu\nu}_{\rm GRAV}+T^{\mu\nu}_{\rm M}$ requires that the gravity  sector zero-point contribution precisely cancel that of the matter fields that it is coupled to. Thus gravity cannot be quantized in isolation but instead is quantized by virtue of its being coupled to a source that is quantized, with the coupling to the source fixing the normalization of the gravitational sector commutation relations. It is this interplay between the gravity and matter sectors that enables the gravity sector to take care of infinities in the matter sector, infinities such as the vacuum energy density and radiative-correction-induced anomalies of the type discussed below that arise because the regularization procedure involves the introduction of a scale symmetry breaking cutoff that would violate the conformal symmetry unless these anomalies are canceled. 

To see how quantization  of conformal gravity works in practice, we note that if we quantize the gravity sector of the conformal theory to lowest order in Planck's constant around flat (viz. the first quantum correction),\footnote{If all mass scales are to come from dynamical symmetry breaking, then since dynamical symmetry breaking is intrinsically quantum-mechanical, all geometric curvature scales must be intrinsically quantum-mechanical too. Thus in \cite{Mannheim2011c,Mannheim2012a} it was proposed that, unlike in Einstein gravity, there be no intrinsic classical gravity at all, with gravity being produced entirely by quantum effects. In such a situation one should expand the theory as a power series in Planck's constant rather than as a power series in the gravitational coupling constant, with there thus being no term of order $\hbar^0$ in the expansion, and with the first non-trivial term being the term of order $\hbar$ given in (\ref{M111}).}  and take the vacuum expectation value of $T^{\mu\nu}_{\rm GRAV}$ in the massless vacuum $|\Omega_0 \rangle$ we obtain a quartically divergent zero-point energy density in the gravity sector of the form \cite{Mannheim2012a}
\begin{equation}
\langle \Omega_0|T^{\mu\nu}_{\rm GRAV}|\Omega_0 \rangle=\frac{2\hbar}{(2\pi)^3} \int_{-\infty}^{\infty}d^3k\frac{Z(k)k^{\mu}k^{\nu}}{\omega_k},
\label{M111}
\end{equation}
where $Z(k=|\bar{k}|)$ is the as yet to be determined gravitational field wave function renormalization constant, as defined \cite{Mannheim2011a,Mannheim2012a} as the coefficient of the delta function in canonical commutation relations for the momentum modes of the gravitational field.  Inserting (\ref{M111}) and (\ref{M103}) into (\ref{M108}) then yields 
\begin{equation}
Z(k)=1.
\label{M112}
\end{equation}
Thus, simultaneously we fix the gravity sector renormalization constant and effect a complete cancellation of the quartically divergent zero-point terms. Moreover, we do not need to introduce any regulators to separately define either $\langle \Omega_0|T^{\mu\nu}_{\rm GRAV}|\Omega_0 \rangle$ or $\langle \Omega_0|T^{\mu\nu}_{\rm M}|\Omega_0 \rangle$, as each term regulates the other as needed to maintain the stationarity condition $\langle \Omega_0|T^{\mu\nu}_{\rm UNIV}|\Omega_0 \rangle=0$, with the cancellation being done mode by mode and not mode sum by mode sum. As long as (\ref{M108}) is maintained order by order in perturbation theory (which it is since both the gravity and matter sectors are renormalizable when conformal), then the mode by mode cancellation will persist, with matrix elements of $T^{\mu\nu}_{\rm UNIV}$ never having a zero-point problem.  In addition, we note that we not only do not need to specify $Z(k)$ a priori,  we actually cannot in fact do so. Rather, $Z(k)$ is determined entirely by the coupling of gravity to matter, with the quantization of matter enforcing the quantization of gravity since the condition  $Z(k)=0$ is not consistent with (\ref{M108}). To underscore that $Z(k)$ cannot be assigned independently but is determined by the structure of the matter source to which gravity is coupled, we note that if the gravitational source consists of $M$ massless gauge bosons and $N$ massless two-component fermions, the vanishing of $\langle \Omega_0|T^{\mu\nu}_{\rm UNIV}|\Omega_0 \rangle$ then entails that $2Z(k)+M-N=0$ \cite{Mannheim2011a}, with gravity adjusting to whatever its source is.

Moreover, since we do not need to introduce any regulators we do not obtain any anomalies such as the trace anomaly (absent any violations of the conformal symmetry the trace of the energy-momentum tensor is zero). Specifically, while conformal invariance Ward identities would be violated by trace anomalies in both $\langle \Omega_0|T^{\mu\nu}_{\rm GRAV}|\Omega_0 \rangle$ and $\langle \Omega_0|T^{\mu\nu}_{\rm M}|\Omega_0 \rangle$, the vanishing of $\langle \Omega_0|T^{\mu\nu}_{\rm UNIV}|\Omega_0 \rangle$ is not a Ward identity condition but a stationarity condition. Since $\langle \Omega_0|T^{\mu\nu}_{\rm UNIV}|\Omega_0 \rangle$ is thus anomaly free, anomalies in $\langle \Omega_0|T^{\mu\nu}_{\rm GRAV}|\Omega_0 \rangle$ and $\langle \Omega_0|T^{\mu\nu}_{\rm M}|\Omega_0 \rangle$ must cancel each other identically, being able to do so because it is that very cancellation that fixes $Z(k)$ in the first place. Moreover, since we can effect a mode by mode cancellation  without needing to look for a regulated sum of modes by regulated sum of modes cancellation, we never have to deal with the trace anomaly at all, and can treat both $T^{\mu\nu}_{\rm GRAV}$ and $T^{\mu\nu}_{\rm M}$ as continuing to retain the tracelessness required by conformal invariance.

The conformal gravity cancellation of zero-point infinities described above is not quite the same as the supersymmetry cancellation, since that cancellation did not address the gravitational zero-point energy problem, to thus leave the issue open.  To clarify the issue, consider the second-order derivative Einstein gravity equation of motion 
\begin{equation}
-\frac{1}{8\pi G}\left(R^{\mu\nu} -\frac{1}{2}g^{\mu\nu}R^{\alpha}_{\phantom{\alpha}\alpha}\right)=T^{\mu\nu}_{\rm M}.
\label{M113}
\end{equation}
If (\ref{M113}) is to be an operator identity, then the two sides of it are to both be quantum-mechanical or to both be classical.  However, since the gravity side is not well-defined quantum-mechanically, one takes it to be classical. But since the matter side is built out of quantum fields, the matter side is quantum-mechanical. To get round this one replaces (\ref{M113}) by a hybrid 
\begin{equation}
-\frac{1}{8\pi G}\left(R^{\mu\nu} -\frac{1}{2}g^{\mu\nu}R^{\alpha}_{\phantom{\alpha}\alpha}\right)_{\rm CL}=\langle \Omega|T^{\mu\nu}_{\rm M}|\Omega \rangle.
\label{M114}
\end{equation}
However, since the matter term in  (\ref{M114}) has a zero-point problem, one must find a mechanism to cancel it, and must do so via the matter side alone. Now while unbroken supersymmetry actually achieves this, as we noted above, broken supersymmetry does not.  But since the gravity side of (\ref{M114}) is finite it cannot be equal to something that is infinite. Thus, in the literature one commonly ignores the fact that gravity is to couple to all forms of energy rather than only to energy differences, and subtracts off the zero-point infinity by hand and replaces (\ref{M114}) by\footnote{CL and FIN respectively denote classical and finite.} 
\begin{equation}
-\frac{1}{8\pi G}\left(R^{\mu\nu} -\frac{1}{2}g^{\mu\nu}R^{\alpha}_{\phantom{\alpha}\alpha}\right)_{\rm CL}=\langle \Omega|T^{\mu\nu}_{\rm M}|\Omega \rangle_{\rm FIN},
\label{M115}
\end{equation}
Thus in treating the contribution of the electron Fermi sea to the stability of white dwarfs or in evaluating the contribution of the cosmic microwave background to cosmology, one uses an energy operator of the generic form $H=\sum(a^{\dagger}(\bar{k})a(\bar{k})+1/2)\hbar\omega_k$, and then by hand discards the $H=\sum \hbar \omega_k/2 $ term. And then, after all this is done, the finite part of $\langle \Omega|T^{\mu\nu}_{\rm M}|\Omega \rangle$ still has an uncanceled and as yet uncontrolled cosmological constant contribution that still needs to be dealt with. The present author is not aware of any formal derivation of (\ref{M115}) starting from a consistent quantum gravity theory, and notes that since it is (\ref{M115}) that is conventionally used in astrophysics and cosmology, it would not appear to yet be on a fully secure footing.\footnote{If one starts with (\ref{M114}) where gravity is classical, any matter sector renormalization anomalies would have to be cancelled within the matter sector alone. Accomplishing this perturbatively has proven to be very difficult. However a non-perturbative possibility has been identified in the literature \cite{Adler1977}, where it was noted that in QED the trace anomaly term is given by $T^{\mu}_{\phantom{\mu}\mu}=(1/4)\beta(\alpha)N[F_{\lambda\sigma}F^{\lambda\sigma}]$ where $N$ denotes normal ordering, with the trace anomaly thus vanishing if the $\beta(\alpha)$ function is zero. In contrast, in the conformal gravity case, gravity itself plays a role in anomaly cancellation, with  $T^{\mu\nu}_{\rm UNIV}$ being anomaly free not just non-perturbatively, but with $Z(k)$ readjusting each time, it is anomaly free order by order in perturbation theory as well.}

In the conformal case not only does the gravity sector zero point cancel the massless fermion quartic divergence, the gravitational zero point continues to cancel the matter sector zero-point contribution when the fermion acquires a mass since the gravity sector zero point readjusts. Specifically, in the event of dynamical symmetry breaking, critical scaling and $\gamma_{\theta}(\alpha)=-1$, one has to take matrix elements of  (\ref{M108}) in the self-consistent, Hartree-Fock vacuum $|\Omega_{\rm M} \rangle$. The quantity $\langle \Omega_{\rm M}|T^{00}_{\rm M}|\Omega_{\rm M}\rangle$ consists of the previously introduced  $\tilde{\epsilon}(M)$ as given in (\ref{L50}) with its dynamically generated $M^2/2g$ term, together with the quartically divergent $\rho_{\rm M}$ as given in (\ref{M103}), as it had originally been removed from (\ref{L43}) since (\ref{L45}) is a vacuum energy density difference. The vanishing of $T^{\mu\nu}_{\rm UNIV}$ then entails that at the minimum where $m=M$ we obtain
\begin{equation}
\langle \Omega_M|T^{00}_{\rm GRAV}|\Omega_M \rangle-\frac{\hbar K^4}{4\pi^2}-\frac{M^4}{16\pi^2\hbar^3}=0.
\label{M116}
\end{equation}
(In (\ref{M116}) we have set $\mu=M$ for convenience.)
From (\ref{M116})  it follows  \cite{Mannheim2012a} that $Z(k)$ is given by\footnote{In \cite{Mannheim2012a} (\ref{M117}) was originally derived via a Feynman contour using the $S_{\nu}(p)$ propagator given in (\ref{L73}). Continuation to the $S_{\mu}(p)$ propagator given in (\ref{L44}) yields (\ref{M117}).} 
\begin{eqnarray}
kZ(k)&=&(k^2+iM^2/\hbar^2)^{1/2}-\frac{iM^2}{4\hbar^2(k^2+iM^2/\hbar^2)^{1/2}} 
\nonumber\\
&+&(k^2-iM^2/\hbar^2)^{1/2}+\frac{iM^2}{4\hbar^2(k^2-iM^2/\hbar^2)^{1/2}}.
\label{M117}
\end{eqnarray}
As we see, $Z(k)$ is again determined by the dynamics, and even though the gravitational modes remain massless, $Z(k)$ adjusts to the fact that the fermion has mass. 

With (\ref{M116}) and (\ref{M117}), we  see that when the symmetry is broken, $\langle \Omega_{\rm M} |T^{00}_{\rm M}|\Omega_{\rm M}\rangle$ adjusts from the purely quartic massless theory (\ref{M103}) by augmenting it with the mass-dependent logarithmic divergence given in (\ref{L47}). This logarithmic divergence is then automatically canceled by the induced and thus dynamically determined vacuum energy density term $M^2/2g$ (dynamical in the sense that it depends on the state in which matrix elements are taken), with gravity then automatically canceling the quartic divergence and the residual finite part, $-M^4/16\pi^2\hbar^3$, of  $\langle \Omega_{\rm M} |T^{00}_{\rm M}|\Omega_{\rm M}\rangle$. Moreover, the cancellation works no matter how big $M^4$ might be, and none of it is observable since it all occurs in the vacuum, i.e. it is due entirely to the occupied negative energy states in the Dirac sea. Specifically, what one measures in actual astrophysical phenomena is not the vacuum but the behavior of the positive energy modes that can  be excited out of it.

To be more specific, we note that since all of the infinities in $T^{\mu\nu}_{\rm GRAV}$ and $T^{\mu\nu}_{\rm M}$  are due to the infinite number of modes in the vacuum sector, if we decompose $T^{\mu\nu}_{\rm GRAV}$ and $T^{\mu\nu}_{\rm M}$ into finite (FIN)  and divergent (DIV) parts according to $T^{\mu\nu}_{\rm GRAV}=(T^{\mu\nu}_{\rm GRAV})_{\rm FIN}+(T^{\mu\nu}_{\rm GRAV})_{\rm DIV}$, $T^{\mu\nu}_{\rm M}=(T^{\mu\nu}_{\rm M})_{\rm FIN}+(T^{\mu\nu}_{\rm M})_{\rm DIV}$, (\ref{M108}) will decompose into 
\begin{eqnarray}
&&(T^{\mu\nu}_{\rm GRAV})_{\rm DIV}+(T^{\mu\nu}_{\rm M})_{\rm DIV}=0,
\label{M118}
\end{eqnarray}
\begin{eqnarray}
&&(T^{\mu\nu}_{\rm GRAV})_{\rm FIN}+(T^{\mu\nu}_{\rm M})_{\rm FIN}=0.
\label{M119}
\end{eqnarray}
All of the infinities are taken care of by (\ref{M118}), and for astrophysics and cosmology we can then use the completely infinity-free (\ref{M119}). In this way for studying white dwarfs or the cosmic microwave background  we can now use  $H=\sum a^{\dagger}(\bar{k})a(\bar{k})\hbar \omega_k$ alone, as the zero-point contribution has already been taken care of by gravity itself and does not appear in (\ref{M119}) at all. Moreover, when we do excite positive energy modes out of the vacuum we will generate a new cosmological constant contribution, and it is this term that is measured in cosmology. Cosmology thus only sees the change in the vacuum energy density due to adding in positive energy modes and does not see the full negative energy mode vacuum energy density itself, i.e. in (\ref{M119})  one is sensitive not to  $\langle \Omega_{\rm M}|T^{\mu\nu}_{\rm M}|\Omega_{\rm M}\rangle$, and not even to  $\langle \Omega_{\rm M}|bT^{\mu\nu}_{\rm M}b^{\dagger}|\Omega_{\rm M}\rangle$, but only to their difference  $\langle \Omega_{\rm M}|bT^{\mu\nu}_{\rm M}b^{\dagger}|\Omega_{\rm M}\rangle -\langle \Omega_{\rm M}|T^{\mu\nu}_{\rm M}|\Omega_{\rm M}\rangle$. Also gravity sees this effect mode by mode, i.e. gravity mode by fermion mode. In contrast, if one uses (\ref{M115}), then gravity sees an entire sum over fermion modes, which is one of the reasons why in the standard Einstein theory the cosmological constant effect is so big. To summarize, if one wants to take care of the cosmological constant problem, one has to take care of the zero-point problem, and when one has a renormalizable theory of gravity, via an interplay with gravity itself one is then able to do so.

Now in order to able to effect (\ref{M118}) and (\ref{M119}) order by order in perturbation theory we need both the gravity and matter sectors to be renormalizable. For a gauge-theory-based matter sector renormalizability is standard, and for the four-fermion interaction that we need in the matter sector in order to control the matter vacuum energy density, renormalizability is realized via $\gamma_{\theta}(\alpha)=-1$. As to the gravity sector, the renormalizability of conformal gravity has been established in \cite{Stelle1977} and \cite{Fradkin1985}. Since the conformal gravity coupling constant $\alpha_g$ is dimensionless, conformal gravity is power-counting renormalizable. Specifically, with $g_{\mu\nu}$ being dimensionless, in an expansion around flat spacetime of the dimension four quantity $C_{\lambda\mu\nu\kappa} C^{\lambda\mu\nu\kappa}$ as a power series in a gravitational fluctuation $h_{\mu\nu}=g_{\mu\nu}-\eta_{\mu\nu}$, each term will contain $h_{\mu\nu}$  a specific number of times together with exactly four derivatives since it is the derivatives that carry the dimension of the $C_{\lambda\mu\nu\kappa} C^{\lambda\mu\nu\kappa}$ term. The term that is quadratic in $h_{\mu\nu}$ will thus give a $1/k^4$ propagator, and each time we work to one more order in $h_{\mu\nu}$ we add an extra $1/k^4$ propagator and a compensating factor of $k^{\mu}k^{\nu}k^{\sigma}k^{\tau}$ in the vertex. With equal numbers of powers of $k^{\mu}$ being added in numerator and denominator, the ultraviolet behavior is not modified, and renormalizability is thereby maintained. Also we note that because of the $k_{\lambda}k_{\mu}k_{\nu}k_{\kappa}$ factor no infrared divergence is generated by the $1/k^4$ propagator.

Additionally, as noted in \cite{Fradkin1985}, in Euclidean space where all the eigenvalues of $g_{\mu\nu}$ are real and positive, in the basis in which  $g_{\mu\nu}$ is diagonal the quantity $C_{\lambda\mu\nu\kappa} C^{\lambda\mu\nu\kappa}$ then consists of a sum of terms each one of which is a positive definite square. Consequently, on taking $\alpha_g$ as defined in (\ref{M105}) to be positive, on every path the exponent in the path integral associated with the Euclidean $iI_{\rm W}$ is negative definite ($-i\alpha_g\int dt_0=-\alpha_g\int dt_4$). The Euclidean path integral is thus well-defined,\footnote{This is not the case for the Euclidean Einstein-Hilbert path integral.} and the theory is renormalizable in the ultraviolet and finite in the infrared. However, one cannot immediately conclude that the conformal gravity Minkowski path integral is well-behaved too because of anomalies and a  possible ghost/unitarity problem that fourth-order derivative theories have been thought to possess. We provided a resolution of the anomaly problem above, and turn now to a resolution of the ghost problem. Our resolution of the ghost problem will enable us to obtain a well-defined Minkowski path integral, one that will require a continuation of the metric into the complex plane. 

\subsection{Conformal Gravity as a Consistent Quantum Gravitational Theory}

As a quantum theory conformal gravity had long been known to be renormalizable ($\alpha_{g}$ being dimensionless), but being fourth order it had long been thought to possess negative norm ghost states that would violate unitarity. This view of conformal gravity is suggested by writing the massless fourth-order  propagator $1/k^4$ as the $M^2\rightarrow 0$ limit
\begin{eqnarray}
\frac{1}{k^4}=\lim_{M^2\rightarrow 0}\left[\frac{1}{M^2}\left(\frac{1}{k^2-M^2}-\frac{1}{k^2}\right)\right].
\label{M120}
\end{eqnarray}
With the second term in (\ref{M120}) having a negative coefficient one immediately anticipates that the theory has states with negative norm. However, from inspection of a c-number propagator alone one cannot determine what quantum-mechanical Green's function the propagator is to correspond to. For this one has to quantize the theory, construct the appropriate Hilbert space, identify appropriate asymptotic boundary conditions,  and then construct the propagator. When Bender and Mannheim did this they found \cite{Bender2008a,Bender2008b} that the quantum Hamiltonian was not Hermitian, but that it instead was  $PT$ symmetric.\footnote{With the metric being $C$ even, the Hamiltonian is actually $CPT$ symmetric, but for our purposes  we will continue to discuss the theory from the $PT$ perspective as that was how the work of \cite{Bender2008a,Bender2008b} was developed. More recently it has been shown that in general one should use $CPT$ rather than $PT$ when considering Hamiltonians that are not Hermitian \cite{Mannheim2015}, though this does not affect the results of \cite{Bender2008a,Bender2008b}.} In such a situation the correct Hilbert space norm is given by the overlap not of the right-eigenvectors of the Hamiltonian with their Dirac conjugates, viz. the Dirac norm $\langle R|R\rangle$, but rather  by the overlap of the right-eigenvectors of the Hamiltonian with its left-eigenvectors, viz. $\langle L|R\rangle$, with the left-eigenvectors being related to the $PT$ conjugates of the right-eigenvectors. And with the $\langle L|R\rangle$ norm being found to  be both finite and non-negative in the fourth-order case, when one uses the $PT$-theory norm one can associate (\ref{M120}) with a unitary theory. (Since with an appropriate operator $A$, one can write $\langle L|=\langle R|A$ and thus $\langle L|R\rangle=\langle R|A|R\rangle$, it is through this $A$ that the minus sign in (\ref{M120}) is generated, rather than through properties of the states themselves.) Thus by recognizing conformal gravity to be a $PT$ theory rather than a Hermitian one, its unitarity can then be secured. 

In addition, with the $1/M^2$ prefactor in (\ref{M120}) actually blowing up in the $M^2\rightarrow 0$ limit, Bender and Mannheim found that $M^2\rightarrow 0$ limit was singular, with the Hamiltonian associated with the pure $1/k^4$ propagator actually not being diagonalizable, but being of Jordan-block form instead.  Since the Hamiltonian is not diagonalizable, it manifestly could not be Hermitian. Thus the ghost problem in fourth-order theories only arose because one tried to treat the theory as though it was a Hermitian theory and as though one could use the standard Dirac norm. Thus the apparent generation of negative Dirac norm states indicates not that the theory violates conservation of probability, but that the Hamiltonian is not Hermitian and the Dirac norm is not the appropriate norm.

While the work of Bender and Mannheim has shown that one can construct a completely consistent quantum Hilbert space for the conformal gravity theory (which is all that matters), recently Woodard has called these results into question \cite{Woodard2015} by claiming that the construction that Bender and Mannheim used does not obey the Correspondence Principle or the usual understanding of  the $\hbar \rightarrow 0$ limit of canonical quantization. Also Woodard claims that the wave functions that Bender and Mannheim used were not normalizable. However, these claims are neither relevant nor correct. Even though the conformal gravity theory does in fact obey the Correspondence Principle (in the way we describe below), whether a theory may or may not obey the Correspondence Principle is not of concern to begin with, since the Correspondence Principle is neither a law of nature nor a complete guide as to what quantum theories are permissible, as the existence of half-integral spin for instance immediately makes manifest. And as to canonical quantization, even though it also is not a law of nature (again witness half-integral spin), Bender and Mannheim did in fact quantize the theory canonically anyway, with Poisson brackets being replaced by commutators in the usual way. What was not standard was that because the theory was a higher-derivative theory, the theory was a constrained theory, with the appropriate Poisson bracket algebra needing to be constructed by the method of Dirac constraints. And then, in order to produce a quantum-mechanical Hamiltonian whose wave functions were indeed normalizable, one had to make an allowed but not ordinarily needed complex symplectic transformation on the Poisson bracket algebra before canonical quantization. This complexification then removed the Ostrogradski instability that the theory would otherwise have had.  With this same complexification, the wave functions were then expressly shown to be normalizable.

To be more specific, we note first that when linearized around flat spacetime according to $g_{\mu\nu}=\eta_{\mu\nu}+h_{\mu\nu}$, through second order the Weyl action given in (\ref{M105}) takes the form \cite{Mannheim2011a} $I_{\rm W}(2)=-(\alpha_g/2)\int d^4x \partial_{\alpha}\partial^{\alpha} K_{\mu\nu}\partial_{\beta}\partial^{\beta} K^{\mu\nu}$, where $K_{\mu\nu}=h_{\mu\nu}-\eta_{\mu\nu}\eta^{\alpha\beta}h_{\alpha\beta}/4$, and where the gauge has been chosen so that $\partial_{\mu}K^{\mu\nu}=0$. Since there are no cross-terms between components of $K_{\mu\nu}$ in $I_{\rm W}(2)$, we can treat each component of $K^{\mu\nu}$ independently. Moreover, since the spatial behavior is not important (all that matters for dynamics is the temporal behavior), we can take the spatial dependence to be momentum modes with frequency $\omega$. On now defining $z=K_{00}$, we can write the action in the $K_{00}$ sector as $I_{\rm PU}=(\gamma/2)\int dt (\ddot{z}^2-2\omega^2\dot{z}^2+\omega^4z^2)$ where we have set $\gamma=-\alpha_g$, to thus put the action in the form of the Pais-Uhlenbeck fourth-order oscillator theory \cite{Pais1950}. To see that we have lost none of the difficulties associated with conformal gravity theory (the difficulties that were then addressed by Bender and Mannheim), we note that the propagator associated with this $I_{\rm PU}$ can immediately be shown \cite{Bender2008a,Bender2008b} to be of the generic form given in (\ref{M120}). 

Noting that  the $I_{\rm PU}$ action is a constrained action ($\dot{z}$ has to serve as the canonical conjugate of both $z$ and $\ddot{z}$), Mannheim and Davidson \cite{Mannheim2000,Mannheim2005} replaced $\dot{z}$ by an independent variable $x$ to give an action $I_{\rm PU}=(\gamma/2)\int dt (\dot{x}^2-2\omega^2x^2+\omega^4z^2)$. Given this action, they then constructed the Hamiltonian  for the theory by the method of Dirac constraints, to obtain $H_{\rm PU}=p_x^2/2\gamma+p_zx+\gamma\omega^2x^2-\gamma\omega^4z^2/2$, where $p_x$ and $p_z$ are the respective Poisson bracket conjugates of $x$ and $z$. Moreover, not only have we not as yet lost any of the difficulties of the quantum theory, the presence of the $-\gamma\omega^4z^2/2$ term in $H_{\rm PU}$ signals an Ostrogradski instability in the classical theory, with $H_{\rm PU}$ being unbounded from below when $\gamma$ is positive and $z$ is real. 

However, Bender and Mannheim pointed out that to draw such a conclusion about the theory is too hasty, since there is no justification for taking $z$ to be real (i.e. for taking $K_{\mu\nu}$ to be real), since nothing in the classical Poisson bracket algebra requires it. Specifically, once one has a classical Poisson bracket algebra, one can make symplectic transformations on it that preserve the Poisson bracket algebra, and nothing restricts these transformations to being real. As long as they do preserve the Poisson bracket algebra, they are fully allowed by classical physics. (Some examples and discussion of complex symplectic transformations that preserve the classical Poisson bracket algebra may be found in \cite{Mannheim2013b,Mannheim2015}). Now ordinarily (i.e. in classical theories that have no Ostrogradski instability), making such complex symplectic transformations has no effect on the theory, and is without any new content. However, things are different in the Pais-Uhlenbeck case, since as one transforms $z$ into the complex plane the asymptotic convergence properties of the theory can change. 

To see what explicitly happens in the Pais-Uhlenbeck case, it is very instructive to consider a path integration of the theory based on the $I_{\rm PU}$ action. To get the path integral to converge we follow the Feynman rule of replacing $\omega^2$ by $\omega^2-i\epsilon$. As noted in \cite{Bender2008b}, this then adds on to each path in the path integral  a term of the form $i\delta I_{\rm PU}=(\gamma/2)\int dt (-2x^2 \epsilon+2\omega^2 \epsilon z^2)$. While the path integral then converges for real $x$ it does not do so for real $z$. The Pais-Uhlenbeck theory path integral thus does not exist if the path integral measure is real (as is to be expected given the Ostrogradski instability). However, it would exist if $z$ (and thus its conjugate $p_z$) are taken to be pure imaginary (in which case the problematic $-\gamma\omega^4z^2/2$ term then would be bounded from below, with $H_{\rm PU}$ containing no $p_z^2$ term that would then become problematic instead). If we divide the complex plane into regions in the shape of the letter $X$, then convergence of the path integral is secured if we put the $x$ variable anywhere in the east or west quadrants of the letter $X$ in the complex $x$ plane (regions that include the real $x$ axis), and if we put the $z$ variable anywhere in the north and south quadrants in the complex $z$ plane (regions that include the pure imaginary $z$ axis). These various regions are known as Stokes wedges, with the boundaries between them being known as Stokes lines. Thus for the Pais-Uhlenbeck oscillator, as we make complex symplectic transformations on the Poisson brackets, in the $(z,p_z)$ sector we cross a Stokes line while doing so, with the convergence properties of the theory then changing radically. On one side of the Stokes line the theory has an Ostrogradski instability, but on the other side it does not.

Now as noted in \cite{Mannheim2013b,Mannheim2015}, as we make symplectic transformations on the classical Poisson brackets we can simultaneously make similarity transformations through the same angles on the quantum commutators in the quantum theory, with the classical and quantum sectors tracking each other identically. Moreover, it was noted in \cite{Bender2008a,Bender2008b} that in the quantum Pais-Uhlenbeck theory one only gets normalizable wave functions if one does indeed transform $z$ across a Stokes line in the complex $z$ plane. Thus it is only in the north or south quadrants in the complex $z$ plane that either the classical or the quantum theories exist, and in these quadrants both the classical and quantum theories are fully consistent, being free of any states with negative energy or any states with either infinite norm or negative norm  \cite{Bender2008a,Bender2008b}. Moreover, the quantization is completely canonical, with classical Poisson brackets that are symplectically transformed through a given complex angle being replaced by quantum commentators that are similarity transformed through the same complex angles. Thus in the north and south Stokes wedges one does have a Correspondence Principle and one does have canonical quantization after all. 

As regards the wave function renormalization issue, we note additionally that if one were to try to associate the propagator given in (\ref{M120}) with an indefinite metric Hilbert space, one would get non-normalizable wave functions, a fact noted by Bender and Mannheim in \cite{Bender2008a} and also by Woodard in \cite{Woodard2015}.  However, precisely because these wave functions are not normalizable, one is not free to quantize the theory with an indefinite metric, and that is what saves the theory, and is what led Bender and Mannheim to the correct Stokes wedges in which one then could consistently quantize the theory with a $PT$-theory norm that was both finite (i.e. normalizable) and non-negative.

To appreciate the point, we note that the commutator relation $[z,p_z]=i\hbar$ is left invariant under $z \rightarrow e^{i\theta}z$, $p_z \rightarrow e^{-i\theta}p_z$. Thus as well as the wave mechanics representation in which we set $z=z$ and $p_z=-i\partial_z$, we can also represent the commutator by $z=e^{i\theta}z$ and $p_z=-e^{-i\theta}i\partial_z$. Now the commutator has to act on functions $\psi(z)$, and we can write $[z,-i\partial_z]\psi(z)=i\hbar \psi(z)$ or we can write $[e^{i\theta}z,-e^{-i\theta}i\partial_z]\psi(e^{i\theta}z)=i\hbar \psi(e^{i\theta}z)$. Both of these representations are only meaningful if the wave function is a good test function. In the Pais-Uhlenbeck case $\psi(z)$ with real $z$ is not a good test function, and it is only on crossing a Stokes line that one finds a domain in the complex plane where  $\psi(e^{i\theta}z)$ is normalizable, just as is needed.

Now for gravity, it might initially appear strange that one could take the metric to be in a Stokes wedge that would permit it to be pure imaginary rather than real. However, recalling that $g^{\mu\alpha}g_{\alpha\nu}=\delta^{\mu}_{\nu}$, we see that this relation is left unchanged if we replace $g_{\alpha\nu}$ by $ig_{\alpha\nu}$ and replace $g^{\mu\alpha}$ by $-ig^{\mu\alpha}$. Moreover, since the Levi-Civita connection contains equal numbers of covariant and contravariant metric tensors, gravity is not sensitive to these replacements, with no gravity experiment to date ever having fixed the overall phase of the metric.\footnote{To fix the phase one would need an experiment involving interference with some other field such as the electromagnetic one.}  Since the Riemann tensor $R_{\lambda\mu\nu\kappa}$ with all four indices covariant has one more covariant $g_{\mu\nu}$  than it has contravariant $g^{\mu\nu}$, under  the replacement of $g_{\alpha\nu}$ by $ig_{\alpha\nu}$, $R_{\lambda\mu\nu\kappa}$ is replaced by $iR_{\lambda\mu\nu\kappa}$. In consequence $C_{\lambda\mu\nu\kappa}C^{\lambda\mu\nu\kappa}$ is replaced by $-C_{\lambda\mu\nu\kappa}C^{\lambda\mu\nu\kappa}$. Hence now, as noted in \cite{Mannheim2016a}, the Euclidean path integral will be well-behaved if rather than be positive, $\alpha_g$ is negative. Hence it is not the Euclidean theory with $\alpha_g$ positive (the case considered in \cite{Fradkin1985}, and also in \cite{Hawking2002} in an analogous situation), but rather that with $\alpha_g$ negative that has the good continuation to a ghost-free Minkowski path integral.

In his paper Woodard showed that it was impossible to fix the fourth-order gravity Ostrogradski instability problem if the spacetime metric is kept real. However, rather than not consider the theory any further, one should first check to see if it would make sense if the metric is taken to be complex, since invariance of the Poisson Bracket algebra permits such a possibility. And as we have seen, the presence of a Stokes line in the complex plane and a continuation across it is all that is needed in order to render the conformal gravity theory completely viable.

With conformal gravity thus being a consistent quantum theory of gravity, one expressly constructed in the four spacetime dimensions for which there is observational evidence,  one does not need to resort to the string theory formulation of quantum gravity. Thus one has no need for supersymmetry (or for extra dimensions for that matter) that are so key to string theory. Also, since conformal gravity has no need to utilize the interplay between spacetime and the fermionic supercharges of supersymmetry that is central to string theory, it has no need to find a way to evade the Coleman-Mandula theorem that would forbid any such interplay for bosonic charges. Moreover, if there is no supersymmetry in nature and if there are no extra dimensions, than rather than being a possible theory of everything, string theory would become a theory of more than everything, containing far more ingredients than there then would be absent supersymmetry and extra dimensions. The economical nature of conformal gravity as a quantum theory of gravity is that by not requiring extra dimensions or supersymmetry, it requires no new spacetime dimensions and requires no new elementary particles beyond those that are already known.

\subsection{Conformal Gravity and the Cosmological Constant Problem}

If one takes the mean-field Lagrangian and couples it to geometry, then just as in the one loop (\ref{L23}) where the mean-field order parameter was found to be coupled to an axial gauge field, on evaluating the analogous one loop fermion Feynman diagram in an external gravitational field, one finds \cite{tHooft2010a,tHooft2010b,tHooft2011} that (cf.  (\ref{M141}) below) the coupling in this case is that of a conformally coupled field, viz. $\partial_{\mu}m(x)\partial^{\mu}m(x)/2-m^2(x)R^{\alpha}_{\phantom{\alpha}\alpha}/12$. However, since we are in a conformal theory, we can make a conformal transform that would bring $m(x)$ in the effective Higgs Lagrangian of (\ref{L56}) to a constant. Thus at $m=M=\mu$, and with $\hbar=1$, when coupled to geometry the effective Higgs Lagrangian of (\ref{L56}) takes the form
\begin{eqnarray}
&&{\cal{L}}_{\rm EFF}=\frac{M^4}{16\pi^2}-\frac{M^2}{512\pi }R^{\alpha}_{\phantom{\alpha}\alpha}.
\label{M121}
\end{eqnarray}

In its coupling to $M^2$  the Ricci scalar appears with the opposite sign to the sign that appears in the Einstein-Hilbert action (compare (\ref{M109}) and (\ref{M141}) below). This then leads to repulsive rather than attractive gravity. In the conformal theory attractive Newtonian gravity arises not from this term but from the $W_{\mu\nu}$ term \cite{Mannheim1989,Mannheim1994}. Since  the Weyl tensor $C_{\lambda\mu\nu\kappa}$ and the conformal gravitational tensor $W_{\mu\nu}$ both vanish in geometries such as Robertson-Walker that are conformal to flat, $W_{\mu\nu}$ plays no role in cosmology, to thus allow cosmological gravity to be repulsive and local gravity to be attractive. This fact was capitalized on in \cite{Mannheim1992} to show that a repulsive cosmological gravity would have no flatness problem. And in \cite{Mannheim1996} it was shown that the theory would have no horizon problem, and that in such a cosmology there would be some cosmic repulsion that would lower the current era value $q_0$ of the deceleration parameter with respect to its value in standard attractive gravity. Specifically in \cite{Mannheim1996} it was shown that even without the $M^4$ term this would reduce $q_0$ from its pure matter inflationary universe value of $q_0=1/2$ to $q_0=0$.  When the cosmological term is included the matter energy-momentum tensor takes the form 
\begin{eqnarray} 
T^{\mu\nu}_{\rm M}&=&i\hbar \bar{\psi}\gamma^{\mu}\partial^{\nu}\psi -\frac{M^2}{256\pi}\left(R^{\mu\nu}-\frac{g^{\mu\nu}}{2}R^{\alpha}_{\phantom{\alpha}\alpha}\right)
-g^{\mu\nu} \frac{M^4}{16\pi^2}.
\label{M122}
\end{eqnarray}
In (\ref{M122})  it is understood that, as discussed above, now only the positive frequency components of the fields are to appear, and not the full vacuum contribution -- i.e. just the finite part of the energy-momentum tensor as given in (\ref{M119}). Then, no matter how big $M$ might be, it was shown \cite{Mannheim2001,Mannheim2003} that $q_0$ was obliged to lie in the narrow range $-1\leq q_0\leq0$, with  the associated luminosity distance $d_{\rm L}$ versus redshift  $z$ relation being  of the form 
\begin{eqnarray} 
d_L=-\frac{c}{H_0}\frac{(1+z)^2}{q_0}\left(1-\left[1+q_0-
\frac{q_0}{(1+z)^2}\right]^{1/2}\right),
\label{M123}
\end{eqnarray}
where $H_0$ is the current value of the Hubble parameter. With $q_0=-0.37$, (\ref{M123}) was found  \cite{Mannheim2003} to provide every bit as good a fit  to the accelerating universe data \cite{Riess1998,Perlmutter1999,Riess2004} as the standard $\Omega_{\rm M}=0.3$, $\Omega_{\Lambda}=0.7$ paradigm. However the fit provided by (\ref{M123}) requires no dark matter or fine tuning at all, with the acceleration coming from the negative effective Newton constant and the negative spatial curvature $k$ that conformal cosmology  possesses. (More technically, it is not that dark matter is excluded, it is just that the contribution of any matter, dark or even luminous, to current Hubble plot era cosmic evolution is highly suppressed in conformal cosmology.) Moreover, conformal cosmology continues  to be accelerating at higher redshift and thus requires none of the fine tuning that would make the standard cosmology only be accelerating at late redshifts. Thus at higher redshift the Hubble plots associated with conformal cosmology and standard cosmology will differ markedly, a potentially testable diagnostic. Finally, we note that with there being no need for dark matter in conformal cosmology, there is no need for supersymmetry to provide any dark matter candidates (not that supersymmetry is currently known to naturally lead to $\Omega_{\rm M}=0.3$, or to $\Omega_{\Lambda}=0.7$ for that matter when it does so).

\subsection{Conformal Gravity and the Dark Matter Problem}

While the Weyl tensor vanishes in geometries that are homogeneous and isotropic, as soon as one introduces localized sources the homogeneity is lost and $W^{\mu\nu}$ of (\ref{M106}) is no longer zero. Despite its somewhat formidable appearance Mannheim and Kazanas \cite{Mannheim1989,Mannheim1994} were able to determine its form exactly and to all orders in classical geometries that are only spherically symmetric about a single point. In particular they found that $B(r)=-g_{00}(r)$ exactly obeys the fourth-order Poisson equation
\begin{eqnarray} 
\nabla^4B(r)=\frac{3}{4\alpha_gB(r)}(T^0_{\phantom{0}0}-T^r_{\phantom{r}r})=f(r).
\label{M124}
\end{eqnarray}
The general solution to this equation is given by 
\begin{eqnarray} 
B(r)&=&-\frac{1}{6}\int_0^r dr^{\prime}f(r^{\prime})\left(3r^{\prime 2}r+\frac{r^{\prime 4}}{r}\right) 
-\frac{1}{6}\int_r^{\infty} dr^{\prime}f(r^{\prime})(3r^{\prime 3}+r^{\prime }r^2)+B_0(r),
\label{M125}
\end{eqnarray}
where $B_0(r)$ obeys $\nabla^4B_0(r)=0$. Since the integration in (\ref{M125}) extends all the way to $r=\infty$, the $B(r)$ potential receives contributions from material both inside and outside any system of interest. According to (\ref{M125}),  a star of radius $r_0$ produces an exterior potential of the form $V^*(r>r_0)=-\beta^*c^2/r+\gamma^*c^2 r/2$ per unit solar mass of star. We thus recover the Newtonian potential while finding that the potential gets modified at large distances, i.e. at precisely the distances where one has to resort to dark matter.\footnote{Recognizing this potential to be in the form of a confining linear potential, and recalling that conformal gravity is perturbatively asymptotically free \cite{Tomboulis1980,Fradkin1985}, we see that conformal gravity has quite a bit of the structure of Yang-Mills theories.} Integrating the  $V^*(r)$ potential over a thin disk of stars with a surface brightness $\Sigma(R)=\Sigma_0\exp(-R/R_0)$ with scale length $R_0$ (the typical configuration for the stars in a spiral galaxy) yields the net local potential produced by the stars in the galaxy itself, and leads to a locally generated contribution to galactic circular velocities of the form \cite{Mannheim2006}
\begin{eqnarray}
v_{{\rm LOC}}^2=
\frac{N^*\beta^*c^2 R^2}{2R_0^3}\bigg{[}I_0\left(\frac{R}{2R_0}
\right)K_0\left(\frac{R}{2R_0}\right)
-I_1\left(\frac{R}{2R_0}\right)
K_1\left(\frac{R}{2R_0}\right)\bigg{]}
+\frac{N^*\gamma^* c^2R^2}{2R_0}I_1\left(\frac{R}{2R_0}\right)
K_1\left(\frac{R}{2R_0}\right),
\label{M126}
\end{eqnarray} 
where $N^*$ is the number of stars in the galaxy. 

There are two contributions due to material outside the galaxy, i.e. due to the rest of the universe. The first is a linear potential term with coefficient $\gamma_0/2=(-k)^{1/2}$ coming from cosmology (associated with the $B_0(r)$ term, and due to writing a comoving Robertson-Walker geometry with negative curvature in the rest frame coordinate system of the galaxy).  The second arises from the integral from $r$ to $\infty$ term in (\ref{M125}) due to cosmological inhomogeneities such as clusters of galaxies, and is of a quadratic potential form with coefficient $\kappa$. When all these contributions are combined, the total circular velocities are given by
\begin{eqnarray} 
v_{{\rm TOT}}^2=v_{{\rm LOC}}^2+\frac{\gamma_0c^2 R}{2}-\kappa c^2 R^2.
\label{M127}
\end{eqnarray}
Mannheim and O'Brien \cite{Mannheim2011b,Mannheim2012b,OBrien2012,Mannheim2013} have applied this formula to the rotation curves of a set of 141 different galaxies and found very good fitting with parameters
\begin{eqnarray}
\beta^*&=&1.48\times 10^5 {\rm cm},\qquad \gamma^*=5.42\times 10^{-41} {\rm cm}^{-1},\qquad
\nonumber\\
\gamma_0&=&3.06\times
10^{-30} {\rm cm}^{-1},\qquad \kappa = 9.54\times 10^{-54} {\rm cm}^{-2},
\label{M128}
\end{eqnarray} 
with no dark matter being needed. Thus even though there is only one free parameter per galaxy, viz. $N^*$, a parameter that is common to all galactic rotation curve fits, and even though there is basically no flexibility, (\ref{M127}) fully captures the essence of the data.

We should note that it was not the dark matter problem the first got the present author interested in conformal gravity. Rather, it was because conformal gravity possessed a symmetry that forbade the presence of any the cosmological constant term at the level of the starting Lagrangian \cite{Mannheim1990}. Moreover, Mannheim and Kazanas set out with the quite limited objective of trying to see whether a theory that was not based on the Einstein-Hilbert action could still lead to a Newtonian potential. It was only on solving the  conformal gravity theory in a static, spherically symmetric geometry that they discovered that the theory not only did indeed support a Newtonian potential, it was accompanied by a linear potential term that they had not anticipated. That this linear potential could then be used to eliminate the need for galactic dark matter is therefore quite non-trivial.

We should also note that in contrast to the conformal gravity fits, dark matter fits to this same set of 141 galaxies requires 282 additional free parameters, viz. two free parameters for each galactic dark matter halo. Now dark matter theory does provide generic forms for the shapes of the halos \cite{Navarro1996,Navarro1997}, but each halo has two free numerical parameters, parameters which for the moment have to be phenomenologically determined by the fitting itself. Thus, with there being no need for dark matter in conformal gravity fits to galactic rotation curves, we again note that there is no need for supersymmetry to provide any dark matter candidates (not that supersymmetry is anyway currently known to naturally lead to values for any of the 282  free halo parameters). Finally, since both (\ref{M123}) and (\ref{M127}) do capture the essence of the astrophysical data to which they were applied, then, if  supersymmetry, dark matter theory, and even string theory, are to be correct, they should be able to derive these formulas for themselves.

We would also like to note that even if a supersymmetric particle is discovered at the LHC, this would not necessarily solve the dark matter problem. Specifically, the so far unsuccessful underground dark matter searches have identified a fairly large exclusion zone in supersymmetric cross section versus supersymmetric mass  plots. For any supersymmetric particles discovered at the LHC to be dark matter they would have to not fall in this exclusion zone, and would, of course, then have to be found in the allowed region.  

As regards conformal gravity, if it is to supplant dark matter then it will have to successfully describe astrophysical phenomena such as  gravitational lensing and the anisotropy structure of the cosmic microwave background. The study of conformal cosmological fluctuation theory given in \cite{Mannheim2012c} provides a first step in this direction. Also, in this same paper a listing of the  challenges that the conformal gravity theory currently faces may be found.

\subsection {Conformal Invariance and the Metrication and Unification of the Fundamental Forces}

With string theory with its supersymmetric underpinnings being capable of addressing both a metrication of all the fundamental forces and a unification of them, it is of interest to see how conformal symmetry fares on these issues where there is no supersymmetry to appeal to and no way to evade the constraints of the Coleman-Mandula theorem. We discuss first metrication, and we shall follow the recent discussion given in \cite{Mannheim2014,Mannheim2016b} where the effects of some generalized geometric connections were considered.

In the presence of some generalized geometric connection $\tilde{\Gamma}^{\lambda}_{\phantom{\alpha}\mu\nu}=\Lambda^{\lambda}_{\phantom{\alpha}\mu\nu}+\delta{\Gamma}^{\lambda}_{\phantom{\alpha}\mu\nu}$ where $\Lambda^{\lambda}_{\phantom{\alpha}\mu\nu}$ is the standard Levi-Civita connection
\begin{eqnarray}
\Lambda^{\lambda}_{\phantom{\alpha}\mu\nu}=\frac{1}{2}g^{\lambda\alpha}(\partial_{\mu}g_{\nu\alpha} +\partial_{\nu}g_{\mu\alpha}-\partial_{\alpha}g_{\nu\mu}),
\label{M129}
\end{eqnarray}
one introduces a generalized spin connection of the form 
\begin{eqnarray}
-\tilde{\omega}_{\mu}^{ab}=-\omega_{\mu}^{ab}+V^{b}_{\lambda}\delta{\Gamma}^{\lambda}_{\phantom{\alpha}\nu\mu}V^{a \nu},
\label{M130}
\end{eqnarray}
where the $V^{a \nu}$ are vierbeins and $\omega_{\mu}^{ab}$ is given by
\begin{eqnarray}
-\omega_{\mu}^{ab}=V^b_{\nu}\partial_{\mu}V^{a\nu}+V^{b}_{\lambda}\Lambda^{\lambda}_{\phantom{\alpha}\mu\nu}V^{a \nu}=\omega_{\mu}^{ba}.
\label{M131}
\end{eqnarray}
In terms of this generalized connection the Dirac action for a massless fermion takes the form
\begin{eqnarray}
I_{\rm D}&=&\frac{1}{2}\int d^4x(-g)^{1/2}i\bar{\psi}\gamma^{a}V^{\mu}_a(\partial_{\mu}+\Sigma_{bc}\tilde{\omega}^{bc}_{\mu})\psi+{\rm H.~c.},
\nonumber\\
\label{M132}
\end{eqnarray}
where $\Sigma_{ab}=(1/8)(\gamma_a\gamma_b-\gamma_b\gamma_a)$ and the $\gamma_a$ refer to a fixed frame.\footnote{As noted in \cite{Mannheim2015}, one should in general replace the Hermitian conjugate term (H. c.) by the $CPT$ conjugate. However, this will have no effect on the results presented here.}

Consider now a  $\delta{\Gamma}^{\lambda}_{\phantom{\alpha}\mu\nu}$ of the form 
\begin{eqnarray}
\delta{\Gamma}^{\lambda}_{\phantom{\alpha}\mu\nu}&=&-\frac{2i}{3}g^{\lambda\alpha}\left(g_{\nu\alpha}A_{\mu} +g_{\mu\alpha}A_{\nu}-g_{\nu\mu}A_{\alpha}\right)
+\frac{1}{2}g^{\lambda\alpha}(Q_{\mu\nu\alpha}+Q_{\nu\mu\alpha}-Q_{\alpha\nu\mu}).
\label{M133}
\end{eqnarray}
Here $Q^{\lambda}_{\phantom{\alpha}\mu\nu}=\Gamma^{\lambda}_{\phantom{\alpha}\mu\nu}-\Gamma^{\lambda}_{\phantom{\alpha}\nu\mu}=-Q^{\lambda}_{\phantom{\alpha}\nu\mu}$ is the antisymmetric Cartan torsion tensor. With $A_{\mu}$ being a vector field, the $A_{\mu}$-dependent connection term is essentially the connection first introduced by Weyl, differing from it only  through the presence of the additional factor of $i$, a factor that enforces $PT$ and $CPT$ symmetry and is crucial for metrication  \cite{Mannheim2014}. Following some algebra the insertion of the full $\tilde{\omega}_{\mu}^{ab}$ into the Dirac action is found to lead to the action \cite{Mannheim2014}
\begin{eqnarray}
I_{\rm D}&=&\int d^4x(-g)^{1/2}i\bar{\psi}\gamma^{a}V^{\mu}_a(\partial_{\mu}+\Sigma_{bc}\omega^{bc}_{\mu}
-iA_{\mu} -i\gamma_5S_{\mu})\psi,
\label{M134}
\end{eqnarray}
where 
\begin{eqnarray}
S^{\mu}&=&\frac{1}{8}(-g)^{-1/2}\epsilon^{\mu\alpha\beta\gamma}Q_{\alpha\beta\gamma}.
\label{M135}
\end{eqnarray}
We recognize $I_{\rm D}$ as describing none other than a fermion coupled to a standard Levi-Civita based spin connection and to chiral electromagnetism. Thus through the use of the generalized spin connection we are able to provide a purely geometric origin for both vector and axial-vector gauge fields. 

Apart from possessing full local vector gauge symmetry [$\psi(x)\rightarrow e^{i\alpha(x)}\psi(x)$, $A_{\mu}(x)\rightarrow A_{\mu}(x)+\partial_{\mu}\alpha(x)$] and full local axial-vector gauge symmetry [$\psi(x)\rightarrow e^{i\gamma_5 \alpha(x)}\psi(x)$, $S_{\mu}(x)\rightarrow S_{\mu}(x)+\partial_{\mu}\alpha(x)$], the action in (\ref{M134}) has another local invariance, namely local conformal invariance, with it being left invariant under  $g_{\mu\nu}(x)\rightarrow e^{2\alpha(x)}g_{\mu\nu}(x)$, $V_{\mu}^a(x)\rightarrow e^{\alpha(x)}V^a_{\mu}(x)$, $\psi(x)\rightarrow e^{-3\alpha(x)/2}\psi(x)$, $A_{\mu}(x)\rightarrow A_{\mu}(x)$, $S_{\mu}(x)\rightarrow S_{\mu}(x)$. (Each of these local transformations has its own $\alpha(x)$ of course.) 
In addition, as noted in \cite{Mannheim2014}, the action also possess two discrete symmetries, namely $PT$ and $CPT$ symmetry, with the factor of $i$ in (\ref{M133}) being needed to secure these invariances for the $A_{\mu}$-dependent 
sector.\footnote{As noted in \cite{Mannheim2014}, if we were not to include the factor of $i$ in (\ref{M133}), the 
$A_{\mu}$-dependent piece of the connection would not couple in the generalized Dirac action at all.}

The extension to the non-Abelian case is direct. If for instance we put the fermions into the fundamental representation of  $SU(N)\times SU(N)$ with $SU(N)$ generators $T^i$ that obey  $[T^i,T^j]=if^{ijk}T^k$, replace $A_{\mu}$ by $g_VT^iA^{i}_{\mu}$, replace $Q_{\alpha\beta\gamma}$ by $g_AT^iQ^i_{\alpha\beta\gamma}$, and thus replace $S_{\mu}$ by $g_AT^iS^{i}_{\mu}$ in the connections, we obtain a locally $SU(N)\times SU(N)$ invariant Dirac action of the form 
\begin{eqnarray}
I_{\rm D}&=&\int d^4x(-g)^{1/2}i\bar{\psi}\gamma^{a}V^{\mu}_a(\partial_{\mu}+\Sigma_{bc}\omega^{bc}_{\mu}
-ig_VT^iA^i_{\mu} -ig_A\gamma_5T^iS^i_{\mu})\psi. 
\label{M136}
\end{eqnarray}
This action is precisely a local chiral Yang-Mills action, and remains locally conformally invariant under $g_{\mu\nu}(x)\rightarrow e^{2\alpha(x)}g_{\mu\nu}(x)$, $V_{\mu}^a(x)\rightarrow e^{\alpha(x)}V^a_{\mu}(x)$, $\psi(x)\rightarrow e^{-3\alpha(x)/2}\psi(x)$, $A^i_{\mu}(x)\rightarrow A^i_{\mu}(x)$, $S^i_{\mu}(x)\rightarrow S^i_{\mu}(x)$, while still being $PT$ and $CPT$ invariant as well. Since the action given in (\ref{M136}) is the standard action that is used to describe the coupling of fermions to Yang-Mills fields and to standard Riemannian geometry, it is the action that is used in particle physics all the time. It thus has a dual characterization -- it can be generated via local gauge invariance or via a generalized geometric connection.

To obtain the form of the kinetic energy operator for the gauge fields and the metric we perform a path integration over the fermion fields (equivalent to a one fermion loop Feynman diagram) using the above Dirac action, to obtain an effective action whose leading term is
\begin{eqnarray}
I_{\rm EFF}&=&\int d^4x(-g)^{1/2}C\bigg{[}\frac{1}{20}\left[R_{\mu\nu}R^{\mu\nu}-\frac{1}{3}(R^{\alpha}_{\phantom{\alpha}\alpha})^2\right]
+\frac{1}{3}G_{\mu\nu}^iG^{\mu\nu}_i+\frac{1}{3}S_{\mu\nu}^iS^{\mu\nu}_i\bigg{]},
\label{M137}
\end{eqnarray}
where $C$ is a log divergent constant.\footnote{The vector piece of $I_{\rm EFF}$ may be found in \cite{tHooft2010a} and the axial-vector piece may be found in \cite{Shapiro2002} and in (\ref{L23}) above.} In (\ref{M137}) we recognize the conformal gravity action with the  $R_{\mu\nu}R^{\mu\nu}-(1/3)(R^{\alpha}_{\phantom{\alpha}\alpha})^2$ term being evaluated here with the Levi-Civita connection alone, and with the rest of the generalized connection emerging as the gauge field sector of a chiral Yang-Mills action.  Thus even though we start with a non-Riemannian connection we finish up with a strictly Riemannian geometry, with all of the non-Riemannian structure  being buried in the gauge fields. As noted in \cite{Mannheim2014}, the reason for this is that a generalized Riemann or Weyl tensor built out of the generalized connection would not be locally conformal invariant, since neither $A_{\mu}(x)$ nor $S_{\mu}(x)$ transform at all under a local conformal transformation. Hence the only allowed action in the pure geometric sector is that based on the Weyl tensor as constructed from the standard Levi-Civita connection alone, with the fermion path integration with a conformal invariant Dirac action having no choice but to produce it in (\ref{M137}).\footnote{As noted in \cite{Mannheim2016b}, since we generate (\ref{M137}) from (\ref{M136}) by a one-loop Feynman diagram, and since one cannot change the Hilbert space in perturbation theory, either the theories based on (\ref{M136}) and (\ref{M137}) both have ghosts or neither does. But  (\ref{M136}) is the standard ghost-free action used in particle physics all the time. Hence the theory based on (\ref{M137}) must be ghost-free, something that was shown in \cite{Bender2008a,Bender2008b} to actually be the case.}

Moreover, now that we have established the generic form needed for the gauge and metric sectors of the theory, and have seen that in this sector there are no cross-terms between any of the various connections in $\tilde{\Gamma}^{\lambda}_{\phantom{\alpha}\mu\nu}$, we now augment the Dirac action with a fundamental  Yang-Mills gauge field ($I_{\rm YM}$) action and a conformal ($I_{\rm W}$) metric sector action of the form
\begin{eqnarray}
&&I_{\rm W}+I_{\rm YM}=\int d^4x(-g)^{1/2}\bigg{[}-2\alpha_g\bigg{(}R_{\mu\nu}R^{\mu\nu}
-\frac{1}{3}(R^{\alpha}_{\phantom{\alpha}\alpha})^2\bigg{)}-\frac{1}{4}G_{\mu\nu}^iG^{\mu\nu}_i-\frac{1}{4}S_{\mu\nu}^iS^{\mu\nu}_i\bigg{]}.
\label{M138}
\end{eqnarray}
This action not only respects full conformal and gauge symmetry, like $I_{\rm D}$ it has a dual characterization -- it  can be generated via local gauge invariance or via a generalized geometric connection. Finally, on adding an $SU(N)\times SU(N)$ invariant four-fermion action 
\begin{eqnarray}
I_{\rm FF}&=&-\int d^4x(-g)^{1/2}\frac{g_{\rm FF}}{2}\bigg{[}\bar{\psi}T^i\psi\bar{\psi}T^i\psi
+\bar{\psi}i\gamma_5T^i\psi\bar{\psi}i\gamma_5T^i\psi\bigg{]},
\label{M139}
\end{eqnarray}
with coupling $g_{\rm FF}$, we can write down the fundamental action for a conformal invariant universe, viz.
\begin{eqnarray}
I_{\rm UNIV}=I_{\rm D}+I_{\rm W}+I_{\rm YM}+I_{\rm FF}.
\label{M140}
\end{eqnarray}
If the dynamics associated with (\ref{M140}) leads to critical scaling and an $I_{\rm FF}$ with dynamical dimension equal to four, the $I_{\rm UNIV}$ action will then provide a fully renormalizable and consistent action for the universe in which all mass is generated in the vacuum by dynamical symmetry breaking.

In addition, we noted that since (\ref{M136}) is the standard action used in physics, the effective action given in (\ref{M137}) must always appear in particle physics at the one fermion loop level.  Now, as noted in \cite{tHooft2010a}, radiative loops due to other standard fields such as scalars and gauge bosons yield a log divergence of the same sign, and thus the fermionically-generated $I_{\rm EFF}$ could not be canceled by other fundamental fields. The infinity in (\ref{M137}) is thus an infinity that supersymmetry could not cancel. 

However, this infinity could be canceled via conformal invariance, and this can be done in two ways. Specifically, since  $I_{\rm W}+I_{\rm YM}$ is fully renormalizable, one could cancel the $C$ term directly by a renormalization counterterm. However, the $C$ term  could also be cancelled non-perturbatively if there is critical scaling. Specifically, we recall that, unlike  the scalar $\Pi_{\rm S}(x)=\langle \Omega|T(\bar{\psi}(x)\psi(x)\bar{\psi}(0)\psi(0))|\Omega\rangle$, in quantum electrodynamics higher-order radiative corrections to the vacuum polarization $\Pi_{\mu\nu}(x)=\langle \Omega|T(\bar{\psi}(x)\gamma_{\mu}\psi(x)\bar{\psi}(0)\gamma_{\nu}\psi(0))|\Omega\rangle$ do not generate higher powers of ${\rm ln}\Lambda^2$, but are all equally linear in  ${\rm ln}\Lambda^2$. With the short-distance behavior of the theory being conformal invariant, one can write (see e.g. \cite{Mannheim1975b}) the short-distance vacuum polarization as the dimension six quantity $\Pi_{\mu\nu}(x)=
f(\alpha){\rm Tr}[\gamma_{\mu}\slashed{x}\gamma_{\nu}(-\slashed{x})]/4\pi^4x^8=f(\alpha)(\eta_{\mu\nu}\partial_{\alpha}\partial^{\alpha}-\partial_{\mu}\partial_{\nu})(1/12\pi^4x^4)$, where $f(\alpha)$ is a power series in $\alpha$. The Fourier transform of $\Pi_{\mu\nu}(x)$ yields just a single ${\rm ln}\Lambda^2$ divergence. Hence, if the respective coefficients of all the perturbative  ${\rm ln}\Lambda^2$ terms sum to zero (viz. $f(\alpha)=0$), this divergence will be canceled completely. The condition that the coefficients do sum to zero requires the coupling constant to be a solution to the Gell-Mann-Low eigenvalue condition, viz. the critical scaling condition. And indeed this is precisely how Johnson, Baker, and Willey were able to make $Z_3$ finite. Thus in the language of perturbation theory, critical scaling uniquely fixes the needed counterterm.

Now while the same analysis would equally apply if there is critical scaling in the axial-vector sector, we have not made a similar analysis for conformal gravity. However, we note that the generation of (\ref{M137}) from (\ref{M136})  involves matrix elements of fermion loops not with scalar insertions of fermion bilinears but with vector, axial-vector and tensor insertions  instead.  Now all of these particular insertions are associated with conservation conditions, and it is thus plausible that the cancellation would hold for conformal gravity too. Then, should it indeed hold, the dynamics associated with $I_{\rm UNIV}$ would not only be renormalizable, non-perturbatively it would even be completely finite.

Moreover, if one goes further and even breaks the conformal invariance by adding a spacetime-dependent mass term $-\int d^4x(-g)^{1/2}\bar{\psi}(x)M(x)\psi(x)$ to the Dirac action, the above $I_{\rm EFF}$ remains intact while being augmented by the "mean-field" action \cite{tHooft2010a,tHooft2010b,tHooft2011}, \cite{Mannheim2012a}
\begin{eqnarray}
 I_{\rm MF}&=&\int d^4x (-g)^{1/2}C\bigg{[}-M^4(x)+\frac{1}{6}M^2(x) R^{\alpha}_{\phantom{\alpha}\alpha}
-(\partial_{\mu}+iA_{\mu})M(x)(\partial^{\mu}-iA^{\mu})M(x)\bigg{]}.
\label{M141}
\end{eqnarray}
Here $C$ is the same log divergent constant as before, with  two last terms in (\ref{M141}) needing to appear jointly in order to maintain local conformal invariance. Again, it does not appear possible for supersymmetry to cancel this infinity as the superpartners are not degenerate with the regular particles. However, as noted above, with critical scaling and $\gamma_{\theta}=-1$, the infinity in  $I_{\rm MF}$ will be canceled.

Now while it is nice to obtain an action such as (\ref{M138}), as given it could not describe the real world since there are no massless axial photons. The axial symmetry thus must be broken spontaneously, and as we have shown in this paper, that is precisely what critical scaling does when $\gamma_{\theta}(\alpha)=-1$. Thus starting from a generalized $SU(N)\times SU(N)$ invariant torsion connection we are led not only to axial-vector gauge bosons, with the non-Abelian equivalent of $\gamma_{\theta}(\alpha)=-1$ we automatically break the associated axial symmetries dynamically. Similarly, since there is only one massless photon and eight massless QCD gauge bosons, any vector symmetries other than $SU(3)\times U(1)$ must be spontaneously broken also, with (\ref{M138}) then being augmented by the mean-field terms that accompany the associated dynamical mass generation.

The  extension of our ideas to grand unified theories  of the strong, electromagnetic and weak interactions is direct since one could endow a generalized connection with all of the needed internal quantum numbers, while not generating any quantum number dependence in the pure metric sector as it is automatically based on the Levi-Civita connection alone. There is however a caveat. The conformal group that underlies conformal invariance is $SO(4,2)$ and its covering group is $SU(2,2)$. The fundamental representation of $SU(2,2)$ is a 4-dimensional spinor representation. Thus, in a conformal invariant world  all fermions must be four-component, with there thus having to be right-handed neutrinos and not just left-handed ones. Intriguingly, the need for right-handed neutrinos is precisely what we had noted earlier. Families of quarks and leptons must thus contain 16 fundamental two-component spinors and not just 15. Hence the smallest grand unified group allowed would be the anomaly-free $SO(10)$, with its fundamental spinor representation being 16-dimensional. Intriguingly, $SO(10)$ contains the chiral weak interaction group $SU(2)_{\rm L}\times SU(2)_{\rm R}\times U(1)$, the need for which we had also noted earlier.
 
If one is to have a chance to achieve coupling constant unification without supersymmetry, one needs some reasonably low lying mass scale beyond those of the standard $SU(3)\times SU(2)_{\rm L}\times U(1)$. Depending on how it is broken the grand unified group  could provide such a scale, though we should note that the $B^0_{\rm s}\rightarrow \mu^{+}+\mu^{-}$ data of \cite{Aaij2013,Khachatryan2015} leave little room for any physics beyond the standard model of any description at current energies. However, we should also note that this whole issue would be moot if the renormalized coupling constant of the grand unified group is itself at a renormalization group fixed point away from the origin. However, the coupling constants would be able to depend on the running scale  if the theory has a non-trivial fixed point for some value of the coupling constant other than the physical one, with the theory tracking to the origin at high energies because of its asymptotic freedom, while tracking to the non-trivial fixed point and spontaneously breaking the symmetry in the infrared. (An alternate possibility was noted in \cite{Mannheim1975} -- when $\gamma_{\rm \theta}(\alpha)=-1$ the fluctuations produced by the then renormalizable four-fermion residual interaction are themselves asymptotically free.)

Without supercharges one is constrained by the Coleman-Mandula theorem, and so one could not unify the strong, electromagnetic and weak interactions with gravity by embedding spacetime and internal symmetries in a common Lie algebra. However, with conformal invariance there is an alternate way to extend unification to include gravity as well.  Specifically, with fermions being in the fundamental representation of the conformal group, consider some general complex transformation on a fermion of the form $\psi \rightarrow \exp(\alpha_{\rm R}+i\alpha_{\rm I}) \psi$, with only $\alpha_{\rm I}$ carrying internal quantum numbers. Then gauging $\alpha_{\rm I}$ gives Yang-Mills while gauging $\alpha_{\rm R}$ gives conformal gravity. Thus starting from the kinetic energy of a free massless fermion in flat spacetime, on imposing all these local gaugings we obtain none other than the Dirac action given in (\ref{M136}). Hence while Yang-Mills theories are obtained by gauging the imaginary part of the phase of the fermion field, gravity is obtained by gauging its real part. In this way spacetime and gravity can be unified, with it being (\ref{M140})  that should be considered as the fundamental action for physics, an action that can be obtained either by local gauging or by geometry, an action that could serve as a candidate theory of everything.

\subsection{Final Comments}

In this paper we have presented arguments to show that conformal invariance can do as well as supersymmetry in addressing some key concerns in particle physics. We thus advocate that conformal symmetry be regarded as a symmetry that is every bit as fundamental to physics as Lorentz invariance and Poincare invariance. Specifically, conformal symmetry is the full symmetry of the light cone, and in the absence of mass all particles must move on the light cone, with conformal symmetry thus being an exact symmetry at the level of the Lagrangian if all mass generation is to come solely from the vacuum. In \cite{Mannheim2012a} we have made the case for local conformal gravity, while in \cite{tHooft2014} 't Hooft has made the case for local conformal symmetry. 

Moreover, we noted above that the action given in (\ref{M136}) is locally conformal invariant, and that a fermion path integration automatically generates the conformal gravity action. However, the action given in (\ref{M136}) is the standard fermion action that is used in particle physics. Thus both conformal invariance and conformal gravity cannot be avoided, and must play some role in physics.

In this paper we have shown that conformal gravity can address the quantum gravity, the cosmological constant, and the dark matter problems. That one theory can address three problems might seem surprising. However, all of these problems have a common origin, namely the extrapolation of the standard Einstein equations  beyond their solar system origins. Specifically, if we extrapolate the Einstein equations to galactic distances and beyond we get the dark matter problem, if we extrapolate to cosmology we get the cosmological constant problem, and if we quantize the theory and extrapolate to short distances far off the mass shell we get renormalization and zero-point problems. Since all of these problems have a common origin, they can equally have a common solution, with conformal gravity potentially being that solution since it provides a very different extrapolation. 

When the present author in the 1970s suggested that one could make the four-fermion interaction renormalizable via dynamical dimensions (as is now established \cite{Mannheim2016c}), it appeared to have the potential to provide a solution to the four-fermion theory of weak interactions that would be an alternative to the spontaneously broken gauge theory solution. However, now we see that when Yang-Mills theories are coupled to gravity, we need both, namely we need  Yang-Mills for scattering amplitudes and we need a renormalizable scalar plus pseudoscalar four-fermion interaction for the vacuum energy density. Then, when we interplay the two, with critical  scaling we find that we can generate dynamical Goldstone and dynamical Higgs bosons, just as needed for a spontaneously broken gauge theory of weak interactions, with there being no need for any elementary Higgs fields at all.

\begin{acknowledgments}
 The author wishes to thank Michael Mannheim and Candost Akkaya for their help in the preparation of the figures, and wishes to thank Dr. Joshua Erlich and Dr. Benjamin Svetitsky for some helpful comments. The author is indebted to Dr. Paul Frampton for informing him of his own work on conformality \cite{Frampton1999}.
\end{acknowledgments}

\appendix
\setcounter{equation}{0}
\def\theequation{A\arabic{equation}}
\section{The Collective Higgs Mode when the Fermion is Massive -- the Calculation}

\subsection{The Basic Equations}

In this appendix we evaluate $I_{\rm cut}$ and $I_{\rm Wick}$ as given in (\ref{L70}) and (\ref{L71}) with $q_{\mu}=(q_0,0,0,0)$. For $I_{\rm cut}$ first it is convenient to remove the $q_0$ dependence from the range of integration, and so we set $p_0=q_0\lambda/2$, $p=q_0\sigma/2$. Following some straightforward algebra, and recalling the extra minus sign in $N(q_0,p,p_0)$ as discussed above, we then obtain
\begin{eqnarray}
I_{\rm cut}&=&-\frac{4i\mu^2}{\pi^3}\int_0^1d\sigma\sigma^2\int_0^{1-\sigma}d\lambda \frac{N_{\rm cut}}{D_{\rm cut}},
\nonumber\\
N_{\rm cut}&=&-(\lambda^2-\sigma^2-1)[(\lambda^2-\sigma^2+1)^2-4\sigma^2]^{1/2}q_0^8,
\nonumber\\
D_{\rm cut}&=&256m^4\mu^4+32m^2\mu^2[(\lambda^2-\sigma^2+1)^2+4\sigma^2]q_0^4
+[(\lambda^2-\sigma^2+1)^2-4\sigma^2]^2q_0^8.
\label{A1}
\end{eqnarray}
With $I_{\rm cut}$ thus behaving as $q_0^8$ at small $q_0$, $I_{\rm cut}$ makes no contribution to the derivative at $q^2=0$ of either $\Pi_{\rm S}(q^2,m)$ (as would be needed for (\ref{L56})) or $\Pi_{\rm P}(q^2,m)$ (as would be needed for (\ref{L66})). With the $\lambda^2-\sigma^2-1$ factor in $N_{\rm cut}$ always being negative in the $I_{\rm cut}$ integration range, $I_{\rm cut}$ is then expressly proportional to a negative number times $i$. The evaluation of $I_{\rm cut}$ at arbitrary $q_0$ can be done numerically, and because of its $q_0$ behavior $I_{\rm cut}$ is quite small, taking the value $-0.000291\mu^2 i$ at the threshold where $q_0^2=2m\mu$. Numerically $I_{\rm cut}$ is found to be a monotonic function of $q_0$, and because $I_{\rm cut}$ is so so small,  we can immediately anticipate that the eventual Higgs boson resonance that we will find above the threshold will have a very narrow width.

With $I_{\rm Wick}$ being evaluated on the $p_4$ axis, for it we set
\begin{eqnarray}
N(q,p)&=&-(p_4^2+p^2+q_0^2/4)
[(p^2+p_4^2-q_0^2/4)^2+p_4^2q_0^2]^{1/2},
\nonumber\\
D(q,p)&=&[(p^2+p_4^2-q_0^2/4)^2+p_4^2q_0^2-m^2\mu^2]^2
+4m^2\mu^2(p^2+p_4^2-q_0^2/4)^2.
\label{A2}
\end{eqnarray}
On setting $p_4=r\cos\theta$, $p=r\sin\theta$, and on then setting $\cos\theta=z$, for $I_{\rm Wick}$ we obtain
\begin{eqnarray}
I_{\rm Wick}&=&\frac{2\mu^2}{\pi^3}\int_0^{\infty}dr r^3\int_0^1dz(1-z^2)^{1/2}
\left[\frac{N(q,r,z)+m^2\mu^2}{D(q,r,z)}\right],
\nonumber\\
N(q,r,z)&=&-(r^2+q_0^2/4)[(r^2-q_0^2/4)^2+r^2z^2q_0^2]^{1/2},
\nonumber\\
D(q,r,z)&=&[(r^2-q_0^2/4)^2+r^2z^2q_0^2-m^2\mu^2]^2
+4m^2\mu^2(r^2-q_0^2/4)^2.
\label{A3}
\end{eqnarray}
Inspection of $D(q,r,z)$ now shows that $D(q,r,z)$ will vanish if $r=q_0/2$, $z=m\mu/rq_0$, i.e. if $z=2m\mu/q_0^2$. Since $z$ is less than one there will always be some $r$ and some $z$ for which $D(q,r,z)$ will vanish if $q_0^2\geq 2m\mu$. We thus identify $q_0^2=q^2=2m\mu$ as a threshold, and anticipate a discontinuity in $I_{\rm Wick}$ if $q^2\geq 2m\mu$. Below we will calculate the discontinuity and show that $I_{\rm Wick}$ with its seemingly real integrand  actually develops an imaginary part when $q^2\geq 2m\mu$, and it is this imaginary part that will then cancel the pure imaginary $I_{\rm cut}$. 

Given (\ref{A3}) we can calculate its $q^2$ derivative at $q^2=0$ algebraically. The $N(q,r,z)/D(q,r,z)$ term yields $-5\mu/128\pi m$ while the $m^2\mu^2/D(q,r,z)$ term yields  $+2\mu/128\pi m$. This then yields the values  $-3\mu/128\pi m$ for $\Pi^{\prime}_{\rm S}(q^2=0) $ and $-7\mu/128\pi m$ for $\Pi^{\prime}_{\rm P}(q^2=0)$ that were used in (\ref{L56}) and (\ref{L66}) above.

For general $q^2$ it is convenient to introduce
\begin{eqnarray}
&&\alpha=\frac{(r^2-q_0^2/4)^2-m^2\mu^2}{r^2q_0^2},
\nonumber\\
&&\beta=\frac{(r^2-q_0^2/4)^2+m^2\mu^2}{r^2q_0^2},
\label{A4}
\end{eqnarray}
so that we can set
\begin{eqnarray}
N(q,r,z)&=&-(r^2+q_0^2/4)rq_0(z^2+\alpha/2+\beta/2)^{1/2},
\nonumber\\
D(q,r,z)&=&r^4q_0^4(z^4+2\alpha z^2+\beta^2).
\label{A5}
\end{eqnarray}
The substitution $z=y/(1+y^2)^{1/2}$ enables us to evaluate the $z$ integrations needed for $I_{\rm Wick}$, according to
\begin{eqnarray}
I_2&=&\int_0^1dz\frac{(1-z^2)^{1/2}}{z^4+2\alpha z^2+\beta^2}
=\int_0^{\infty} dy \frac{1}{y^4(1+2\alpha+\beta^2)+2y^2(\alpha+\beta^2)+\beta^2}
\nonumber\\
&=&\frac{\pi}{4\beta(\alpha^2-\beta^2)^{1/2}}[(\alpha+\beta^2+(\alpha^2-\beta^2)^{1/2})^{1/2}
-(\alpha+\beta^2-(\alpha^2-\beta^2)^{1/2})^{1/2}],
\label{A6}
\end{eqnarray}
and 
\begin{eqnarray}
I_1&=&\int_0^1dz\frac{(1-z^2)^{1/2}(z^2+\alpha/2+\beta/2)^{1/2}}{z^4+2\alpha z^2+\beta^2}
\nonumber\\
&=&\int_0^{\infty} dy \frac{1}{2^{1/2}(1+y^2)^{1/2}}
\frac{(2y^2+(\alpha+\beta)(1+y^2))^{1/2}}{y^4(1+2\alpha+\beta^2)+2y^2(\alpha+\beta^2)+\beta^2}
\nonumber\\
&=&-\frac{i}{8^{1/2}\beta^2(\alpha+\beta)(\alpha-\beta)^{1/2}}
[(\alpha+\beta)(\alpha^2-\beta^2)^{1/2}(F_-+F_+)
+(\alpha^2+\alpha\beta-2\beta^2)(F_--F_+)],
\label{A7}
\end{eqnarray}
where
\begin{eqnarray}
F_{\pm}=\int_0^{\phi} d\theta \frac{1}{(1-j_{\pm}\sin^2\theta)(1-k\sin^2\theta)^{1/2}},
\label{A8}
\end{eqnarray}
with 
\begin{eqnarray}
j_{\pm}=\frac{1+2\alpha+\beta^2}{\alpha+\beta^2\pm(\alpha^2-\beta^2)^{1/2}},
\qquad
\phi=i~{\rm arcsinh}(\infty),\qquad k=\frac{2+\alpha+\beta}{\alpha+\beta}.
\label{A9}
\end{eqnarray}
The substitution $\sin\theta=i\tan\nu$ allows us to rewrite (\ref{A8}) as 
\begin{eqnarray}
F_{\pm}&=&i\int_0^{\pi/2}\frac{d\nu}{\cos\nu} \frac{1}{(1+j_{\pm}\tan^2\nu)(1+k\tan^2\nu)^{1/2}}
\nonumber\\
&=&-\frac{i}{j_{\pm}-1}K(1-k)+\frac{ij_{\pm}}{j_{\pm}-1}E(1-j_{\pm},1-k),
\nonumber\\
\label{A10}
\end{eqnarray}
where $K(1-k)$ and $E(1-j_{\pm},1-k)$ are the complete elliptic integrals
\begin{eqnarray}
K(1-k)&=&\int_0^{\pi/2}d\nu \frac{1}{(1-(1-k)\sin^2\nu)^{1/2}},
\nonumber\\
E(1-j_{\pm},1-k)&=&\int_0^{\pi/2}d\nu 
\frac{1}{(1-(1-j_{\pm})\sin^2\nu)(1-(1-k)\sin^2\nu)^{1/2}}.
\label{A11}
\end{eqnarray}
Finally, in terms of all these expressions we can write $\Pi_{\rm S}(q^2,m)$ as
\begin{eqnarray}
\Pi_{\rm S}(q^2,m)&=&\frac{2\mu^2}{\pi^3}\int_0^{\infty}dr r^3\bigg{[}-\frac{(r^2+q_0^2/4)I_{1}}{r^3q_0^3}
+\frac{m^2\mu^2I_{2}}{r^4q_0^4}\bigg{]}+I_{\rm cut}.
\label{A12}
\end{eqnarray}
Then with the physical mass being given by $M$, to find any Higgs boson we need to look for zeros of the finite
\begin{eqnarray}
\hat{\Pi}_{\rm S}(q^2,M)=\Pi_{\rm S}(q^2,M)-g^{-1}.
\label{A13}
\end{eqnarray}
where $g^{-1}$ is given in (\ref{L49}).

While it does not appear to be possible to do the integration in (\ref{A12}) analytically, the utility of (\ref{A12}) is that we can extract an analytic expression for the discontinuity from it. However, before doing so we first evaluate $\hat{\Pi}_{\rm S}(q^2,M)$ below the $q^2=2M\mu$ threshold. At $q^2=0$ we can evaluate $\hat{\Pi}_{\rm S}(q^2=0,M)$ analytically to obtain the value $\mu^2/4\pi^2=0.025330\mu^2$. As we increase $q^2$, via numerical integration we find that ${\rm Re}[\hat{\Pi}_{\rm S}(q^2,M)]$ decreases monotonically, reaching a value of $0.003373\mu^2$ at $q^2=2M\mu$. We can thus anticipate that it will vanish a little beyond the threshold. 

\subsection{The Discontinuity}

As we had noted earlier, above the threshold the $I_1$ and $I_2$ terms in (\ref{A12}) becomes undefined at $r=q_0/2$. To avoid this we must either move $q_0$ off the real axis or keep $q_0$ real and deform the $r$-integration contour. To implement the former we look for  $\hat{\Pi}_{\rm S}(q^2,M)$ to vanish at some $q_0= q_{\rm R}-i\Gamma$, with a necessarily positive $\Gamma$ of dimension $(M\mu)^{1/2}$ if the Higgs boson is indeed to be a resonance. With the quantity $r^2-q_0^2/4$ that appears in $\alpha$ and $\beta$ then becoming $r^2-(q_{\rm R}-i\Gamma)^2/4$ near the resonance, to implement the more convenient latter procedure, for $q_0>q_{\rm R}$ only we split the (\ref{A12}) integral into two parts, a real part, $I_{\rm R}$, that consists of an integration involving  two intervals $r\in (0,q_{\rm R}/2-\Gamma)$ and $r\in (q_{\rm R}/2+\Gamma,\infty)$, and a complex part $I_{\rm com}$  along a semicircle in the upper half $r$ plane of radius $\Gamma $ in which $r=q_{\rm R}/2+\Gamma e^{i\theta}$ where $\theta\in (\pi,0)$. Then we can solve for the real and imaginary parts of $\hat{\Pi}_{\rm S}(q^2,M)=0$ to fix both the position and the width of the resonance at some $q_0=q_{\rm R}-i\Gamma$, $q^2=q_{\rm R}^2-\Gamma^2-2iq_{\rm R}\Gamma$.

Now before we solve for the location of the Higgs boson we do not know whether it will in fact turn out to be a narrow resonance. Thus we shall take $\Gamma $  to be small, and then self-consistently discover that the solution is one in which it is in fact small. With small $\Gamma$ we can evaluate (\ref{A12}) on the semicircle in the upper half $r$ plane by making a Taylor series expansion. With the measure for the integration on the semicircle being given  by  $dr=i\Gamma e^{i\theta}d\theta$, to lowest order in $\Gamma$ we only need  to evaluate the integrand in (\ref{A12}) to zeroth order in $\Gamma$.  On inserting $r=q_{\rm R}/2+\Gamma e^{i\theta}$ in (\ref{A7})  this yields 
\begin{eqnarray}
I_1&\rightarrow& \int_0^{\infty} dy \frac{q_{\rm R}^8y}{(1+y^2)^{1/2}(y^2(q_{\rm R}^4-4M^2\mu^2)-4M^2\mu^2)^2}
\nonumber\\
&=&\bigg{(}\frac{q_{\rm R}^4(1+y^2)^{1/2}}{2(4M^2\mu^2+(4M^2\mu^2-q_{\rm R}^4)y^2)} 
+\frac{q_{\rm R}^2}{4(q_{\rm R}^4-4M^2\mu^2)^{1/2}}
 {\rm ln}\left(\frac{q_{\rm R}^2+(q_{\rm R}^4-4M^2\mu^2)^{1/2}(1+y^2)^{1/2}}{q_{\rm R}^2-(q_{\rm R}^4-4M^2\mu^2)^{1/2}(1+y^2)^{1/2}} \right)\bigg{)}\bigg{|}_0^{\infty}
\nonumber\\
&=&\frac{i\pi q_{\rm R}^2}{4(q_{\rm R}^4-4M^2\mu^2)^{1/2}}-\frac{q_{\rm R}^4}{8M^2\mu^2}-\frac{q_{\rm R}^2}{4(q_{\rm R}^4-4M^2\mu^2)^{1/2}}
{\rm ln}\left(\frac{q_{\rm R}^2+(q_{\rm R}^4-4M^2\mu^2)^{1/2}}{q_{\rm R}^2-(q_{\rm R}^4-4M^2\mu^2)^{1/2}}\right).
\label{A14}
\end{eqnarray}
Similarly, for (\ref{A6}) we obtain 
\begin{eqnarray}
I_2&\rightarrow &\int_0^{\infty} dy \frac{q_{\rm R}^8}{(y^2(q_{\rm R}^4-4M^2\mu^2)-4M^2\mu^2)^2}
\nonumber\\
&=&\bigg{(}\frac{q_{\rm R}^8y}{8M^2\mu^2(4M^2\mu^2+(4M^2\mu^2-q_{\rm R}^4)y^2)} 
+\frac{q_{\rm R}^8}{32M^3\mu^3(q_{\rm R}^4-4M^2\mu^2)^{1/2}}
{\rm ln}\left(\frac{2M\mu+(q_{\rm R}^4-4M^2\mu^2)^{1/2}y}{2M\mu-(q_{\rm R}^4-4M^2\mu^2)^{1/2}y} \right)\bigg{)}\bigg{|}_0^{\infty}
\nonumber\\
&=&\frac{i\pi q_{\rm R}^8}{32M^3\mu^3(q_{\rm R}^4-4M^2\mu^2)^{1/2}}.
\label{A15}
\end{eqnarray}
Finally, on inserting (\ref{A14}) and (\ref{A15}) into (\ref{A12}) and doing the now trivial $\theta$ integration from $\theta=\pi$ to $\theta=0$, the real part of $I_{\rm com}$ is found to evaluate to
\begin{eqnarray}
{\rm Re}[I_{\rm com}]&=&\frac{q_{\rm R}^3\Gamma}{4\pi^3M^2} +\frac{\mu^2 q_{\rm R}\Gamma}{2\pi^3(q_{\rm R}^4-4M^2\mu^2)^{1/2}}
{\rm ln}\left(\frac{q_{\rm R}^2+(q_{\rm R}^4-4M^2\mu^2)^{1/2}}{q_{\rm R}^2-(q_{\rm R}^4-4M^2\mu^2)^{1/2}} \right),
\label{A16}
\end{eqnarray}
while the imaginary part evaluates to 
\begin{eqnarray}
{\rm Im}[I_{\rm com}]&=&\frac{i\mu\Gamma q_{\rm R}(q_{\rm R}^2-2M\mu)^{1/2}}{4\pi^2M(q_{\rm R}^2+2M\mu)^{1/2}}.
\label{A17}
\end{eqnarray}

As we see, there is an explicit  branch point at $q_{\rm R}^2=2M\mu$ in $\Pi_{\rm S}(q^2,M)$ just as we had anticipated. (There is no branch point at  $q^2=-2M\mu$  since (\ref{A12}), (\ref{A16}), and (\ref{A17}) only hold for timelike $q_{\mu}$.) The discontinuity structure exhibited  in (\ref{A16}) is reminiscent of that obtained for $\Pi_{\rm S}(q^2,M)$ in the NJL model as given in (\ref{L29}), where there is also a  threshold branch point. Also we note that even though the imaginary parts given in (\ref{A14}) and (\ref{A15}) are actually singular at the branch point, their coefficients are such that when they combine in (\ref{A17}) the singularity is canceled. Since singularities of this sort are not allowed, their cancellation in (\ref{A17}) provides a nice internal check on our calculation. With this cancellation, rather than diverge at the threshold  ${\rm Im}[I_{\rm com}]$ actually vanishes there. With $I_{\rm cut}$ not vanishing there the Higgs boson must thus lie above threshold. With $I_{\rm cut}$ and ${\rm Im}[I_{\rm com}]$ having opposite signs, a cancellation between them can thus be effected above threshold, with the resulting sign of $\Gamma$ then indeed being the positive one required by unitarity. With $q_{\rm R}$ being fixed by a cancellation between ${\rm Re}[I_{\rm Wick}]$ and ${\rm Re}[I_{\rm com}]$, the imaginary part cancellation then fixes the magnitude of $\Gamma$.

\subsection{Numerical Results}

For the actual numerical work we must evaluate not $\Pi_{\rm S}(q^2,M)$ itself but $\hat{\Pi}_{\rm S}(q^2,M)=\Pi_{\rm S}(q^2,M)-g^{-1}$, as only the latter quantity is finite. We shall use a hat notation to indicate that we now refer quantities to $\hat{\Pi}_{\rm S}(q^2,M)$ rather than to $\Pi_{\rm S}(q^2,M)$. Since we have broken the evaluation of $\hat{\Pi}_{\rm S}(q^2,M)$ into a low section, $\hat{I}_{\rm Wick}({\rm low})$, where $r<q_{\rm R}/2-\Gamma$, a high section, $\hat{I}_{\rm Wick}({\rm high})$, where $r>q_{\rm R}/2+\Gamma$, and a semicircle section $I_{\rm com}$, then since the integration that fixes $g^{-1}$ in (\ref{L48}) involves the full $r\in (0,\infty)$ range, we need to include the contribution, $\hat{I}_{\rm ggap}$,  to $g^{-1}$ in the region $r \in (q_{\rm R}/2-\Gamma,q_{\rm R}/2+\Gamma)$. This contribution is readily found to evaluate to
\begin{eqnarray}
\hat{I}_{\rm ggap}&=&\frac{2\mu^2 q_{\rm R}^3\Gamma}{\pi^2(q_{\rm R}^4+16M^2\mu^2)}.
\label{A18}
\end{eqnarray}
With this addition the full $\hat{\Pi}_{\rm S}(q^2,M)$ is given by
\begin{eqnarray}
&&\hat{\Pi}_{\rm S}(q^2,M)=\hat{I}_{\rm Wick}({\rm low})+\hat{I}_{\rm Wick}({\rm high})
+\hat{I}_{\rm ggap}+{\rm Re}[I_{\rm com}]+{\rm Im}[I_{\rm com}]+I_{\rm cut}.
\label{A19}
\end{eqnarray}
With everything now being well-defined, we can proceed to solve the condition $\hat{\Pi}_{\rm S}(q^2,M)=0$, and  numerically find that  $\hat{\Pi}_{\rm S}(q^2,M)$ vanishes at 
\begin{eqnarray}
q_{\rm R}&=&1.480(M\mu)^{1/2},\qquad\Gamma=0.017i(M\mu)^{1/2},\qquad
q^2=(2.189-0.051i)M\mu.
\label{A20}
\end{eqnarray}
In this solution the six terms  in (\ref{A19}) respectively evaluate to $0.004710$, $-0.008832$, $0.001610$, $0.002517$, $0.000406i$, $-0.000400i$ (in units of $\mu^2$), as given to six decimal places, with $\hat{\Pi}_{\rm S}(q^2,M)$ thus vanishing to five.

We thus self-consistently confirm that $q_{\rm R}$ is indeed close to threshold where $q_{\rm R}=1.414(M\mu)^{1/2}$, and that $\Gamma$ is indeed small and that its sign had correctly been chosen. Near the resonance pole $\hat{\Pi}_{\rm S}(q^2,M)$ behaves as
\begin{eqnarray}
\hat{\Pi}_{\rm S}(q^2,M)&=&(q^2-(q_{\rm R}-i\Gamma)^2)
 (-0.021662+0.000484i),
\label{A21}
\end{eqnarray}
with $T_{\rm S}(q^2,M)=1/(g^{-1}-\Pi_{\rm S}(q^2,M))$ thus having the Breit-Wigner structure
\begin{eqnarray}
T_{\rm S}(q^2,M)=\frac{46.141+1.030i}{q^2-2.2189M\mu+0.051iM\mu}.
\label{A22}
\end{eqnarray}

\end{document}